\documentclass[preprint,3p]{elsarticle}
\usepackage{amsfonts,amssymb}
\usepackage{mathtools}
\usepackage{graphicx}
\usepackage{wrapfig}
\usepackage{caption}
\usepackage{feynmp-auto}
\usepackage{comment}
\makeatletter
\newcommand*\wt[1]{\mathpalette\wthelper{#1}}
\newcommand*\wthelper[2]{%
        \hbox{\dimen@\accentfontxheight#1%
                \accentfontxheight#11.15\dimen@
                $\m@th#1\widetilde{#2}$%
                \accentfontxheight#1\dimen@
        }%
}

\newcommand*\accentfontxheight[1]{%
       \fontdimen5\ifx#1\displaystyle
                \textfont
        \else\ifx#1\textstyle
                \textfont
        \else\ifx#1\scriptstyle
                \scriptfont
        \else
                \scriptscriptfont
        \fi\fi\fi3
}
\makeatother

\newcommand*{\momentumarrow}[4]{%
    \fmfcmd{style_def marrow#1
    expr p = drawarrow subpath (0.3, 0.7) of p shifted #3 #2 withpen pencircle scaled 0.4;
    enddef;}
    \fmf{marrow#1,tension=0}{#4}}

\newcommand*{\tr}{\ensuremath{\mathbf{tr}}}
\newcommand*{\sgn}{\ensuremath{\mathrm{sgn}}}
\renewcommand*{\Im}{\ensuremath{\mathrm{Im}}}
\renewcommand*{\Re}{\ensuremath{\mathrm{Re}}}

\newcommand*{\p}{\ensuremath{{\mathbf{p}}}}
\newcommand*{\q}{\ensuremath{{\mathbf{q}}}}
\renewcommand*{\k}{\ensuremath{{\mathbf{k}}}}
\renewcommand*{\l}{\ensuremath{{\mathbf{l}}}}

\begin{document}
\unitlength = 1mm

\journal{Annals of Physics}

\begin{frontmatter}
\title{Berry Fermi Liquid Theory}
\author[uchicago,stanford]{Jing-Yuan Chen}%\corref{corrauthor}\cortext[corrauthor]{Corresponding author}\ead{chjy@uchicago.edu}
\author[uchicago]{Dam Thanh Son}%\ead{dtson@uchicago.edu}
\address[uchicago]{Kadanoff Center for Theoretical Physics, University of Chicago, Chicago, IL 60637, USA}
\address[stanford]{Stanford Institute for Theoretical Physics, Stanford University, CA 94305, USA}

\begin{abstract}
  We develop an extension of the Landau Fermi liquid theory to systems
  of interacting fermions with non-trivial Berry curvature.  We propose
  a kinetic equation and a constitutive relation for the
  electromagnetic current that together encode the linear response of
  such systems to external electromagnetic perturbations, to leading
  and next-to-leading orders in the expansion over the frequency and
  wave number of the perturbations. We analyze the Feynman diagrams
  in a large class of interacting quantum field theories and show
  that, after summing up all orders in perturbation theory, the
  current-current correlator exactly matches with the result obtained
  from the kinetic theory.
\end{abstract}


\end{frontmatter}

\tableofcontents

\

\section{Introduction}

One of the main cornerstones of condensed matter physics is Landau's
Fermi liquid theory. This theory, proposed by Landau in late
1950s~\cite{landau1957theory,landau1957oscillations}, describes
low-energy dynamics of normal Fermi liquids in terms of
quasiparticles, whose interactions are specified by a set of Landau's
parameters. The theory was constructed by Landau phenomenologically
at first, and subsequently found its theoretical justification from the
diagrammatic approach, where an analysis of the infrared singularities
of Feynman diagrams reveals, in particular, the precise connection
between the Landau parameters and the four-point vertex, evaluated in
a particular kinematic 
regime~\cite{landau1959theory,Nozieres:1962zz,Luttinger:1962zz,abrikosov1975methods}. 
More recently, Landau's Fermi liquid theory has been reinterpreted in the
language of the renormalization
group~\cite{Polchinski:1992ed,Shankar:1993pf}. Except for a possible
instability in the Bardeen-Cooper-Schrieffer (BCS) channel, Landau's
Fermi liquid theory provides a truly universal low-energy effective
description of Fermi system with short-ranged interactions.

It has been known for some time, however, that effects related to the
Berry curvature of the fermion in momentum space are beyond the
purview of Landau's theory. For reviews of these effects see
Refs.~\cite{Chang:2008zza,Xiao:2009rm}. It was shown in
Ref.~\cite{Sundaram:1999zz} that the semiclassical equation of motion
of a wave packet should involve an anomalous velocity. One of the
consequences of this modification is the anomalous Hall
effect~\cite{KarplusLuttinger}. For non-interacting fermions in (2+1)
dimensions, the fractional part of the anomalous Hall coefficient can
be related to the Berry phase that the fermion obtains when it moves
around the Fermi disk~\cite{Haldane:2004zz}. Other examples include the
chiral anomaly in (3+1) dimensions and effects associated with it, in
particular the chiral magnetic effect or the anomalous contribution to
magnetoresistance~\cite{Son:2012wh,Son:2012bg,Stephanov:2012ki,Son:2012zy}.

Most treatments of fermionic systems with Berry curvature so far
neglected the interactions between the fermions. This leaves one with
the question: which ones of the results, obtained for non-interacting
fermions, survive when interactions are taken into account? For
example, does the anomalous Hall coefficient continue to be equal to
the Berry phase of the quasiparticle around the Fermi disk, or there
are corrections due to the interactions?

In this paper, we address the question of how the Berry phase of the
fermionic quasiparticle makes appearance in the Landau's Fermi liquid
theory. We derive, by resumming all Feynman diagrams, a linearized
kinetic equation, capable of giving the linear electromagnetic
response in a Fermi liquid with Berry curvature, up to the
next-to-leading order in the expansion over momentum. This linear
theory is sufficient for the anomalous Hall effect and the chiral
magnetic effect. From the kinetic equation, we found that the
anomalous Hall coefficient does not coincide with the Berry phase, but
contains in addition to the Berry phase a contribution
coming from the emergent electric dipole moment of the quasiparticles, which we
show to be in general non-zero, as well as a contribution from the quasiparticle
interactions. In this work, we limit ourselves to
the linearized theory in the clean limit. We hope to extend the scope of the theory 
in future work; for instance, to include interaction effects  
into the previous studies of Hall response in Fermi gas in the presence of impurities 
\cite{sinitsyn2007semiclassical, sugimoto2007gauge, nagaosa2010anomalous}.

In previous literature, the work that has most overlap with ours is
Ref.~\cite{shindou2008gradient} where the interplay between Berry
curvature and interaction has been studied in a very general
context. The authors of Ref.~\cite{shindou2008gradient} showed, via
the Keldysh formalism, that the quasiparticles' motion has an
anomalous velocity due to the Berry curvature, as in the
non-interacting case, but the content of the Berry curvature is
modified by interactions. There are four main differences between
Ref.~\cite{shindou2008gradient} and this work. First, in contrast to
Ref.~\cite{shindou2008gradient}, we study linear response which
does not see the effect of the anomalous velocity in the Boltzmann equation. Rather,
as we will show, in linear response, the physical effects of Berry curvature only show up in (the
non-quasiparticle contribution to) the current, an effect not computed
in Ref.~\cite{shindou2008gradient}. Second, while both Ref.~\cite{shindou2008gradient}  and us
considered new contributions to the Hall conductivity from quasiparticle
excitations, we identify some contributions missed from Ref.~\cite{shindou2008gradient}.
Third, we are able to take into
account the effects of the quasiparticle collisions and the finite
quasiparticle lifetime, and show they do not affect the Hall response. 
Last, we are able to answer the question
whether interesting transport phenomena such as the anomalous Hall
effect involve Fermi surface contribution only, or involve Fermi sea
contribution as well.

The structure of the paper is as follows. In Section
\ref{sect_kinetic} we first review the Landau Fermi liquid theory,
more precisely, the linearized version of the theory. We then propose
a kinetic theory that would capture the full linear electromagnetic
response of a Fermi liquid with Berry curvature. In Section
\ref{sect_QFT} we show, by a careful analysis of Feynman diagrams,
that the kinetic theory reproduces correctly the linear response in
the field theory to all orders in perturbation theory. The analysis
identifies the parameters of the kinetic theory with objects in the field
theory. Section \ref{sect_discussion} contains final discussions.
Appendices A and B are devoted to details about quasiparticle
collisions. Appendices C and D contain certain technical details of
our diagrammatic analysis.

\section{The Kinetic Formalism of Berry Fermi Liquid}
\label{sect_kinetic}

The main problem considered in this paper is that of computing the
linear response of an interacting system to an external
electromagnetic (EM) field $A_\mu$ which varies slowly in space and
time. In quantum field theory, the computation of linear response
corresponds to computing the current-current correlation. Remarkably,
it is shown~\cite{landau1959theory,abrikosov1975methods} that, for
Fermi liquids in the long wavelength limit, the quantum
field-theoretical computation, to all orders of the perturbation
theory, can be arranged in a way that admits a classical
interpretation as a kinetic theory---the famous Landau Fermi liquid
theory~\cite{landau1957theory,landau1957oscillations}. The purpose of
this paper is to extend the Landau Fermi liquid theory into ``Berry
Fermi liquid theory'' which incorporate effects of Berry curvature.
Before we present the Berry Fermi liquid theory, we first briefly
review how linear response is computed in Landau Fermi liquid theory.

Let us start with a system of fermions, interacting through a
finite-ranged interaction, in $d$ spatial dimensions with $d\geq 2$ (or
$(d+1)$ spacetime dimensions). We assume the ground state at chemical
potential $\epsilon_F$ is a Fermi liquid, with a sharp Fermi surface
(FS). (We assume that the Kohn-Luttinger instability
\cite{Kohn:1965zz} occurs at an energy scale much smaller than any
scales of interest.) The low energy excitations are fermionic
quasiparticles or quasiholes near the FS. For simplicity we assume
one, non-degenerate, FS, i.e., each momentum $\k$ near the FS corresponds
to only one quasiparticle.

We now perturb this system by a small external EM field $A_\mu$.
Physically, this causes a deformation of the FS, which can also be
viewed as creating quasiparticles and quasiholes, which in Landau's
theory are described by the quasiparticle distribution function
$\delta\!f(\p; x)$ with $\p$ near the FS. In the linear response
theory we keep $\delta\! f$ to linear order of $A_\mu$.

The Landau Fermi liquid theory matches with quantum field theory at
long wavelength. If we Fourier transform $-i\hbar\partial_{x^\mu}$ to
$q_\mu$, then in the long-wavelength limit under consideration,
$A_\mu$ and $\delta f$ only have $q$ modes with $q \ll p_F$ and $q \ll
\hbar/r_{int}$, where $p_F$ is the size scale of the FS (there is no notion
of ``Fermi momentum'' since we do not assume rotational symmetry), 
and $r_{int}$ is the range of interaction between quasiparticles (this is 
why we assumed finite-ranged interactions). In practice, we keep
$\hbar\,\partial_x$, or equivalent $q$, to leading order in Landau
Fermi liquid theory.

It can be shown that the collision (decay included) rate of quasiparticles is suppressed 
beyond leading order in $q$, due to the limited availability of decay channels. 
In particular, the suppression is by an extra order of $q$ for $d\geq 3$
\cite{landau1957oscillations,Luttinger:1961zz,abrikosov1975methods}, 
and by an extra $q \ln q$ for $d=2$ 
\cite{baym1978physics,chubukov2003nonanalytic,chubukov2005singular}. Thus, quasiparticle collision 
can be neglected in Landau Fermi liquid theory.

The computation of linear response in Landau Fermi liquid theory
proceeds in two steps. One first computes $\delta\!f$ as a linear
function of $A$ by solving the Boltzmann equation, and then expresses
(the quantum expectation of) the induced current $\delta J^\mu$ as a
linear function of $\delta\!f$, and hence of $A$. In Landau's Fermi liquid
theory, the energy of a single quasiparticle has the form
\begin{eqnarray}
\epsilon(\p; x) = E(\p) + \int_\k \mathcal{U}(\p, \k) \ \delta\! f(\k; x)
\label{LFL_quasiparticle_energy}
\end{eqnarray}
where $\int_\k \equiv \int d^d k / (2\pi\hbar)^d$. Here $E(\p)$ is the kinetic energy of the
quasiparticle, and $\mathcal{U}(\p, \k)$, even under exchange of $\p$
and $\k$, parametrizes the contact interaction between two
quasiparticles of momenta $\p$ and $\k$. (If the system has rotational
symmetry, the Landau Fermi liquid parameters are obtained by putting
$\p$ and $\k$ on the Fermi surface and expanding $\mathcal U$ in
angular harmonics in the angle between $\p$ and $\k$.) Both $E$ and
$\mathcal{U}$ are microscopic inputs into Landau's theory.
Landau's Fermi liquid theory postulates a collisionless Boltzmann equation
(as we assume clean limit in this paper),
\begin{equation}
  \frac{\partial f(\p; x)}{\partial t} 
  + \frac{\partial\epsilon(\p;x)}{\partial p_i} \frac{\partial f(\p;x)}{\partial x^i}
  + \left( E_i(x) + B_{ij}(x) \frac{\partial\epsilon(\p;x)}{\partial p_j} - \frac{\partial\epsilon(\p;x)}{\partial x^i}\right)
     \frac{\partial f(\p;x)}{\partial p_i} = 0.
\end{equation}
where $E_i=F_{i0}=\partial_{x^i} A_0-\partial_{t} A_i$ is the electric field, $B_{ij}=F_{ij}=\partial_{x^i} A_j-\partial_{x^j} A_i$ is the magnetic field, and we have absorbed the electric charge into the field potential $A$. Writing $f(\p;x)=\theta(\epsilon_F-E(\p)) + \delta\!f(\p;x)$ and linearizing over $\delta\!f$ and $A$, one finds
\begin{eqnarray}
v^\mu(\p) \: \partial_{x^\mu} \delta\! f(\p; x) = \delta(\epsilon_F-E(\p)) \: v^i(\p) \: \left(F_{i0}(x) - \partial_{x^i} \epsilon(\p; x)\right)
\label{LFL_Boltzmann_Eq_coordinates}
\end{eqnarray}
where $v^0 \equiv 1$, $v^i(\p) \equiv \partial_p^i E(\p) \equiv \partial E(\p)/\partial p_i$, and $x^0\equiv t$. Notice that, due to the delta function on the right hand side, Eq.~\eqref{LFL_Boltzmann_Eq_coordinates} involves only the FS, but not, say, the whole Fermi sea. Performing the Fourier transformation $-i\partial_{x^\mu} \rightarrow q_\mu$, where, in our convention, $-q_0=q^0$ is the energy, while $q_i=q^i$ is the momentum, the Boltzmann equation then reads
\begin{eqnarray}
\delta\!f(\p; q) = \delta(\epsilon_F-E) \frac{v^i}{v^\mu q_\mu - i\epsilon} \left( -iF_{i0}(q) - q_i \int_\k \mathcal{U}(\p, \k) \: \delta\!f(\k; q) \right)
\label{LFL_Boltzmann_Eq}
\end{eqnarray}
where $F_{\mu\nu}(q) = 2iq_{[\mu} A_{\nu]}$. This is an integral equation
from which one can find $\delta\!f$ in terms of $A$. It follows from
Eq.~(\ref{LFL_Boltzmann_Eq}) that the the coefficient of linear dependence
between $\delta\!f$ and $A$ is finite in the limit $q\to 0$ and $|\q|/q^0$ fixed. 
In this paper we count this as \emph{zeroth} order (leading order) in $q$. Note that we placed an $i\epsilon$ prescription in the denominator; its sign is such that $q^0$ appears as $q^0+i\epsilon$. This corresponds to the retarded boundary condition that at infinite past the system is in its ground state.

Now suppose we have solved for $\delta f$ as a linear function of $A$ from \eqref{LFL_Boltzmann_Eq}. Then the induced current in Landau Fermi liquid theory is given by
\begin{eqnarray}
\delta J^\mu(x) = \int_\p \left( v^\mu(\p) \: \delta\!f(\p; x) + \delta(\epsilon_F - E(\p)) \delta^\mu_i v^i(\p) \: (\epsilon(\p; x)-E(\p)) \right).
\label{LFL_current}
\end{eqnarray}
The first term is simply the current created by the quasiparticles that were excited. The second term, by recognizing $\delta(\epsilon_F-E) \: v^i = -\partial_p^i \theta(\epsilon_F-E)$ and integrating by parts over $\p$, is the current due to quasiparticles in the Fermi sea having their velocity perturbed by interactions with the excited quasiparticles $\partial_p^i (\epsilon-E)$. (Although ``quasiparticles in the Fermi sea'' are generally not well-defined far from the FS, from the expression \eqref{LFL_current} we clearly see only those quasiparticles near the FS are involved.) This is the procedure of computing linear response in Landau Fermi liquid theory.

The Landau Fermi liquid theory has achieved great success in
describing various physical phenomena. However, there are many
interesting physical phenomena related to the effects of Berry
curvature that are beyond the scope of Landau Fermi liquid theory.
One of such phenomena is the anomalous Hall effect. It is easy to
see, however, that the anomalous Hall effect is formally one order (in 
external momentum) higher than those included in the standard Landau Fermi liquid
theory. Indeed, in the anomalous Hall effect, $\delta J_H \sim \sigma_H
F$ where $\sigma_H$ is independent of $q$ and $F$ is linear in $q$. This
requires linear response computation at first order in $q$, yet in the
conventional Landau Fermi liquid theory the responses are computed to
zeroth order in $q$. Thus, if one wants to extend the Fermi liquid
theory to include effects of the Berry curvature, to be consistent
one needs to include all effects to first order in external momentum (in particular,
quasiparticle collisions also need to be taken into account, but as we will see, they are ``uninteresting''). We will
call this extended kinetic formalism the ``Berry Fermi liquid
theory''.

The kinetic formalism of Berry Fermi liquid theory consists of two
parts: the Boltzmann equation, and the expression of the current in
terms of the distribution function. We will also find that the
consistency of the theory requires certain relationships between the
chemical potential dependence of the Fermi velocity and the Landau interaction 
potential, and between the chemical potential dependence of the
Hall conductivity tensor (to be defined later) and the Berry curvature of
the fermionic quasiparticle. We will present this formalism in this
Section. In the next Section, we will show this kinetic formalism
exactly matches with quantum field theory (QFT) computation to all
orders in diagrammatic expansion, for a large class of QFTs.

We assume that there is no external field violating spacetime translational symmetry except for the present external EM field. We assume the EM $U(1)$ charge conservation is not broken by the ground state. We do not assume the presence of any other symmetry.

\subsection{Boltzmann Equation}
\label{ssect_kinetic_BE}

We now present our proposal for the Berry Fermi liquid theory, postponing
its theoretical justification to the next Section of the paper.
In a Berry Fermi liquid, as in the usual Fermi liquid theory, the energy of a quasiparticle with momentum $\p$ near the FS depends on the occupation at other momenta. To first order in $A$ and first order in $\partial_x$, the energy is
\begin{eqnarray}
\epsilon(\p; x) = E(\p) - \mu^{\mu\nu}(\p) \frac{F_{\mu\nu}(x)}{2} + \int_\k \left(\mathcal{U}(\p, \k) \ \delta\!f(\k; x) + \mathcal{V}^\nu(\p, \k) \ \partial_{x^\nu} \delta\!f(\k; x)\right).
\label{BFL_quasiparticle_energy}
\end{eqnarray}
Compared to \eqref{LFL_quasiparticle_energy}, here $\mu^{\mu\nu}$, antisymmetric in $\mu\nu$, is the EM dipole moment of the quasiparticles (the purely spatial components $\mu^{ij}$ correspond to the magnetic dipole moment and the mixed components $\mu^{i0}$ to the electric dipole moment), and $\mathcal{V}^\nu(\p, \k)$, odd under exchange of $\p$ and $\k$, is the gradient interaction potential between quasiparticles. The function $\mathcal{V}^\nu(\p,\k)$ is the additional function parametrizing the dependence of the energy of the quasiparticle with momentum $\p$ on the gradient of the distribution function at $\k$. Since we are performing a gradient expansion of the interaction between two quasiparticles, our assumption of interaction being finite-ranged is needed.

Extended to sub-leading order in spacetime derivative, the linearized (in $\delta f$ and $A$) Boltzmann equation now includes collision term. Although we need to include collision for completeness, we emphasize it is ``uninteresting'' towards the focus of this paper as it does not contribute to interesting physics such as the anomalous Hall effect, as we will show later in Appendix A.

The collision term is different from that in classical Boltzmann equation, and must be obtained quantum mechanically. The collisionful Boltzmann equation we find is to modify \eqref{LFL_Boltzmann_Eq_coordinates} by the replacement $v^i(\p) \partial_{x^i} \rightarrow v^i(\p) \partial_{x^i} - \delta(\epsilon_F-E(\p))\int_\k \mathcal{C}(\p, \k) \: \partial_{t}^2$ on both sides, yielding
\begin{eqnarray}
&& v^\mu(\p) \: \partial_{x^\mu} \delta\!f(\p; x) \ - \ \delta(\epsilon_F-E(\p))\int_\k \mathcal{C}(\p, \k) \ \partial_t^2 \delta\!f(\k; x) \nonumber \\[.2cm]
&=& \delta(\epsilon_F-E(\p)) \: \left( v^i(\p) \: F_{i0}(x) - v^i(\p)\partial_{x^i} \epsilon(\p; x) + \int_\k \mathcal{C}(\p, \k)\:\delta(\epsilon_F-E(\k)) \ \partial_t^2 \epsilon(\k; x)\right).
\label{BFL_Boltzmann_Eq}
\end{eqnarray}
Here $\mathcal{C}(\p, \k)$, symmetric under exchange of $\p$ and $\k$, is the effective collision kernel defined on the FS. It has the following properties (which we will show when we perform the QFT derivation in the next Section):
\begin{itemize}
\item
Collisions do not change the total number of fermionic excitations, i.e.
\begin{eqnarray}
\int_\p \delta(\epsilon_F-E(\p)) \ \mathcal{C}(\p, \k) =0.
\label{Collision_conservation}
\end{eqnarray}
\item
$\mathcal{C}(\p, \k)$ is not regular over the FS. It can be separated into a positive ``quasiparticle decay'' piece that is non-vanishing only when $\p=\k$ on the FS, plus a piece that is non-vanishing for general values of $\p$ and $\k$.
\end{itemize}
The $\partial_t^2$ in the collision term has been long known. Recall that in the ``thermal regime'' where temperature $T\gg \partial_t$, linearizing the classical Boltzmann collision term yields the scaling of $T^2$. But here we are in the ``quantum regime'' where temperature is negligible, $T\ll \partial_t^2$; according to Landau's semi-classical argument~\cite{landau1957oscillations}, in this regime the scaling should be replaced by $\partial_t^2$. Luttinger also has a field theory power counting argument \cite{Luttinger:1961zz}; we will adopt this method in Appendix A.

We have to emphasize that such $\partial_t^2$ parametrization of collision only holds for $d\geq 3$. In $d=2$ the collision term cannot be parametrized in any simple form~\cite{chubukov2003nonanalytic,chubukov2005singular}, as we will discuss in Appendix B. Fortunately, our main focus -- the computation of the Hall current -- is not undermined by this failure of parametrizing collisions in $d=2$, because collision does not contribute to it.

The Boltzmann equation can be solved in principle, order by order in $q$. First in \eqref{BFL_Boltzmann_Eq} we Fourier transform $-i\partial_{x^\mu}$ into $q_\mu$. Let us separate $\delta\!f=\delta\!f_0 + \delta\!f_1$, where the subscript labels the order in $q$. Then the Boltzmann equation \eqref{BFL_Boltzmann_Eq} reads (the collision term only holds for $d\geq 3$)
\begin{eqnarray}
\delta\!f_0(\p; q) = \delta(\epsilon_F-E) \frac{v^i}{v^\mu q_\mu - i\epsilon} \left( -iF_{i0}(q) - q_i \int_\k \mathcal{U}(\p, \k) \: \delta\!f_0(\k; q) \right),
\label{BFL_Boltzmann_Eq_0}
\end{eqnarray}
\begin{eqnarray}
\delta\!f_1(\p; q) &=& \delta(\epsilon_F-E) \frac{v^i \: q_i}{v^\mu q_\mu - i\epsilon} \left( \mu^{\mu\nu} \frac{F_{\mu\nu}(q)}{2} - \int_\k \left(\mathcal{U}(\p, \k) \: \delta\!f_1(\k; q) + \mathcal{V}^\nu(\p, \k) \ iq_\nu\, \delta\!f_0(\k; q) \right) \right) \nonumber \\[.2cm]
&& + \ \delta(\epsilon_F-E) \frac{i(q_0)^2}{v^\mu q_\mu - i\epsilon} \int_\k \mathcal{C}(\p, \k) \left(\delta\!f_0(\k; q) + \delta(\epsilon_F-E(\k)) \int_\l \mathcal{U}(\k, \l) \: \delta\!f_0(\l; q) \right)
\label{BFL_Boltzmann_Eq_1}
\end{eqnarray}
at zeroth and first order in $q$ respectively. We will prove these two equations from QFT in Section \ref{ssect_QFT_BE}. Note that the zeroth order Boltzmann equation \eqref{BFL_Boltzmann_Eq_0} is that in Landau Fermi liquid theory.

One may have noticed that there is no reference to Berry curvature in the Boltzmann equation. Notably, the Berry curvature $b^{ij}$ should induce an anomalous velocity $b^{ij} F_{j\nu} v^\nu$~\cite{Sundaram:1999zz}. However, since \eqref{BFL_Boltzmann_Eq} is of order $A$, the effect of anomalous velocity will be order $A^2$, which we assumed to be negligible. (If one works beyond linear response, and assumes stable quasiparticle, the anomalous velocity term would be present~\cite{shindou2008gradient}.) Other effects of Berry curvature are negligible in the Boltzmann equation for the same reason.

\subsection{Current}
\label{ssect_kinetic_current}

At equilibrium there is some equilibrium current $J_{eq.}^\mu$. In most systems at equilibrium only the charge density $J_{eq.}^0$ is non-zero, while $J_{eq.}^i=0$. As we perturb the system, extra current $\delta J^\mu$ of order $A$ will be induced. In our formalism, we propose
\begin{eqnarray}
\delta J^\mu(x) &=& \int_\p \left( v^\mu(\p) \: \delta\!f(\p; x) + \mu^{\mu\nu}(\p) \: \partial_{x_\nu} \delta\!f(\p; x) + \delta(\epsilon_F - E(\p)) \delta^\mu_i v^i(\p) \: (\epsilon(\p; x)-E(\p)) \right)\nonumber \\[.2cm]
&& + \: \sigma^{\mu\nu\lambda} \frac{F_{\nu\lambda}(x)}{2}\,.
\label{BFL_current}
\end{eqnarray}
Inside the integral on the right-hand side of \eqref{BFL_current} are three terms. The first term is the transport current due to the deformation of the FS. The second term is the magnetization / electric polarization current~\cite{Chen:2014cla}. The third term, as in Landau Fermi liquid theory, is the current due to the quasiparticles in the Fermi sea getting an extra interaction-induced velocity $\partial_p^i (\epsilon-E)$, and then rewritten via integrating $\p$ by parts. All these three terms only involve $\p$ near the FS, as desired.

In the last term of \eqref{BFL_current}, $\sigma^{\mu\nu\lambda}$, totally antisymmetric in $\mu\nu\lambda$, is the Hall conductivity tensor. As we will discuss in Section \ref{ssect_kinetic_sigma_EF}, it has very interesting relation to the FS, and that is how Berry curvature enters the formalism.

Now in \eqref{BFL_current} we Fourier transform $-i\partial_{x^\mu}$ into $q_\mu$. At zeroth and first order in $q$ respectively, the current reads
\begin{eqnarray}
\delta J_0^\mu (q) &=& \int_\p \left( v^\mu \: \delta\!f_0(\p; q) + \delta(\epsilon_F - E) \delta^\mu_i v^i \! \int_\k \mathcal{U}(\p, \k) \: \delta\!f_0(\k; q) \right),
\label{BFL_current_0}
\end{eqnarray}
\begin{eqnarray}
\delta J_1^\mu (q) &=& \int_\p \left[ v^\mu \: \delta\!f_1(\p; q) + \mu^{\mu\nu} \: iq_\nu\, \delta\!f_0(\p; q) \phantom{\frac{F_\mu}{F_\nu}} \right. \nonumber \\[.2cm]
&& \hspace{.7cm} \left. + \: \delta(\epsilon_F - E) \delta^\mu_i v^i \left( -\mu^{\nu\lambda} \frac{F_{\nu\lambda}(q)}{2} + \int_\k \left(\mathcal{U}(\p, \k) \: \delta\!f_1(\k; q) + \mathcal{V}^\nu(\p, \k) \ iq_\nu\, \delta\!f_0(\k; q) \right) \right) \right] \nonumber \\[.2cm]
&& + \: \sigma^{\mu\nu\lambda} \frac{F_{\nu\lambda}(q)}{2}.
\label{BFL_current_1}
\end{eqnarray}
Notice $\delta J_0^\mu$ is that in Landau Fermi liquid theory. We will prove these two equations from QFT in Section \ref{ssect_QFT_current}. In the proof, we will also discuss the microscopic contributions to $\mu^{\mu\nu}$. The magnetic dipole moment is generally non-zero; in the presence of interactions~\cite{shindou2008gradient}, the electric dipole moment will also be non-zero in general, as we will see in the proof.

Although we call $\sigma^{\mu\nu\lambda}$ the Hall conductivity
tensor, it is not the full Hall conductivity in linear
response. The full Hall conductivity also receives contributions from
the $\p$ integral, and depends on the ratio $|\q|/q^0$.
For example, in order to find the Hall conductance in transport measurements,
we should compute the local Hall conductivity in the limit of $q_j=0$ and $q^0\rightarrow 0$ \cite{cooper1997thermoelectric}.
It is convenient to choose the gauge $A_0=0$, so that $F_{j0}=-iq_0 A_j$. 
From the Boltzmann equation we have $\delta f_0 = -\delta(\epsilon_F-E) v^j A_j$,
which leads to the anomalous Hall current
\begin{align}
\delta J_H^i = \left(\sigma^{ij0} -\int_\p \delta(\epsilon_F-E) \ 2v^{[i} \mu^{j]0} + \int_{\p, \k} \delta(\epsilon_F-E(\p)) \, \delta(\epsilon_F-E(\k)) \ v^{[i}(\p) \: v^{j]}(\k) \ \mathcal{V}^0(\p, \k) \right) F_{j0} \nonumber \\[.2cm]
(q^j=0, \ q^0 \mbox{ small})
\label{BFL_AHE_uniform}
\end{align}
(although $\delta f_1$ is 
non-zero due to collisions, we already mentioned that collisions do not contribute 
to the Hall current, as shown in Appendix A). Thus, the full
Hall conductivity, in this limit which corresponds to Hall conductance in transport experiments, receives 
contribution not only from the $\sigma$ tensor, but also the electric dipole moment $\mu^{j0}$ \cite{shindou2008gradient} as well as the 
(temporal) gradient interaction $\mathcal{V}^0$ of the quasiparticles -- both are generally 
non-zero in the presence of interaction, as we will show in the 
QFT derivation. Similarly, if we take the other order of limits, $q_0=0$ first and $q_j$ small 
so that $F_{j0}=i q_j A_0$, and compute the local Hall conductivity, we will find
\begin{align}
\delta J_H^i = \left(\sigma^{ij0} + \int_\p \delta(\epsilon_F-E) \ \underline{\mu}^{ij}\right) F_{j0} \hspace{1.5cm} (q^0=0, \ q^j \mbox{ small})
\label{BFL_AHE_static}
\end{align}
where $\underline{\mu}^{ij}$ is related to the magnetic dipole moment $\mu^{ij}$ via the recursion relation
\begin{align}
\underline{\mu}^{ij}(\p) = \mu^{ij}(\p) - \int_\k \delta(\epsilon_F-E(\k)) \ \mathcal{U}(\p, \k) \ \underline{\mu}^{ij}(\k).
\end{align}
So again the $\sigma$ tensor does not give the full Hall conductivity.

\subsection{Chemical Potential Dependence of the Single Particle Kinetic Energy}
\label{ssect_kinetic_E_EF}

We have separated the quasiparticle distribution into an equilibrium
part $\theta(\epsilon_F-E)$ and an excitation part $\delta\!f$.
Within Fermi liquid theory, such a separation is ambiguous:
the same state may equally well be described either by starting with a
slightly lower chemical potential and exciting some quasiparticles
above the FS, or by starting with a slightly higher chemical potential
and exciting some quasiholes below the FS. Clearly, for the theory to
be self-consistent, all these different descriptions of the same state
must be equivalent. For this, the following relationship between $E$
at different chemical potentials must hold:
\begin{eqnarray}
\frac{\partial E(\p)}{\partial \epsilon_F} = \int_\k \mathcal{U}(\p, \k) \frac{\partial}{\partial \epsilon_F} \theta(\epsilon_F-E(\k)) = \int_\k \mathcal{U}(\p, \k) \left(1- \frac{\partial E(\k)}{\partial \epsilon_F} \right) \delta(\epsilon_F-E(\k)).
\label{BFL_dE_dEF}
\end{eqnarray}
This can be physically understood from
\eqref{BFL_quasiparticle_energy}, setting $F_{\mu\nu}=0$ and
$\partial_x \delta f=0$. Furthermore, we will prove it from QFT in
Section \eqref{sssect_QFT_int_Prop_on_EF}. Taking $\partial_p^i$ of
\eqref{BFL_dE_dEF}, we obtain the chemical potential dependence of
$v^i(\p)$ on the FS.

Strictly speaking, the reasoning above only applies when the FS
changes continuously with the chemical potential. If the system
undergoes a quantum phase transition at some $\epsilon_F$, around which the FS develops new
disconnected components, as illustrated in the figure:
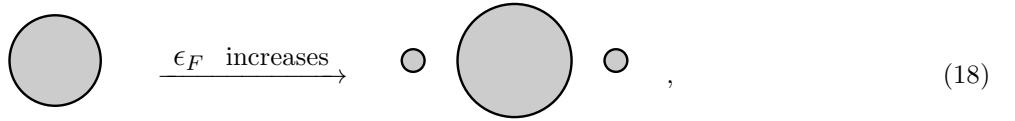
\begin{eqnarray}
\parbox{15mm}{
\begin{fmffile}{zzz-FS_low_E_F}
\begin{fmfgraph*}(15, 15)
\fmfleftn{l}{3}\fmfrightn{r}{3}
\fmf{phantom}{l2,o,r2}
\fmfv{d.sh=circle,d.f=20,d.si=0.8h}{o}
\end{fmfgraph*}
\end{fmffile}
}
\hspace{.5cm} \xrightarrow{\ \mbox{\large $\epsilon_F$ \ } \mbox{increases} \ } \hspace{.1cm} 
\parbox{40mm}{
\begin{fmffile}{zzz-FS_high_E_F}
\begin{fmfgraph*}(40, 15)
\fmfleftn{l}{3}\fmfrightn{r}{3}
\fmf{phantom}{l2,ol,oll,o,orr,or,r2}
\fmfv{d.sh=circle,d.f=20,d.si=h}{o}
\fmfv{d.sh=circle,d.f=20,d.si=0.2h}{ol}
\fmfv{d.sh=circle,d.f=20,d.si=0.2h}{or}
\end{fmfgraph*}
\end{fmffile}
},
\label{FS_discont_change}
\end{eqnarray}
then the formula \eqref{BFL_dE_dEF} not necessarily holds.

\subsection{Chemical Potential Dependence of the Hall Conductivity Tensor}
\label{ssect_kinetic_sigma_EF}

The Hall conductivity tensor in \eqref{BFL_current} seems to have no
reference to the FS. But in fact the Hall conductivity tensor is
related to the FS via the Berry curvature in a very interesting
manner. We will distinguish two cases. In the first case, either $d=2$, or $d>2$ and
the Berry curvature is an exact 2-form on the FS, so that the system has no
anomaly-related transport effects. (The anomaly-related transport
effects include, e.g., the chiral magnetic effect in $(3+1)d$, but not
the anomalous Hall effect in $(2+1)d$.) Then we turn to the case in $d>2$ with
non-exact Berry curvature on the FS, so that the system exhibits
anomaly-related transport effects~\cite{Son:2012wh,Stephanov:2012ki}.

\subsubsection{Without Anomaly-Related Transport}
\label{sssect_kinetic_sigma_EF_no_anomaly}

Let us review the story in Berry Fermi gas. In Fermi gas, particles are stable, so one can define the Berry connection $a^j$ and Berry curvature $b^{ij}$ for all particles in the Fermi sea:
\begin{eqnarray}
a^j(\p) \equiv (-i\hbar) \: \mathfrak{u}^\dagger_\alpha(\p) \: \partial_p^j \mathfrak{u}^\alpha(\p),
\end{eqnarray}
\begin{eqnarray}
b^{ij}(\p) \equiv 2\partial_p^{[i} a^{j]}(\p) = (-2i\hbar) \: \partial_p^{[i} \mathfrak{u}^\dagger_\alpha(\p) \: \partial_p^{j]} \mathfrak{u}^\alpha(\p) 
\end{eqnarray}
where $\mathfrak{u}^\alpha(\p)$ is the spinor or Bloch state of the fermion (recall that $\partial_p^i\equiv \partial/\partial p_i$). The Berry curvature induces an anomalous velocity~\cite{Sundaram:1999zz} and a change of the classical phase space measure~\cite{Xiao:2005qw, Duval:2005vn}, leading to Hall conductivity tensor of the form
\begin{eqnarray}
\sigma^{\mu\nu\lambda} = \sigma_{o}^{\mu\nu\lambda} + 3 \int_\p \theta(\epsilon_F-E(\p)) \: v^{[\mu}(\p) \: b^{\nu\lambda]}(\p)
\label{BFL_sigma_const_Fermi_sea}
\end{eqnarray}
where $a^0=0, \ b^{0\mu}=0$. Here $\sigma_{o}^{\mu\nu\lambda}$ is the contribution from valence bands / Dirac sea, and is independent of $\epsilon_F$. The second term seems like a Fermi sea property, but as observed by Haldane~\cite{Haldane:2004zz}, one can integrate $\p$ by parts and get
\begin{eqnarray}
\sigma^{\mu\nu\lambda} = \sigma_{o}^{\mu\nu\lambda} + 6 \int_\p \delta(\epsilon_F-E(\p)) \: \delta^{[\mu}_0 v^\nu(\p) \: a^{\lambda]}(\p),
\label{BFL_sigma_const_FS_1}
\end{eqnarray}
so that the kinetic part of the Hall conductivity tensor is actually a FS property; notice the kinetic part has no ${ijk}$ components, but only ${ij0}$ ones. For $d>2$ (recall $d$ is the number of spatial dimensions), there is another way to integrate \eqref{BFL_sigma_const_Fermi_sea} by parts, also promoted by Haldane~\cite{Haldane:2004zz}. Using $v^0\equiv 1 = \partial_p^k p^k/d$, we have
\begin{eqnarray}
\sigma^{ij0} &=& \sigma_{o}^{ij0} + \frac{3}{d-2} \int_\p \theta(\epsilon_F-E(\p)) \ b^{[ij}(\p) \ \partial_p^{k]} p_k \nonumber \\[.2cm]
&=& \sigma_{o}^{ij0} + \frac{3}{d-2}\int_\p \delta(\epsilon_F-E(\p)) \: b^{[ij}(\p) v^{k]}(\p) \: p_k \nonumber \\[.2cm]
&& \hspace{.85cm} + \: \frac{6}{d-2}\int_\p \partial_p^{[k} \left( \delta(\epsilon_F-E(\p)) \: v^i(\p) \: a^{j]}(\p) \: p_k\right)
\label{BFL_sigma_const_FS_2}
\end{eqnarray}
and $\sigma^{ijk}=\sigma_o^{ijk}$. The last line is a boundary term that is non-vanishing if the fermion is in a lattice and the FS intersects the boundary of our choice of first Brillouin zone~\cite{Haldane:2004zz} (because $p_k$ jumps by a reciprocal lattice vector when we identify the opposite boundaries of the first Brillouin zone). The advantage of \eqref{BFL_sigma_const_FS_2} over \eqref{BFL_sigma_const_FS_1} is that it involves $b^{ij}$ instead of the gauge dependent $a^i$ (except for the boundary term); as we will see later, this makes \eqref{BFL_sigma_const_FS_2} more convenient for generalization to include anomaly-related transport effects.

Now we turn to the $\epsilon_F$ dependence of $\sigma^{\mu\nu\lambda}$ in Berry Fermi liquid. In the presence of interaction, the picture of quasiparticles is only valid near the FS, so whether the $\epsilon_F$ dependence of $\sigma^{\mu\nu\lambda}$ can be expressed as a FS property becomes important at conceptual level: It determines, in order to study linear response to EM field at long wavelength, whether knowing the system is a Fermi liquid at low energy is enough, or we have to know more beyond the low energy behaviors. Our conclusion is, the former is true -- the fact that the system is a Fermi liquid is enough. More exactly, we will show in Section \ref{sssect_QFT_dsigma_dEF} that if the FS changes continuously with the chemical potential, then
\begin{eqnarray}
\frac{d\sigma^{\mu\nu\lambda}}{d\epsilon_F} = \frac{d}{d\epsilon_F} \ 6 \int_\p \delta(\epsilon_F-E(\p)) \: \delta^{[\mu}_0 v^\nu(\p) \: a^{\lambda]}(\p),
\label{BFL_dsigma_dEF_1}
\end{eqnarray}
where the Berry connection $a^i$ is now defined by the spinor / Bloch state $\mathfrak{u}(\p)$ of an on-shell quasiparticle near the FS (and $a^0=0$ as usual). For $d>2$, the above is equivalent to
\begin{eqnarray}
\frac{d\sigma^{ij0}}{d\epsilon_F} &=& \frac{d}{d\epsilon_F} \ \frac{3}{d-2}\int_\p \delta(\epsilon_F-E(\p)) \: b^{[ij}(\p) v^{k]}(\p) \: p_k \nonumber \\[.2cm]
&& + \ \frac{d}{d\epsilon_F} \ \frac{6}{d-2}\int_\p \partial_p^{[k} \left( \delta(\epsilon_F-E(\p)) \: v^i(\p) \: a^{j]}(\p) \: p_k\right),
\label{BFL_dsigma_dEF_2e}
\end{eqnarray}
\begin{eqnarray}
\frac{d\sigma^{ijk}}{d\epsilon_F} &=& 0.
\label{BFL_dsigma_dEF_2b}
\end{eqnarray}
Thus, we conclude that in Berry Fermi liquid, \eqref{BFL_sigma_const_FS_1} and \eqref{BFL_sigma_const_FS_2} still hold as in Berry Fermi gas. Although we demonstrated in Section \ref{ssect_kinetic_current} that $\sigma^{\mu\nu\lambda}$ is not the full Hall conductivity, those remaining contributions are nevertheless always FS integrals. Therefore the full conductivity is always equal to a chemical potential independent part (as long as the FS changes continuously) plus a FS integral.

In Berry Fermi gas, $\sigma_{o}^{\mu\nu\lambda}$ is topological~\cite{Ishikawa:1986wx, Haldane:2004zz}. It would be interesting to study if it is still topological in Berry Fermi liquid. In particular, it is unknown to us whether $\sigma_{o}^{\mu\nu\lambda}$ can have a jump when the FS develops new disconnected components, as in the example \eqref{FS_discont_change}.

Note that what we presented above does not mean the FS Berry curvature contributions in Fermi gas and Fermi liquid are the same. If we explicitly carry out the chemical potential derivative in \eqref{BFL_dsigma_dEF_1}, we find
\begin{eqnarray}
\frac{d\sigma^{\mu\nu\lambda}}{d \epsilon_F} = 3\int_\p \delta(\epsilon_F-E(\p)) \left( v^{[\mu}(\p) \: b^{\nu\lambda]}(\p) - \frac{\partial E(\p)}{\partial \epsilon_F} \delta^{[\mu}_0 \: b^{\nu\lambda]}(\p)- 2\delta^{[\mu}_0 v^\nu(\p) \: b^{\lambda]F}(\p) \right).
\label{BFL_dsigma_dEF_detail}
\end{eqnarray}
Here $b^{\lambda F}$ is the mixed Berry curvature of momentum and chemical potential:
\begin{eqnarray}
b^{kF}(\p) \equiv (-i\hbar) \left( \partial_p^k \mathfrak{u}^\dagger_\alpha(\p) \: \frac{\partial \mathfrak{u}^\alpha(\p)}{\partial \epsilon_F} - \frac{\partial \mathfrak{u}^\dagger_\alpha(\p)}{\partial \epsilon_F} \: \partial_p^k \mathfrak{u}^\alpha(\p) \right)
\end{eqnarray}
and $b^{0F}=0$; it satisfies the Bianchi identity $\partial b^{\nu\lambda}/\partial \epsilon_F = 2 \partial_p^{[\lambda} b^{\nu] F}$. The second and third terms of \eqref{BFL_dsigma_dEF_detail} are from interactions.

\subsubsection{With Anomaly-Related Transport in $d>2$}
\label{sssect_kinetic_sigma_EF_with_anomaly}

For $d>2$, when the Berry curvature is not an exact 2-form on the FS, the system has anomaly-related transport effects.

Let us first review the effects in Berry Fermi gas. The expression \eqref{BFL_sigma_const_Fermi_sea} still holds, and we start from there. Now we have to take extra care when rewriting it via integration by parts. More precisely, $a^i$ cannot be continuously defined over the entire FS, so the expression \eqref{BFL_sigma_const_FS_1} is not so useful. The alternative expression \eqref{BFL_sigma_const_FS_2} promoted by \cite{Haldane:2004zz} is still useful as long as we take into account the ``Berry curvature defects'' where $\partial_p^{[k} b^{ij]} \neq 0$ (e.g. monopoles in $d=3$):
\begin{eqnarray}
\sigma^{ij0} &=& \sigma_{o}^{ij0} + \frac{3}{d-2}\int_\p \theta(\epsilon_F-E(\p)) \: b^{[ij}(\p) \ \partial_p^{k]} p_k \nonumber \\[.2cm]
&=& \sigma_o^{ij0} - \frac{3}{d-2}\int_\p \theta(\epsilon_F-E(\p)) \partial_p^{[k} b^{ij]}(\p) \: p_k  \nonumber \\[.2cm]
&& \hspace{.8cm} + \ \frac{3}{d-2}\int_\p \delta(\epsilon_F-E(\p)) \: b^{[ij}(\p) v^{k]}(\p) \: p_k \nonumber \\[.2cm]
&& \hspace{.8cm} + \ \frac{6}{d-2}\int_\p \partial_p^{[k} \left( \delta(\epsilon_F-E(\p)) \: v^i(\p) \: a^{j]}(\p) \: p_k\right).
\label{BFL_sigma_const_FS_anom_e}
\end{eqnarray}
The boundary term in the last line is explained below \eqref{BFL_sigma_const_FS_2}; although $a^i$ is not continuously defined over the FS, it can be continuously defined around where the FS intersects the boundary of the first Brillouin zone. The defects lie along where $\partial_p^{[k} b^{ij]} \neq 0$, and they are generically $d-3$ dimensional. In this paper we assume there is no defect in the vicinity of the FS, and thus the second term is left unchanged under small continuous variation of $\epsilon_F$. In this spirit, we can combine the $\sigma_{o}^{ij0}$ term and the $\partial_p^{[k} b^{ij]}$ term and call their sum $\sigma_a^{ij0}$. A similar integration by parts can be carried out in the spatial components~\cite{Son:2012wh,Son:2012zy}:
\begin{eqnarray}
\sigma^{ijk} &=& \sigma_{o}^{ijk} + 3\int_\p \theta(\epsilon_F-E(\p)) \: b^{[ij}(\p) \ \partial_p^{k]} E(\p) \nonumber \\[.2cm]
&=& \sigma_o^{ijk} - 3\int_\p \theta(\epsilon_F-E(\p)) \partial_p^{[k} b^{ij]}(\p) \: E(\p) \nonumber \\[.2cm]
&& \hspace{.8cm} + \ 3\int_\p \delta(\epsilon_F-E(\p)) \: b^{[ij}(\p) v^{k]}(\p) \: \epsilon_F
\label{BFL_sigma_const_FS_anom_b}
\end{eqnarray}
(the second and third terms separately vanish in the absence of Berry curvature defect), whose $\partial_p^{[k} b^{ij]}$ term is again independent of small continuous variation of $\epsilon_F$, and again we can combine the $\sigma_o^{ijk}$ term and the $\partial_p^{[k} b^{ij]}$ term and call their sum $\sigma_a^{ijk}$. The simplest example of \eqref{BFL_sigma_const_FS_anom_e} with Berry curvature defect is the anomalous Hall effect in Weyl metals~\cite{yang2011quantum}; the simplest example of \eqref{BFL_sigma_const_FS_anom_b} is the chiral magnetic effect~\cite{Son:2012wh,Stephanov:2012ki,Son:2012zy}.

For Berry Fermi liquid, the gauge invariant \eqref{BFL_dsigma_dEF_detail} still holds even when the Berry curvature is not exact on the FS, and we start from there. In Section \ref{sssect_QFT_dsigma_dEF} we will show \eqref{BFL_dsigma_dEF_detail} is equivalent to
\begin{eqnarray}
\frac{d\sigma^{ij0}}{d\epsilon_F} &=& \frac{d}{d\epsilon_F} \ \frac{3}{d-2}\int_\p \delta(\epsilon_F-E(\p)) \: b^{[ij}(\p) v^{k]}(\p) \: p_k \nonumber \\[.2cm]
&& + \ \frac{d}{d\epsilon_F} \ \frac{6}{d-2}\int_\p \partial_p^{[k} \left( \delta(\epsilon_F-E(\p)) \: v^i(\p) \: a^{j]}(\p) \: p_k\right),
\label{BFL_dsigma_dEF_e}
\end{eqnarray}
\begin{eqnarray}
\frac{d\sigma^{ijk}}{d\epsilon_F} &=& \frac{d}{d\epsilon_F} \ 3\int_\p \delta(\epsilon_F-E(\p)) \: b^{[ij}(\p) v^{k]}(\p) \: \epsilon_F
\label{BFL_dsigma_dEF_b}
\end{eqnarray}
as long as there is no Berry curvature defect near the FS; they reduce to \eqref{BFL_dsigma_dEF_2e}\eqref{BFL_dsigma_dEF_2b} if the Berry curvature is exact on the FS. Thus, for Berry Fermi liquid we can write
\begin{eqnarray}
\sigma^{ij\lambda} &=& \sigma_a^{ij\lambda} + 3\int_\p \delta(\epsilon_F-E(\p)) \: b^{[ij}(\p) v^{k]}(\p) \: P^\lambda_k (\p) \nonumber \\[.2cm]
&& \hspace{.85cm} + \ 6\int_\p \partial_p^{[k} \left( \delta(\epsilon_F-E(\p)) \: v^i(\p) \: a^{j]}(\p) \: P^\lambda_k\right),
\label{BFL_sigma_const_FS}
\end{eqnarray}
where $P^0_k\equiv p_k/(d-2)$ and $P^l_k\equiv \epsilon_F \delta^l_k$ (the second line vanishes if $\lambda$ is spatial). Here $\sigma_a^{\mu\nu\lambda}$ is independent of $\epsilon_F$ for generic values of $\epsilon_F$; but it depends on $\epsilon_F$ at special values of $\epsilon_F$ where some Berry curvature defect is brought across the Fermi level. Moreover, as before, it is unknown whether $\sigma_a^{\mu\nu\lambda}$ can have a jump in situations like \eqref{FS_discont_change}. In \eqref{BFL_sigma_const_FS_anom_e} and \eqref{BFL_sigma_const_FS_anom_b} for Fermi gas, we are able to separate $\sigma_a^{\mu\nu\lambda}$ into $\sigma_o^{\mu\nu\lambda}$ plus a Berry curvature defect term inside the Fermi sea. Such separation is in general not well-defined for Fermi liquid.

A final subtlety needs to be addressed. If we shift the definition of $\vec{p}$ by a constant vector, or shift the definitions of $E(\p)$ and $\epsilon_F$ together by a constant value, no physics should change. However, the kinetic term in \eqref{BFL_sigma_const_FS}, due to its $P^\lambda_k$ factor, does not necessarily satisfy this property in the presence of anomaly-related transport effects. There is no inconsistency here, as our starting point \eqref{BFL_dsigma_dEF_detail} does not have this problem. This just implies that, if we perform such shifts, $\sigma_a^{\mu\nu\lambda}$ also needs to be shifted such that $\sigma^{\mu\nu\lambda}$ remains unchanged. In Berry Fermi gas, this can be verified explicitly in \eqref{BFL_sigma_const_FS_anom_e} and \eqref{BFL_sigma_const_FS_anom_b}.

\subsection{Correspondence between Kinetic Theory and Field Theory}

In the next Section we will provide a diagrammatic derivation of the
Berry Fermi liquid theory. Here we summarize the identification of
the various quantities appearing in the Berry Fermi liquid theory and
objects in the resummed perturbation theory.
\begin{itemize}
\item
$\delta f(\p; x)$, the quasiparticle distribution, is given by the FS singular part of the perturbed Wigner function, as introduced in Section \ref{ssect_QFT_BE}.
\item
$\delta J^\mu(x)$, the induced current, is the quantum expectation \eqref{current_QFT}.
\item
$E(\p)$, the kinetic energy of a quasiparticle, is defined by the full propagator \eqref{chiu_near_FS} near the FS. Its chemical potential dependence is given by \eqref{dE_dEF}.
\item
$\mathfrak{u}_\alpha(\p)$, the spinor / Bloch state of a quasiparticle, needed to define the Berry curvature, is defined in \eqref{G_expression} and above \eqref{u_derivative_on_shell}. Its chemical potential dependence is given below \eqref{du_dEF}.
\item
$\mu^{\mu\nu}(\p)$, the EM dipole moment of a quasiparticle, is given by \eqref{EM_dipole_def}, and discussed in detail in Section \ref{sssect_QFT_EM_dipole}.
\item
$\mathcal{U}(\p, \k)$, the contact interaction energy between two quasiparticles, is given by \eqref{dE_dEF}.
\item
$\mathcal{V}^\mu(\p, \k)$, the gradient interaction energy between two quasiparticles, is given by \eqref{grad_int_def}.
\item
$\mathcal{C}(\p, \k)$, the near-FS effective collision kernel between two quasiparticles (in $d\geq 3$), is defined in \eqref{C_kernal_def}, whose details are discussed in Section \ref{sssect_QFT_Collision} and Appendix A.
\item
$\sigma^{\mu\nu\lambda}$, the Hall conductivity tensor, is defined in \eqref{sigma_def}. Its chemical potential dependence is given by the Berry curvature around the FS, as shown in Section \ref{sssect_QFT_dsigma_dEF}.
\end{itemize}
$E(\p)$ and $\mathcal{U}(\p, \k)$ are familiar parameters in Landau Fermi liquid theory, while the other parameters $\mu^{\mu\nu}(\p), \mathcal{V}^\mu(\p, \k), \mathcal{C}(\p, \k)$ and $\sigma^{\mu\nu\lambda}$ are new.

\section{Diagrammatic Proof of the Kinetic Formalism}
\label{sect_QFT}

In this Section we will prove the kinetic formalism presented above by analyzing the quantum field theory (QFT) to all orders in perturbation theory. Before we go into any details, we sketch the idea behind our proof as the following. Our goal is to compute linear response, i.e. the induced current $\delta J$ as a linear function of the electromagnetic (EM) connection $A$, to first order in the external momentum $q$ carried in $A$. The evaluation of $\delta J^\mu$ can be separated, formally, into two parts:
\begin{itemize}
\item
The first part follows from Cutkosky cut, and corresponds to the quasiparticle contributions to $\delta J$, that is, the first line of \eqref{BFL_current}. This part involves the excitation and collision of quasiparticles, described by the Boltzmann equation \eqref{BFL_Boltzmann_Eq}.
\item
The second part is contributions which do not follow from Cutkosky cut. This part is the non-quasiparticle, or ground state, contribution, which gives rise to the Hall conductivity tensor in \eqref{BFL_current}, whose chemical potential dependence is given by the Berry curvature on the Fermi surface (FS).
\end{itemize}
This is our simple sketch of the idea. Now we shall present the proof in full detail.

First we state the assumptions about our QFT and its ground state and low energy spectrum.

Our QFT consists of a multi-component fermionic field $\psi^\alpha$, charged under electromagnetism (EM). The index $\alpha$ can be a spinor index if $\psi$ is a Dirac spinor, or in general labels different bands. The fermionic field may interact via massive fields (generically denoted as $\phi$) or/and via finite-ranged self-interactions. But we assume any field other than $\psi$ to be EM neutral, and the EM couplings to $\psi$ only take place in the non-interacting terms of $\psi$ in the Lagrangian, but not in any interacting terms. Thus, there can be EM couplings such as $A\psi^\dagger \psi$ (including $F\psi^\dagger \psi$) and $AA\psi^\dagger \psi$, but there is no EM coupling like $A\phi\psi^\dagger \psi$ or $A\psi^\dagger \psi \psi^\dagger \psi$.

We assume the system is under chemical potential $\epsilon_F$ for the fermionic field and at negligible temperature (but high enough to avoid the Kohn-Luttinger instability~\cite{Kohn:1965zz}). We assume the EM $U(1)$ gauge invariance is not broken by the ground state. We assume there is no band degeneracy near the Fermi surface (FS), and for simplicity, we assume the Fermi level crosses only one band of the spectrum of the fermionic field. We assume the only low energy excitations are quasiparticles of this band. Thereby the system is said to be a Fermi liquid at low energy.

We assume spacetime translational symmetry is not broken by anything except for the present external EM field. We do not assume any symmetry otherwise. In this Section we set $\hbar$ to $1$.

The proof is organized as follows. We first discuss the properties of a single full propagator and a pair of full propagators. Then we introduce the irreducible 2-particle interaction vertex, and discuss its relation to the chemical potential dependence of the full propagator (from which the chemical potential dependence \eqref{BFL_dE_dEF} of the kinetic energy follows). Then we introduce the properties of the bare EM coupling vertex. Next we present the recursion relation satisfied by the full EM coupling vertex; we will also extract the implications of the Ward-Takahashi identity, and present a simple proof to Luttinger's theorem. Having had all these preliminaries, we are ready to prove the main results. We first show the Boltzmann equation \eqref{BFL_Boltzmann_Eq_0}\eqref{BFL_Boltzmann_Eq_1} follows from the recursion relation satisfied by the full EM coupling vertex. Then we compute the quantum expectation of the current and show it takes the form \eqref{BFL_current_0}\eqref{BFL_current_1} given in the kinetic theory. Finally we study the chemical potential dependence \eqref{BFL_dsigma_dEF_detail} of the Hall conductivity tensor. Along the way, we will also discuss the microscopic ingredients of the EM dipole moment. %As a bonus, we obtain an alternative diagrammatic proof to the Coleman-Hill theorem~\cite{Coleman:1985zi} for QFTs restricted to our assumptions.

\subsection{Propagator}
\label{ssect_QFT_prop}

\subsubsection{Single Propagator}

\begin{center}
\begin{fmffile}{zzz-G}
\begin{fmfgraph*}(35, 15)
\fmfleftn{l}{3}\fmfrightn{r}{3}
\fmf{fermion,label.side=left,label=$p$}{r2,l2}
\fmfdot{l2,r2}
\fmflabel{$\alpha$}{l2}
\fmflabel{$\beta$}{r2}
\end{fmfgraph*}
\end{fmffile}
\end{center}

The full propagator (all QFT quantities are time ordered unless otherwise specified) $iG^\alpha_{\ \beta}(p)$ is a matrix in the components of the fermionic field. We let the energy $p^0=-p_0=0$ on the FS. The assumption of Fermi liquid amounts to the assumptions of the form of $iG$ at small $p^0$. By general analytic properties of fermionic propagators \cite{Luttinger:1961zz, abrikosov1975methods} (in particular, the property that $G$ must be Hermitian at $p^0=0$), and the specific requirement that at low energy there is one species of stable quasiparticle, as $p^0\rightarrow 0$ the full propagator of our assumed Fermi liquid should take the form
\begin{eqnarray}
iG^\alpha_{\ \beta}(p) \simeq \frac{i u^\alpha(p) u^\dagger_\beta(p)}{\chi_u(p)} + \sum_w \frac{i w^\alpha(p) w^\dagger_\beta(p)}{\chi_w(p)}.
\label{G_expression}
\end{eqnarray}
The eigenvector $u$ is the band that crosses the Fermi level, with singular eigenvalue whose inverse is of the form
\begin{eqnarray}
\chi_u(p) = \frac{p^0 - \xi(\p) + i\epsilon\: \sgn\,\xi(\p)}{Z(\p)} \ + \ \cdots
\label{chiu_near_FS}
\end{eqnarray}
where $\xi(\p) \equiv E(\p) - \epsilon_F$ and $(\cdots)$ are terms of higher suppression in $p^0$; the quasiparticle renormalization factor $Z(\p)$ should be understood as the inverse of the coefficient of $p^0$ in $\chi_u$ (with rotational invariance, $Z$ can depend on $|\p|$; without rotational invariance, it may depend on all components of $\p$). The $w$'s are all other eigenvectors, and their eigenvalues $1/\chi_w$ are regular and nearly real.

For Landau Fermi liquid theory, \eqref{chiu_near_FS} is enough, but for Berry Fermi liquid we need to know one order higher in $p^0$, i.e. work up to $(p^0)^2$ order in the $(\cdots)$ terms. This will be handled later in Section \ref{sssect_QFT_Collision} and Appendix A. One should also worry about whether the diagonalization \eqref{G_expression} fails as we consider one order higher in $p^0$; using the method in Appendix A one can easily see this problem occurs only at two orders higher in $p^0$, so in this paper we do not need to worry about this.

In Section \ref{ssect_kinetic_sigma_EF}, we used the quasiparticle spinor / Bloch state $\mathfrak{u}(\p)$; it refers to $u(p)$ with $p$ on-shell and near the FS, i.e. $p^0=\xi(\p) \rightarrow 0$. The $p$ derivatives of $u(p)$ and $\mathfrak{u}(\p)$ are related by
\begin{eqnarray}
\left. (\partial_p^\mu+v^\mu \partial_{p^0}) u(p) \right|_{p \ on-shell} = \partial_p^\mu \mathfrak{u}(\p).
\label{u_derivative_on_shell}
\end{eqnarray}
Here $\partial_p^0 \equiv \partial / \partial p_0 = -\partial / \partial p^0 = \partial_{p_0} = - \partial_{p^0}$ while $\partial_p^i \equiv \partial / \partial p_i = \partial / \partial p^i  = \partial_{p_i}= \partial_{p^i}$, and recall that $v^0\equiv 1$ and $v^i\equiv \partial_p^i E$ as introduced in Section \ref{sect_kinetic}.

The following identity, which follows from the product rule of derivative, is useful in this paper:
\begin{eqnarray}
\partial_p^\mu (G^{-1})^\alpha_{\ \beta} \ u^\beta = - (G^{-1})^\alpha_{\ \beta} \ \partial_p^\mu u^\beta + \partial_p^\mu \chi_u u^\alpha + \chi_u \partial_p^\mu u^\alpha
\label{IBPtrick}
\end{eqnarray}
for $p$ near the FS; note that
\begin{eqnarray}
-Z \partial_p^\mu \chi_u = -Z\: u^\dagger_\alpha \ \partial_p^\mu (G^{-1})^\alpha_{\ \beta} \ u^\beta = v^\mu + \left(\mbox{terms vanish on the FS}\right).
\label{chiu_derivative}
\end{eqnarray}
There is a similar identity for $u^\dagger_\alpha \ \partial_p^\mu(G^{-1})^\alpha_{\ \beta}$.

Now we look at the momentum derivative of the full propagator:
\begin{eqnarray}
\partial_p^\nu iG^\alpha_{\ \beta}(p) &=& iG^\alpha_{\ \alpha'} \ \partial_p^\nu (iG^{-1})^{\alpha'}_{\ \beta'} \ iG^{\beta'}_{\ \beta} \ - \ iZ u^\alpha u^\dagger_\beta \ i\pi \delta(p^0-\xi) \ \partial_p^\nu \sgn\,\xi \nonumber \\[.2cm]
&=& iG^\alpha_{\ \alpha'} \ \partial_p^\nu (iG^{-1})^{\alpha'}_{\ \beta'} \ iG^{\beta'}_{\ \beta} - iZ u^\alpha u^\dagger_\beta \ i\delta_{FS} \ \delta^\nu_i v^i,
\label{G_derivative_indices}
\end{eqnarray}
where
\begin{eqnarray}
\delta_{FS}(p) \equiv 2\pi\delta(p^0) \: \delta(\xi(\p)).
\end{eqnarray}
The presence of the second term in \eqref{G_derivative_indices} is because as $p^i$ varies across the FS, the $p^0$ pole in $1/\chi_u$ moves across the real axis. This abrupt change is not captured by the first term. The expression of the second term can be obtained by principle function decomposition $(x \pm i\epsilon)^{-1}= \mathcal{P} x^{-1} \mp i\pi\delta(x)$.

To avoid having too many fermion component indices in equations in this paper, we introduce two notations: single fermion linear space and double fermion linear space. Consider $G^\alpha_{\ \beta}$. In single fermion linear space, $^\alpha_{\ \beta}$ are viewed as two indices, so $G$ is viewed as a matrix in single fermion linear space. In double fermion linear space, $^\alpha_{\ \beta}$ together is viewed as one index, so $G$ is viewed as a vector in the the double fermion linear space. In our proof, only a few index contractions are to be understood in single fermion linear space, most are understood in double fermion linear space. To distinguish them, we will enclose objects contracted in single fermion linear space by curly brackets $\{ \ \}$, while do not enclose objects contracted in double fermion linear space by anything. For example, according to \eqref{chiu_derivative}, $v^\mu$ can be expressed as
\begin{eqnarray}
-\frac{v^\mu}{Z} = \left. u^\dagger_\alpha \: \partial_p^\mu (G^{-1})^\alpha_{\ \beta} \: u^\beta \right|_{p \ on \ FS}= \left. \left\{ u^\dagger \: \partial_p^\mu G^{-1} \: u \right\} \right|_{p \ on \ FS}= \left. (uu^\dagger)^T \partial_p^\mu G^{-1} \right|_{p \ on \ FS}
\label{v_in_QFT}
\end{eqnarray}
in explicit index notation, single fermion notation, and double fermion notation respectively. We will introduce more about double fermion notation in Section \ref{sssect_QFT_int_vertex}.

In the double fermion notation introduced above, \eqref{G_derivative_indices} can be expressed compactly as
\begin{eqnarray}
\partial_p^\nu iG = i\Delta_0 \ \partial_p^\nu iG^{-1} + (Zuu^\dagger) \delta_{FS} \delta^\nu_i v^i,
\label{G_derivative}
\end{eqnarray}
where
\begin{eqnarray}
i(\Delta_0)^{\alpha \ \gamma}_{\ \delta, \ \beta}(p) \equiv iG^\alpha_{\ \beta}(p) \ iG^\gamma_{\ \delta}(p)
\end{eqnarray}
is a matrix in double fermion notation. Since the momentum argument in both $iG$'s is the same, $i\Delta_0$ has a double pole in $p^0$ when all of its four indices are in the $u$ band.

\subsubsection{Double Propagator}

An important step towards the QFT foundation of the Landau Fermi liquid theory is the observation that, the semiclassical notion of ``deformation $\delta f$ of the FS'' originates from the pole structure of the double propagator $iG(p+q/2) iG(p-q/2)$, for $q$ small and $p$ near the FS~\cite{landau1959theory,abrikosov1975methods}. Now we make a similar analysis, but with non-trivial $u^\alpha(p)$, and work to first order in $q$.

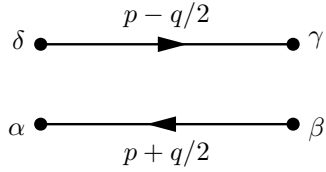
\begin{center}
\begin{fmffile}{zzz-GG}
\begin{fmfgraph*}(35, 20)
\fmfleftn{l}{5}\fmfrightn{r}{5}
\fmf{fermion,label.side=left,label=$p-q/2$}{l4,r4}
\fmf{fermion,label.side=left,label=$p+q/2$}{r2,l2}
\fmfdot{l2,r2,l4,r4}
\fmflabel{$\delta$}{l4}
\fmflabel{$\alpha$}{l2}
\fmflabel{$\gamma$}{r4}
\fmflabel{$\beta$}{r2}
\end{fmfgraph*}
\end{fmffile}
\end{center}

Consider the product of two full fermionic propagators, as drawn above, with arbitrary $p$ and small $q$; more exactly, $q\ll p_F$ where $p_F$ is the size scale of the FS. To first order in $q$, we express the product in the form
\begin{eqnarray}
iG^\alpha_{\ \beta}(p+q/2) \ iG^\gamma_{\ \delta}(p-q/2) &=& \left( i\Delta_0(p) + i\Delta'_0(p; q) + i\Delta^r_1(p; q) + i\Delta^s_1(p; q) + i\Delta'_1(p; q) \right)^{\alpha \ \gamma}_{\ \delta, \ \beta} \nonumber \\[.2cm]
&& - \ {D_1}^{\alpha \ \gamma}_{\ \delta, \ \beta}(p; q),
\label{doubleprop_parametrization}
\end{eqnarray}
where the subscripts $0$ or $1$ denote the order in $q$. Here $\Delta_0, \Delta^r_1, \Delta^s_1$ are regular as $q\rightarrow 0$. In particular, $\Delta_0$ has been introduced in \eqref{G_derivative}, and $\Delta^r_1$ and $\Delta^s_1$ follow from the two terms of \eqref{G_derivative_indices} when expanding $G(p\pm q/2)$ in $q$:
\begin{eqnarray}
(\Delta^r_1)^{\alpha \ \gamma}_{\ \delta, \ \beta}(p; q) \equiv -i \frac{q_\lambda}{2} \left( \left\{ G \ \partial_p^\lambda G^{-1} \ G\right\}^\alpha_{\ \beta} G^\gamma_{\ \delta} - G^\alpha_{\ \beta} \left\{ G \ \partial_p^\lambda G^{-1} \ G\right\}^\gamma_{\ \delta} \right),
\end{eqnarray}
\begin{eqnarray}
(\Delta^s_1)^{\alpha \ \gamma}_{\ \delta, \ \beta}(p; q) \equiv \frac{q_\lambda}{2} \delta^\lambda_k v^k \: Z \delta_{FS} \left(u^\alpha u^\dagger_\beta \: G^\gamma_{\ \delta} - G^\alpha_{\ \beta} \: u^\gamma u^\dagger_\delta\right).
\end{eqnarray}
Clearly both $\Delta^r_1$ and $\Delta^s_1$ vanish when all indices are projected onto one band.

When there is no FS, the expansion of the double propagator as $i\Delta_0+i\Delta^r_1+i\Delta^s_1$ is legitimate. When FS is present, such naive expansion in $q$ misses contributions that are related to the pole structure difference across the FS. These extra contributions are denoted by $\Delta'_0$ and $\Delta'_1$, which are singular as $q\rightarrow 0$. Explicitly, they are given by
\begin{eqnarray}
\Delta'_0 = \Delta' \ (uu^\dagger) (uu^\dagger)^T,
\end{eqnarray}
\begin{eqnarray}
\Delta'_1 = \Delta' \ iq_\mu \mathcal{A}^\mu,
\end{eqnarray}
where we defined
\begin{eqnarray}
\Delta'(p; q) \equiv Z^2(\p) \: \delta_{FS}(p) \frac{v^i(\p) \ q_i}{v^\mu(\p) \ q_\mu - i \epsilon \: \sgn q^0},
\label{Delta_prime}
\end{eqnarray}
which is familiar from Landau Fermi liquid theory~\cite{landau1959theory, abrikosov1975methods} when $u^\alpha$ is one-component (i.e. $u=1$ trivially). Also, we introduced the abbreviation
\begin{eqnarray}
(\mathcal{A}^\mu)^{\alpha \ \gamma}_{\ \delta, \ \beta} \equiv \frac{-i}{2}\left(\partial_p^\mu u^\alpha u^\dagger_\delta - u^\alpha \partial_p^\mu u^\dagger_\delta\right) (u^\gamma u^\dagger_\beta) - \frac{-i}{2}(u^\alpha u^\dagger_\delta) \left(\partial_p^\mu u^\gamma u^\dagger_\beta - u^\gamma \partial_p^\mu u^\dagger_\beta\right).
\end{eqnarray}
Below we present the derivation for $\Delta'_0$ and $\Delta'_1$.

Let us focus on the double $u$-band term in the double propagator:
\begin{eqnarray}
\frac{i\left(Zu^\alpha u^\dagger_\beta\right)(p+q/2)}{p^0+q^0/2-\xi(\p+\q/2)+i\epsilon\: \sgn\,\xi(\p+\q/2)} \ \frac{i\left(Zu^\gamma u^\dagger_\delta\right)(p-q/2)}{p^0-q^0/2-\xi(\p-\q/2)+i\epsilon\: \sgn\,\xi(\p-\q/2)}.
\label{ubandproduct}
\end{eqnarray}
What are missing in the naive expansion over $q$ are the contributions when the two $i\epsilon$ prescriptions in the denominators take opposite signs. To extract these missing pieces, we perform principle function decomposition $(x \pm i\epsilon)^{-1}= \mathcal{P} x^{-1} \mp i\pi\delta(x)$ and keep those terms which are non-vanishing only when $\sgn\, \xi(\p\pm\q/2)$ are opposite. Such terms are
\begin{eqnarray}
&& \left[ i\pi \ \frac{\sgn\,\xi(\p-\q/2) - \sgn\,\xi(\p+\q/2)}{-q^0 + \xi(\p+\q/2)-\xi(\p-\q/2)} \ \frac{\delta(p^0+q^0/2-\xi(\p+\q/2)) + \delta(p^0-q^0/2-\xi(\p-\q/2))}{2} \right. \nonumber \\[.2cm]
&& \hspace{0.1cm} \left. - \phantom{\frac{1}{1}} \!\!\!\! (i\pi)^2 2\theta(-\sgn\,\xi(\p-\q/2)\sgn\,\xi(\p+\q/2)) \delta(p^0+q^0/2-\xi(\p+\q/2))\delta(p^0-q^0/2-\xi(\p-\q/2)) \right] \nonumber \\[.2cm]
&& \times \ i\left(Zu^\alpha u^\dagger_\beta\right)(p+q/2) \ i\left(Zu^\gamma u^\dagger_\delta\right)(p-q/2).
\end{eqnarray}
Expanding the generalized functions in the square bracket in $q$, we have
\begin{eqnarray}
&& \left[ i\pi \frac{-2 \delta(\xi(\p)) v^i q_i}{v^\mu(\p) q_\mu} \delta(p^0-\xi(\p)) - (i\pi)^2 2|v^i(\p) q_i| \delta(\xi(\p)) \delta(p^0-\xi(\p)) \delta(v^\mu(\p) q_\mu) \ + \ \mathcal{O}(q^2) \right] \nonumber \\[.2cm]
&& \times \ i\left(Zu^\alpha u^\dagger_\beta\right)(p+q/2) \ i\left(Zu^\gamma u^\dagger_\delta\right)(p-q/2).
\end{eqnarray}
Now we recognize the square bracket is nothing but $-i\Delta'/Z^2$ expressed in principle function decomposition. Finally we expand the two $(Zuu^\dagger)$'s to zeroth and first order in $q$, we obtain the expression for $i\Delta'_0+i\Delta'_1$ presented above.

Developing along this line of thinking, one is led to the standard formalism of Cutkosky cut~\cite{Cutkosky:1960sp}, which we will discuss in Appendix A. In particular, see Eq.~\eqref{Delta_from_Cutkosky} for the derivation of $\Delta'$ from Cutkosky-cutting the double propagator.

What is $D_1$ in \eqref{doubleprop_parametrization}? The step \eqref{ubandproduct} is not quite right, for it completely ignored the $(\cdots)$ terms in \eqref{chiu_near_FS}. While it is legitimate to do so at leading order in $q$ (in Landau's theory), at first order in $q$ there are missed contributions, which we call $D_1$ and corresponds to quasiparticle decay. We will postpone its discussion to Section \ref{sssect_QFT_Collision}, when we discuss the quasiparticle decay term along with other quasiparticle collision terms.

\subsection{Interaction and the Chemical Potential Dependence of Propagator}
\label{ssect_QFT_int}

\subsubsection{$q$-2PI Interaction Vertex}
\label{sssect_QFT_int_vertex}

Let $i\wt{V}^{\alpha \ \gamma}_{\ \delta, \ \beta}(p, k; q)$ be the full $q$-2PI (defined below) interaction vertex, with two incoming fermions of momenta and indices $(p-q/2, \delta)$ and $(k+q/2, \beta)$, and two outgoing fermions with momenta and indices $(k-q/2, \gamma)$ and $(p+q/2, \alpha)$, as drawn below.
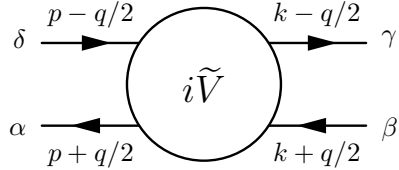
\begin{center}
\begin{fmffile}{zzz-wtV}
\begin{fmfgraph*}(45, 20)
\fmfleftn{l}{5}\fmfrightn{r}{5}
\fmf{phantom}{l4,lv4,rv4,r4}
\fmf{phantom}{r2,rv2,lv2,l2}
\fmf{fermion,tension=0.3,label.side=left,label=$p-q/2$}{l4,lv4}
\fmf{fermion,tension=0.3,label.side=left,label=$k+q/2$}{r2,rv2}
\fmf{fermion,tension=0.3,label.side=left,label=$k-q/2$}{rv4,r4}
\fmf{fermion,tension=0.3,label.side=left,label=$p+q/2$}{lv2,l2}
\fmflabel{$\delta$}{l4}
\fmflabel{$\alpha$}{l2}
\fmflabel{$\gamma$}{r4}
\fmflabel{$\beta$}{r2}
\fmf{phantom}{l3,o,r3}
\fmf{phantom,label.dist=0,label={\Large $i\wt{V}$}}{l3,r3}
\fmfv{d.sh=circle,d.f=empty,d.si=0.45w}{o}
\end{fmfgraph*}
\end{fmffile}
\end{center}
(In this paper, external propagators without a solid dot at the end are always stripped off.) Here $q$-2PI means that $i\wt{V}$ is a sum of connected, 1PI (with respect to the fermion only) interaction diagrams, such that in each diagram there does not exist two internal fermion propagators whose momenta are dictated by momentum conservation to differ by $q$. Equivalently, for each diagram, one cannot find two internal fermion propagators cutting which will disconnect the diagram into two parts, such that the external lines of $(p-q/2, \delta)$ and $(p+q/2, \alpha)$ are on one part, while the external lines of $(k+q/2, \beta)$ and $(k-q/2, \gamma)$ are on the other part. For example, in the four diagrams below (internal fermionic propagators always mean full propagators), the two on the left are $q$-2PI, while the two on the right are not.
\begin{center}
\begin{fmffile}{zzz-q2PI-1}
\begin{fmfgraph*}(35, 20)
\fmfleftn{l}{2}\fmfrightn{r}{2}
\fmf{fermion}{l2,v2,v3,r2}
\fmf{fermion}{r1,v4,v1,l1}
\fmf{dashes,left=0.1,tension=0}{v4,v2}
\fmf{dashes,right=0.1,tension=0}{v1,v3}
\end{fmfgraph*}
\end{fmffile}
\begin{fmffile}{zzz-q2PI-2}
\begin{fmfgraph*}(35, 20)
\fmfleftn{l}{2}\fmfrightn{r}{2}
\fmf{fermion,tension=2}{l2,v1,l1}
\fmf{fermion,tension=1.5}{r1,v2,v3,r2}
\fmf{phantom,tension=2}{r1,v2}
\fmf{phantom,tension=2}{v3,r2}
\fmf{dashes,tension=2}{v1,v4}
\fmf{phantom}{v2,v5,v6,v4}
\fmf{phantom}{v3,v7,v8,v4}
\fmf{dashes,tension=0}{v2,v5}
\fmf{dashes,tension=0}{v3,v7}
\fmffreeze
\fmf{fermion,left=0.7}{v4,v7}
\fmf{fermion}{v7,v5}
\fmf{fermion,left=0.7}{v5,v4}
\fmf{dashes,left=0.15}{v7,v4}
\end{fmfgraph*}
\end{fmffile}
\hspace{1cm}
\begin{fmffile}{zzz-not-q2PI-1}
\begin{fmfgraph*}(35, 20)
\fmfleftn{l}{2}\fmfrightn{r}{2}
\fmf{fermion}{l2,v2,v3,v6,r2}
\fmf{fermion}{r1,v5,v4,v1,l1}
\fmf{dashes,left=0.1,tension=0}{v4,v2}
\fmf{dashes,right=0.1,tension=0}{v1,v3}
\fmf{dashes,tension=0}{v5,v6}
\end{fmfgraph*}
\end{fmffile}
\begin{fmffile}{zzz-not-q2PI-2}
\begin{fmfgraph*}(35, 20)
\fmfleftn{l}{2}\fmfrightn{r}{2}
\fmf{fermion,tension=2}{l2,v1,l1}
\fmf{fermion,tension=1.5}{r1,v2,v3,r2}
\fmf{phantom,tension=2}{r1,v2}
\fmf{phantom,tension=2}{v3,r2}
\fmf{dashes,tension=2}{v1,v4}
\fmf{phantom}{v2,v5,v6,v4}
\fmf{phantom}{v3,v7,v8,v4}
\fmf{dashes,tension=0}{v2,v5}
\fmf{dashes,tension=0}{v3,v7}
\fmffreeze
\fmf{fermion,left=0.7}{v4,v7}
\fmf{fermion}{v7,v5}
\fmf{fermion,left=0.7}{v5,v4}
\fmf{dashes,left}{v5,v7}
\end{fmfgraph*}
\end{fmffile}
\end{center}
For Fermi liquid, the q-2PI interaction vertex $\wt{V}(p, k; q)$ is generally regular and analytic in low orders in $q$ near $q=0$. Keeping zeroth and first order, we write $\wt{V}(p, k; q) = \wt{V}_0(p, k) + \wt{V}_1(p, k; q)$. The method in Appendix A estimates the non-analyticity to occur at third order.

The $q$-2PI interaction vertex is the building block of the full interaction vertex $iV$. The latter is a geometric series given by the recursion relation
\begin{eqnarray}
\parbox{30mm}{
\begin{fmffile}{zzz-V-recursion}
\begin{fmfgraph*}(25, 15)
\fmfleftn{l}{5}\fmfrightn{r}{5}
\fmf{phantom}{l4,lv4,rv4,r4}
\fmf{phantom}{r2,rv2,lv2,l2}
\fmf{fermion,tension=0}{l4,lv4}
\fmf{fermion,tension=0}{r2,rv2}
\fmf{fermion,tension=0}{rv4,r4}
\fmf{fermion,tension=0}{lv2,l2}
\fmf{phantom}{l3,o,r3}
\fmfv{d.sh=circle,d.f=empty,d.si=0.43w}{o}
\fmf{phantom,label.dist=0,label=$iV$}{l3,r3}
\end{fmfgraph*}
\end{fmffile}
}
= \hspace{.5cm}
\parbox{30mm}{
\begin{fmffile}{zzz-V-recursion-wtV}
\begin{fmfgraph*}(25, 15)
\fmfleftn{l}{5}\fmfrightn{r}{5}
\fmf{phantom}{l4,lv4,rv4,r4}
\fmf{phantom}{r2,rv2,lv2,l2}
\fmf{fermion,tension=0}{l4,lv4}
\fmf{fermion,tension=0}{r2,rv2}
\fmf{fermion,tension=0}{rv4,r4}
\fmf{fermion,tension=0}{lv2,l2}
\fmf{phantom}{l3,o,r3}
\fmfv{d.sh=circle,d.f=empty,d.si=0.43w}{o}
\fmf{phantom,label.dist=0,label=$i\wt{V}$}{l3,r3}
\end{fmfgraph*}
\end{fmffile}
}
+\hspace{.5cm}
\parbox{50mm}{
\begin{fmffile}{zzz-V-recursion-wtV-V}
\begin{fmfgraph*}(45, 15)
\fmfleftn{l}{5}\fmfrightn{r}{5}
\fmf{fermion}{l4,lm4,rm4,r4}
\fmf{fermion}{r2,rm2,lm2,l2}
\fmf{phantom,tension=2}{l4,lm4}
\fmf{phantom,tension=2}{rm4,r4}
\fmf{phantom,tension=2}{lm2,l2}
\fmf{phantom,tension=2}{r2,rm2}
\fmf{phantom}{l3,ol}
\fmf{phantom}{or,r3}
\fmf{phantom,tension=0.8}{ol,or}
\fmfv{d.sh=circle,d.f=empty,d.si=0.25w}{ol}
\fmfv{d.sh=circle,d.f=empty,d.si=0.25w}{or}
\fmfv{label.dist=0,label=$i\wt{V}$}{ol}
\fmfv{label.dist=0,label=$iV$}{or}
\end{fmfgraph*}
\end{fmffile}
}
\label{full_int_vertex}
\end{eqnarray}
The full interaction vertex is singular in the $q\rightarrow 0$ limit, due to the presence of $\Delta'$ in the double propagators, as well as the presence of collision factors to be discussed in Section \ref{sssect_QFT_Collision}.

Before we proceed, we say a bit more about the double fermion notation. Consider an object, perhaps with spacetime indices, $\left(X^{\alpha \ \gamma}_{\ \delta, \ \beta}\right)^{\mu\nu\rho\dots}(p, k; q)$. This object is a matrix in the double fermion linear space. We now introduce its transpose:
\begin{eqnarray}
\left(X^{\alpha \ \gamma}_{\ \delta, \ \beta}\right)^{\mu\nu\rho\dots}(p, k; q) = \left(\left(X^{\gamma \ \alpha}_{\ \beta, \ \delta}\right)^{\mu\nu\rho\dots}(k, p; -q) \right)^T.
\end{eqnarray}
Diagrammatically, the transpose in double fermion linear space corresponds to ``turning the diagram $180$ degrees''; note that the spacetime indices are unaffected by the transpose. Finally, we introduce the convention that, for objects like $X$ which involve two momenta $p$ and $k$, the contraction with another object implies a momentum integral, for example
\begin{eqnarray}
\left(X^{\mu\nu\rho\dots} Y\right)^\alpha_{\ \delta}(p; q) \equiv \int_k \left(X^{\alpha \ \gamma}_{\ \delta, \ \beta}\right)^{\mu\nu\rho\dots}(p, k; q) \ Y^\beta_{\ \: \gamma}(k; q)
\end{eqnarray}
where $\int_k \equiv \int d^{d+1} k / (2\pi)^{d+1}$.

Now, by definition of $i\wt{V}$, we see it satisfies $i\wt{V} = (i\wt{V} )^T$, and similarly for all the $\Delta$'s and $D_1$ in \eqref{doubleprop_parametrization}. We will need these transpose properties when we derive the current in Section \ref{ssect_QFT_current}.

\subsubsection{Full Interaction Vertex}

Using the double fermion notation introduced above, we expand the recursion relation \eqref{full_int_vertex} to zeroth and first order in $q$. At zeroth order,
\begin{eqnarray}
iV_0 = i\wt{V}_0 + i\wt{V}_0 \left(i\Delta_0+i\Delta'_0\right) iV_0 = i\bar{V}_0 + i\bar{V}_0 \ i\Delta'_0 \ iV_0
\end{eqnarray}
where we defined the geometric series $i\bar{V}_0$ via the recursion relation
\begin{eqnarray}
i\bar{V}_0 = i\wt{V}_0 + i\wt{V}_0 \ i\Delta_0 \ i\bar{V}_0.
\end{eqnarray}
$i\bar{V}_0$ can be understood as $iV$ in the limit $q^0 \rightarrow 0$, $q^i/q^0 \rightarrow 0$ (because $\Delta'$ vanishes in the $q^i/q^0 \rightarrow 0$ limit), and is closely related to Landau's contact interaction potential $\mathcal{U}$~\cite{landau1959theory,abrikosov1975methods}, as we will see later. 

At first order,
\begin{eqnarray}
iV_1 &=& i\wt{V}_1 + i\wt{V}_1\left(i\Delta_0+i\Delta'_0\right) iV_0  + i\wt{V}_0 \left(i\Delta^r_1+i\Delta^s_1+i\Delta'_1\right) iV_0 + i\wt{V}_0 \left(-C_1\right) iV_0 \nonumber \\[.2cm]
&& + \ i\wt{V}_0\left(i\Delta_0+i\Delta'_0\right) iV_1.
\label{full_int_vertex_1}
\end{eqnarray}
Here ${C_1}^{\alpha \ \gamma}_{\ \delta, \ \beta}(p, k; q)$ is the quasiparticle collision term. Let us explain it now.

\subsubsection{Quasiparticle Collision}
\label{sssect_QFT_Collision}

Let us emphasize our comment in Section \ref{ssect_kinetic_BE} again: Collision is a single band ($u$ band) effect that is ``uninteresting'', as our main interest is multi-component effects such as Berry curvature. More particularly, in Appendix A it is shown that collision has no contribution to the antisymmetric part of the current-current correlation (which include interesting physics such as anomalous Hall effect and chiral magnetic effect). Here we are including collision just for completeness.

The quasiparticle collision term ${C_1}^{\alpha \ \gamma}_{\ \delta, \ \beta}(p, k; q)$ is defined as
\begin{eqnarray}
-C_1(p, k; q) &\equiv& -D_1(p; q) \ (2\pi)^{d+1}\delta^{d+1}(p-k) \ - \ C^{ph}_1(p, k; q) \ - \ C^{pp}_1(p, k; q).
\end{eqnarray}
The decay term $D_1$ in $C_1$ is from \eqref{doubleprop_parametrization} but left unexplained there. Where do $D_1$, $C^{ph}_1$ and $C^{pp}_1$ come from? Recall that in \eqref{doubleprop_parametrization} we could not naively expand the two propagators in $q$ individually; there are terms non-analytic in $q$ to be carefully taken care of. Similarly, here in the recursion relation for $iV$, we cannot naively expand the $i\wt{V}$'s and the double propagators individually. The non-analytic contributions that are missed from such naive expansion are $D_1$, $C^{ph}_1$ and $C^{ph}_1$.
 
Formally, the three terms in the definition of $-C_1(p, k; q)$ correspond to the following three pairs of Cutkosky-cut sub-diagrams:
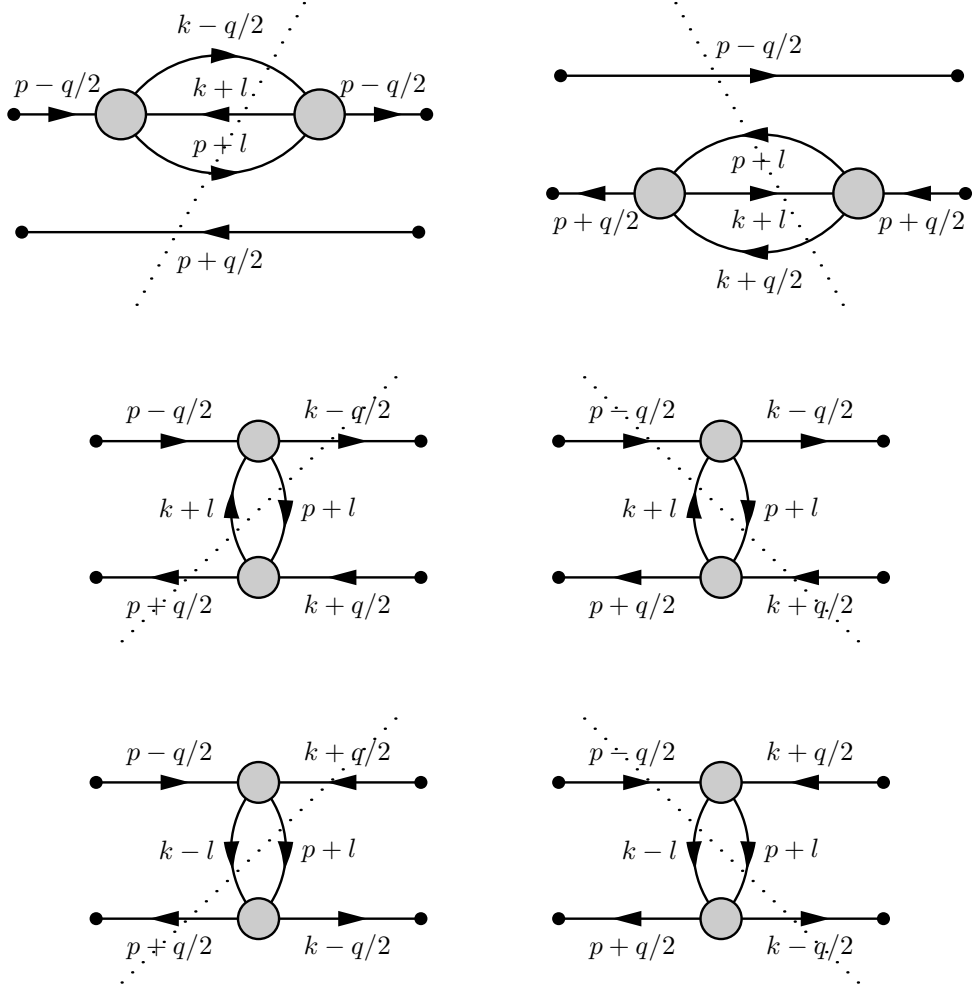
\begin{center}
\begin{fmffile}{zzz-Decay1}
\begin{fmfgraph*}(55, 40)
\fmfleftn{l}{9}\fmfrightn{r}{9}
\fmf{phantom}{r6,rm6,m6,lm6,l6}
\fmf{phantom}{l9,m9,r9}
\fmf{phantom}{r1,m1,l1}
\fmf{phantom,tension=2}{l1,m1}
\fmf{phantom,tension=2}{r9,m9}
\fmf{fermion,tension=0.3,label.side=left,label=$p+q/2$}{r3,l3}
\fmf{fermion,tension=0.3,label.side=left,label=$p-q/2$}{rm6,r6}
\fmf{fermion,tension=0.3,label.side=left,label=$p-q/2$}{l6,lm6}
\fmfdot{r3,l3,r6,l6}
\fmfv{d.sh=circle,d.f=20,d.si=0.12w}{rm6}
\fmfv{d.sh=circle,d.f=20,d.si=0.12w}{lm6}
\fmf{fermion,tension=0.2,label.side=right,label=$k+l$}{rm6,lm6}
\fmf{fermion,tension=0,right=0.6,label.side=left,label=$p+l$}{lm6,rm6}
\fmf{fermion,tension=0,left=0.6,label.side=left,label=$k-q/2$}{lm6,rm6}
\fmf{dots,tension=0}{m9,m1}
\end{fmfgraph*}
\end{fmffile}
\hspace{1cm}
\begin{fmffile}{zzz-Decay2}
\begin{fmfgraph*}(55, 40)
\fmfleftn{l}{9}\fmfrightn{r}{9}
\fmf{phantom}{r4,rm4,m4,lm4,l4}
\fmf{phantom}{l9,m9,r9}
\fmf{phantom}{r1,m1,l1}
\fmf{phantom,tension=2}{r1,m1}
\fmf{phantom,tension=2}{l9,m9}
\fmf{fermion,tension=0.3,label.side=left,label=$p-q/2$}{l7,r7}
\fmf{fermion,tension=0.3,label.side=left,label=$p+q/2$}{r4,rm4}
\fmf{fermion,tension=0.3,label.side=left,label=$p+q/2$}{lm4,l4}
\fmfdot{r4,l4,r7,l7}
\fmfv{d.sh=circle,d.f=20,d.si=0.12w}{rm4}
\fmfv{d.sh=circle,d.f=20,d.si=0.12w}{lm4}
\fmf{fermion,tension=0.2,label.side=right,label=$k+l$}{lm4,rm4}
\fmf{fermion,tension=0,right=0.6,label.side=left,label=$p+l$}{rm4,lm4}
\fmf{fermion,tension=0,left=0.6,label.side=left,label=$k+q/2$}{rm4,lm4}
\fmf{dots,tension=0}{m9,m1}
\end{fmfgraph*}
\end{fmffile}
\end{center}

\vspace{0.2cm}

\begin{center}
\begin{fmffile}{zzz-PHExchange1}
\begin{fmfgraph*}(45, 35)
\fmfleftn{l}{5}\fmfrightn{r}{5}
\fmf{phantom}{l4,m4,r4}
\fmf{phantom}{r2,m2,l2}
\fmf{fermion,tension=0.3,label.side=left,label=$p-q/2$}{l4,m4}
\fmf{fermion,tension=0.3,label.side=left,label=$k+q/2$}{r2,m2}
\fmf{fermion,tension=0.3,label.side=left,label=$k-q/2$}{m4,r4}
\fmf{fermion,tension=0.3,label.side=left,label=$p+q/2$}{m2,l2}
\fmfdot{r2,l2,r4,l4}
\fmfv{d.sh=circle,d.f=20,d.si=0.12w}{m4}
\fmfv{d.sh=circle,d.f=20,d.si=0.12w}{m2}
\fmf{fermion,tension=0,left=0.4,label.side=left,label=$k+l$}{m2,m4}
\fmf{fermion,tension=0,left=0.4,label.side=left,label=$p+l$}{m4,m2}
\fmf{dots}{r5,l1}
\end{fmfgraph*}
\end{fmffile}
\hspace{1cm}
\begin{fmffile}{zzz-PHExchange2}
\begin{fmfgraph*}(45, 35)
\fmfleftn{l}{5}\fmfrightn{r}{5}
\fmf{phantom}{l4,m4,r4}
\fmf{phantom}{r2,m2,l2}
\fmf{fermion,tension=0.3,label.side=left,label=$p-q/2$}{l4,m4}
\fmf{fermion,tension=0.3,label.side=left,label=$k+q/2$}{r2,m2}
\fmf{fermion,tension=0.3,label.side=left,label=$k-q/2$}{m4,r4}
\fmf{fermion,tension=0.3,label.side=left,label=$p+q/2$}{m2,l2}
\fmfdot{r2,l2,r4,l4}
\fmfv{d.sh=circle,d.f=20,d.si=0.12w}{m4}
\fmfv{d.sh=circle,d.f=20,d.si=0.12w}{m2}
\fmf{fermion,tension=0,left=0.4,label.side=left,label=$k+l$}{m2,m4}
\fmf{fermion,tension=0,left=0.4,label.side=left,label=$p+l$}{m4,m2}
\fmf{dots}{l5,r1}
\end{fmfgraph*}
\end{fmffile}
\end{center}

\vspace{0.2cm}

\begin{center}
\begin{fmffile}{zzz-PPExchange1}
\begin{fmfgraph*}(45, 35)
\fmfleftn{l}{5}\fmfrightn{r}{5}
\fmf{phantom}{l4,m4,r4}
\fmf{phantom}{r2,m2,l2}
\fmf{fermion,tension=0.3,label.side=left,label=$p-q/2$}{l4,m4}
\fmf{fermion,tension=0.3,label.side=right,label=$k-q/2$}{m2,r2}
\fmf{fermion,tension=0.3,label.side=right,label=$k+q/2$}{r4,m4}
\fmf{fermion,tension=0.3,label.side=left,label=$p+q/2$}{m2,l2}
\fmfdot{r2,l2,r4,l4}
\fmfv{d.sh=circle,d.f=20,d.si=0.12w}{m4}
\fmfv{d.sh=circle,d.f=20,d.si=0.12w}{m2}
\fmf{fermion,tension=0,right=0.4,label.side=right,label=$k-l$}{m4,m2}
\fmf{fermion,tension=0,left=0.4,label.side=left,label=$p+l$}{m4,m2}
\fmf{dots}{r5,l1}
\end{fmfgraph*}
\end{fmffile}
\hspace{1cm}
\begin{fmffile}{zzz-PPExchange2}
\begin{fmfgraph*}(45, 35)
\fmfleftn{l}{5}\fmfrightn{r}{5}
\fmf{phantom}{l4,m4,r4}
\fmf{phantom}{r2,m2,l2}
\fmf{fermion,tension=0.3,label.side=left,label=$p-q/2$}{l4,m4}
\fmf{fermion,tension=0.3,label.side=right,label=$k-q/2$}{m2,r2}
\fmf{fermion,tension=0.3,label.side=right,label=$k+q/2$}{r4,m4}
\fmf{fermion,tension=0.3,label.side=left,label=$p+q/2$}{m2,l2}
\fmfdot{r2,l2,r4,l4}
\fmfv{d.sh=circle,d.f=20,d.si=0.12w}{m4}
\fmfv{d.sh=circle,d.f=20,d.si=0.12w}{m2}
\fmf{fermion,tension=0,right=0.4,label.side=right,label=$k-l$}{m4,m2}
\fmf{fermion,tension=0,left=0.4,label.side=left,label=$p+l$}{m4,m2}
\fmf{dots}{l5,r1}
\end{fmfgraph*}
\end{fmffile}
\end{center}
The gray blobs represent full interaction vertices $iV$. The two cut sub-diagrams for $-D_1$ involve a quasiparticle decaying into two quasiparticles and a quasihole (or a hole decaying into two holes and a particle). The cut sub-diagrams for $-C^{ph}_1$ involve the exchange of an on-shell particle-hole pair, while the cut sub-diagrams for $-C^{pp}_1$ involve the exchange of an on-shell particle-particle (or hole-hole) pair. The computation of cut diagrams is explained in Appendix A; there, we will also argue that these three pairs are the only cut sub-diagrams that contribute at order $q$.

For $d\geq 3$ spatial dimensions, $C_1$ should scale as $\sim (q^0)^2$. This can be seen by counting the availability of collision channels constraint by energy and momentum conservation in the presence of FS~\cite{landau1959theory, abrikosov1975methods}. As we show in Appendix A, in $d\geq 3$ we can parametrize the cut sub-diagrams above by
\begin{eqnarray}
-D_1(p; q) = - \gamma(\p) \ \delta_{FS}(p) \ Z(\p)^3 \ (uu^\dagger)(p) \: (uu^\dagger)^T(p) \ \frac{|q^0| (q^0)^2}{(v^\mu(\p) q_\mu)^2},
\label{Decay_parametrization}
\end{eqnarray}
\begin{eqnarray}
-C^{ph}_1(p, k; q) = 2\lambda^{ph}(\p, \k) \ \delta_{FS}(p) \ \delta_{FS}(k) \ (Z^2 uu^\dagger)(p) (Z^2 uu^\dagger)^T(k) \ \frac{|q^0|(q^0)^2}{(v(\p)^\mu q_\mu) (v(\k)^\mu q_\mu)},
\label{Cph_parametrization}
\end{eqnarray}
\begin{eqnarray}
-C^{pp}_1(p, k; q) = - \lambda^{pp}(\p, \k) \ \delta_{FS}(p) \ \delta_{FS}(k) \ (Z^2 uu^\dagger)(p) (Z^2 uu^\dagger)^T(k) \ \frac{|q^0|(q^0)^2}{(v(\p)^\mu q_\mu) (v(\k)^\mu q_\mu)}.
\label{Cpp_parametrization}
\end{eqnarray}
(We have omitted the $i\epsilon$ prescription accompanying $v^\mu q_\mu$ in the denominator; in time-ordered correlation its sign should be $-\sgn(q^0)$, i.e. $\sgn(q_0)$, as usual.)
 In particular, the parameter $\gamma(\p)$, defined near the FS, is positive and regular, and is related to the imaginary part of the fermion self-energy via
\begin{eqnarray}
\chi_u(p) = \frac{p^0 - \xi(\p) + i\epsilon\: \sgn\,\xi(\p)}{Z(\p)} + i\frac{3}{2}\gamma(\p) \ p^0 |p^0| + (\mbox{higher orders in } p^0)
\label{chiu_near_FS_good}
\end{eqnarray}
as explained in Appendix A. The other two parameters, $\lambda^{ph}(\p, \k)$ and $\lambda^{pp}(\p, \k)$, defined near the FS, are both positive and regular, and symmetric under exchange of $\p$ and $\k$. Moreover, from the computation in Appendix A, we have the relation
\begin{eqnarray}
\int_k \frac{-v^\mu(\k) q_\mu}{Z(\k)} (C^{ph}_R)_1(p, k; q) = 2\frac{v^\mu(\p) q_\mu}{Z(\p)}(D_R)_1(p; q) = 2\int_k \frac{v^\mu(\k) q_\mu}{Z(\k)} (C^{pp}_R)_1(p, k; q).
\label{C_D_relation}
\end{eqnarray}
In terms of the parameters $\gamma$, $\lambda^{ph}$ and $\lambda^{pp}$, this reads
\begin{eqnarray}
\int_{k} \ \delta_{FS}(k) \ Z(\k) \ \lambda^{ph}(\p, \k) = \gamma(\p) = \int_{k} \ \delta_{FS}(k) \ Z(\k) \ \lambda^{pp}(\p, \k).
\label{gamma_lambda_relation}
\end{eqnarray}
In particular, this relation implies
\begin{eqnarray}
\int_k C_1(p, k; q) \ \frac{uu^\dagger(k)}{Z(\k)} v^\mu(\k) q_\mu = 0.
\label{Collision_WTId}
\end{eqnarray}
This is related to the Ward-Takahashi identity, as we will see in Section \ref{sssect_QFT_WTId}. More physically, it is related to the fact that collisions do not change the total number of fermionic excitations, as discussed above \eqref{Collision_conservation}.

Piecing up the above, the collision factor $C_1$ can be written as
\begin{eqnarray}
-C_1(p, k; q) &=& -|q^0| \ \frac{(Zuu^\dagger)(p) \: q^0}{v^\mu(\p) q_\mu} \ \delta_{FS}(p) \ \mathcal{C}(\p, \k) \ \delta_{FS}(k) \frac{(Zuu^\dagger)^T(k) \: q^0}{v^\mu(\k) q_\mu}
\end{eqnarray}
where $\mathcal{C}(\p, \k)$, symmetric in $\p, \k$, is microscopically defined when both $\p$ and $\k$ are on the FS:
\begin{eqnarray}
\delta_{FS}(p) \ \mathcal{C}(\p, \k) \ \delta_{FS}(k) &\equiv& \delta_{FS}(p) Z(\p) \gamma(\p) \: (2\pi)^{d+1}\delta^{d+1}(p-k) \nonumber \\[.2cm]
&& + \ \delta_{FS}(p) Z(\p) \left(-2\lambda^{ph}(\p, \k) + \lambda^{pp}(\p, \k) \right) \delta_{FS}(k) Z(\k)
\label{C_kernal_def}
\end{eqnarray}
(since $\mathcal{C}$ is defined only when $\p, \k$ are on the FS, we can ``remove'' $\delta_{FS}(p)\delta_{FS}(k)$ from the first term on the right-hand-side unambiguously) and it satisfies
\begin{eqnarray}
\int_k \mathcal{C}(\p, \k) \ \delta_{FS}(k) = 0.
\label{Collision_WTId_alt}
\end{eqnarray}
This is the collision effect $\mathcal{C}$ appearing in the kinetic theory in Section \ref{ssect_kinetic_BE}.

For $d=2$ spatial dimensions, $D_1$, $C^{ph}_1$ and $C^{pp}_1$ cannot be parametrized in any simple form, as discussed in Appendix B. Moreover, they are not order $q$; they also involve order $q \ln q$ terms, which are less suppressed than order $q$. The failure of the parametrization raises problem in the computation for e.g. the longitudinal current in $d=2$. But as shown in Appendix A, collisions do not contribute to the anomalous Hall current and the chiral magnetic current, so our main discussion about them is not undermined. Also, despite that there is no simple parametrization in $d=2$, \eqref{Collision_WTId} must still hold as it is dictated by the Ward-Takahashi identity.

\subsubsection{Chemical Potential Dependence of Propagator}
\label{sssect_QFT_int_Prop_on_EF}

Having defined the $q$-2PI vertex $i\wt{V}$, we are ready to find the chemical potential dependence of the propagator. The procedure below is analogous to \cite{Nozieres:1962zz}, but allowing multi-component $u^\alpha$.

We define the notation
\begin{eqnarray}
\partial^F \equiv \partial/\partial \epsilon_F - \partial/\partial p^0.
\end{eqnarray}
The subtraction of $\partial /\partial p^0$ is because our $p^0$ is defined such that $p^0=0$ at the FS, and we want $\partial^F$ to extract the effects of physically shifting the FS; for example, $\partial^F (p^0-(E-\epsilon_F)) = \partial^F E = \partial E/ \partial \epsilon_F$. The FS dependence of the propagator can be derived in analogy to \eqref{G_derivative}, but with $\partial_p^\nu$ replaced with $\partial^F$:
\begin{eqnarray}
\partial^F iG = i\Delta_0 \ \partial^F iG^{-1} - (Zuu^\dagger) \left(1-\partial^F E \right) \delta_{FS}
\label{G_FSderivative}
\end{eqnarray}
where the expression of $\partial^F G^{-1}$, and hence $\partial^F E$, are to be derived below. The second term in \eqref{G_FSderivative} relied on the assumption that when the chemical potential changes, the FS changes continuously, so \eqref{G_FSderivative} (and hence the discussion below) does not apply to discrete values of $\epsilon_F$ around which the FS develops new disconnected components, as in the example \eqref{FS_discont_change}.

Let $G_{bare}$ be the bare fermion propagator and $G_{bare}^{-1}$ is its inverse ignoring the $i\epsilon$. In the kinetic energy sector of the bare Lagrangian, $\epsilon_F$ always appears as $i\partial_{x^0} + \epsilon_F$, therefore $\partial^F G_{bare}^{-1} = 0$. Because $G^{-1}=G_{bare}^{-1} - \Sigma$ where $\Sigma$ is the self-energy, we get $\partial^F G^{-1} = -\partial^F \Sigma$. Diagrammatically, one can see in the presence of interaction, when the propagator is varied, the self-energy varies as $-\delta i\Sigma = i\wt{V}_0 \ \delta iG$. Therefore
\begin{eqnarray}
\partial^F iG^{-1} = -\partial^F i\Sigma = i\wt{V}_0 \ \partial^F iG.
\end{eqnarray}
Substituting \eqref{G_FSderivative} into the above yields
\begin{eqnarray}
\partial^F iG^{-1} = -\partial^F i\Sigma = - i\bar{V}_0 \ (Zuu^\dagger) \left(1-\partial^F E \right) \delta_{FS}.
\label{Ginv_FSderivative}
\end{eqnarray}
Recall that $\bar{V}_0$ is defined by the recursion relation $i\bar{V}_0 = i\wt{V}_0 + i\wt{V}_0 \ i\Delta_0 \ i\bar{V}_0$.

Let us focus on the change of the $u$-band eigenvalue of $G^{-1}$, given by
\begin{eqnarray}
\partial^F \chi_u = (uu^\dagger)^T \partial^F G^{-1}.
\end{eqnarray}
Now take $p$ near the FS. We can expand this in powers of $p^0$. In particular, by comparison with \eqref{chiu_near_FS}, we shall identify the coefficients at zeroth and first order in $p^0$ as
\begin{eqnarray}
\partial^F \chi_u = \left(-\frac{\partial^F E}{Z} - (E - \epsilon_F) \: \partial^F \frac{1}{Z} \right) + p^0 \: \partial^F \frac{1}{Z} + \mathcal{O}((p^0)^2).
\end{eqnarray}
To make this in parallel with \eqref{chiu_derivative}, we shall define $v^F\equiv \partial^F E$. Note that $v^F$ has nothing to do with ``Fermi velocity'' (in this paper there is no notion of Fermi velocity, as we did not assume rotational symmetry).

Now, for $p$ near the FS, we can read-off:
\begin{eqnarray}
\partial^F Z(\p) &=& \left.\partial_{p^0} \left( -Z^2 (uu^\dagger)^T \partial^F G^{-1} \right) \right|_{p \: on-shell} \nonumber \\[.2cm]
&=& \left. Z(\p) \ \partial_{p^0} \int_k (Zuu^\dagger)^T(p) \bar{V}_0(p, k) (Zuu^\dagger)(k) \ \left(1-\partial^F E(\k) \right)\delta_{FS}(k) \ \: \right|_{p \: on-shell}
\end{eqnarray}
and
\begin{eqnarray}
&& \partial^F E(\p) = \left. -(Zuu^\dagger)^T \partial^F G^{-1} \right|_{p \: on-shell} = \int_k \mathcal{U}(\p, \k) \ \left(1-\partial^F E(\k) \right)\delta_{FS}(k), \nonumber \\[.2cm]
&& \mathcal{U}(\p, \k) \equiv \left. (Zuu^\dagger)^T(p) \ \bar{V}_0(p, k) \ (Zuu^\dagger)(k) \right|_{p, k \: on-shell}.
\label{dE_dEF}
\end{eqnarray}
Thus we have proven \eqref{BFL_dE_dEF}. At the same time we found the microscopic expression for $\mathcal{U}$, which is the same as that in \cite{abrikosov1975methods} except here we need to contract with the four $u$'s. $\mathcal{U}$ is even under the exchange of $\p, \k$, because $\bar{V}_0=(\bar{V}_0)^T$.

We can also find the change of the eigenvector $u$ for $p$ near the FS. Up to an unimportant complex phase, we have
\begin{eqnarray}
\partial^F u^\alpha(p) = \sum_w \frac{w^\alpha}{-\chi_w} \left\{ w^\dagger \partial^F G^{-1} u \right\}, \ \ \ \ \ \ \ \partial^F u^\dagger_\alpha(p) = \sum_w \left\{ u^\dagger \partial^F G^{-1} w \right\}\frac{w^\dagger_\alpha}{-\chi_w},
\label{du_dEF}
\end{eqnarray}
where $\partial^F G^{-1}$ is given by \eqref{Ginv_FSderivative}. $\partial^F \mathfrak{u}$ is related to $\partial^F u$ in a way similar to \eqref{u_derivative_on_shell}, with $v^\mu$ replaced by $v^F=\partial^F E$. It appears in the kinetic formalism through \eqref{BFL_dsigma_dEF_detail}, which we will prove in Section \ref{sssect_QFT_dsigma_dEF}.

\subsection{Electromagnetic Coupling and the Ward-Takahashi Identity}
\label{ssect_QFT_EM_WTId}

\subsubsection{Bare Electromagnetic Vertices}

By our assumptions about the QFT, the EM field $A$ only couples to the kinetic sector of $\psi$ but not to the interacting sector. Due to the smallness of $A$, we only need to consider the EM vertices $A\psi^\dagger\psi$ and $A^2\psi^\dagger\psi$. We denote by $(i\wt{\Gamma}^\alpha_{\ \delta})^\mu(p; q)$ the bare $A\psi^\dagger\psi$ EM vertex with incoming fermion of momentum $p-q/2$ and fermion component index $\delta$, and outgoing fermion of momentum $p+q/2$ and index $\alpha$. We denote by $(i\wt{\Xi}^\alpha_{\ \delta})^{\mu\nu}(p; q, q')$ the bare $AA\psi^\dagger\psi$ vertex with incoming fermion of momentum $p-(q+q')/2$ and index $\delta$, outgoing fermion of momentum $p+(q+q')/2$ and index $\alpha$, and photons of incoming momenta $q$ (with vector index $\mu$) and $q'$ (with vector index $\nu$).
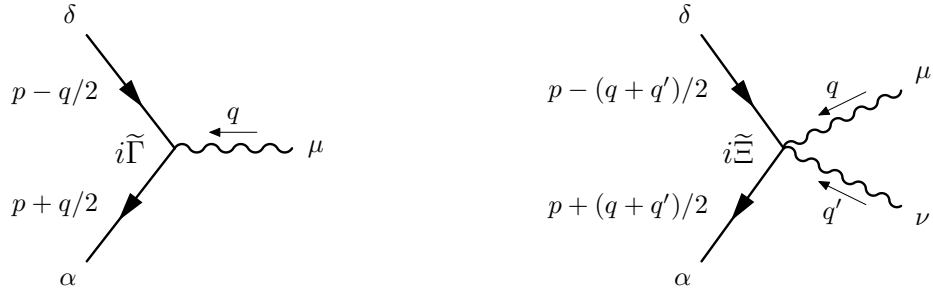
\begin{center}
\vspace{.5cm}
\hspace{1cm}
\begin{fmffile}{zzz-wtGamma}
\begin{fmfgraph*}(30,30)
\fmfleftn{l}{5}\fmfrightn{r}{5}
\fmf{fermion}{v,l1}
\fmf{fermion}{l5,v}
\fmf{photon,tension=1.5,label.dist=9,label=$q$}{r3,v}
\fmfv{label.dist=11,label={\large $i\wt{\Gamma}$}}{v}
\fmfv{label.dist=-12,label=$p+q/2$}{l2}
\fmfv{label.dist=-12,label=$p-q/2$}{l4}
\fmflabel{$\alpha$}{l1}
\fmflabel{$\delta$}{l5}
\fmflabel{$\mu$}{r3}
\momentumarrow{a}{up}{6}{r3,v}
\end{fmfgraph*}
\end{fmffile}
\hspace{4.5cm}
\begin{fmffile}{zzz-wtXi}
\begin{fmfgraph*}(30,30)
\fmfleftn{l}{5}\fmfrightn{r}{5}
\fmf{fermion}{v,l1}
\fmf{fermion}{l5,v}
\fmf{photon,tension=0.7,label.dist=9,label=$q' \hspace{-1.5mm}$}{r2,v}
\fmf{photon,tension=0.7,label.dist=9,label=$q \hspace{-1mm}$}{r4,v}
\fmfv{label.dist=11,label={\large $i\wt{\Xi}$}}{v}
\fmfv{label.dist=-10,label=$p+(q+q')/2$}{l2}
\fmfv{label.dist=-10,label=$p-(q+q')/2$}{l4}
\fmflabel{$\alpha$}{l1}
\fmflabel{$\delta$}{l5}
\fmflabel{$\mu$}{r4}
\fmflabel{$\nu$}{r2}
\momentumarrow{a}{down}{6}{r2,v}
\momentumarrow{b}{up}{6}{r4,v}
\end{fmfgraph*}
\end{fmffile}
\vspace{.5cm}
\end{center}
Since they are bare quantities, both of them are regular and analytic in $q$ (also $q'$ for $\wt{\Xi}$). To first order in $q$ (and $q'$, at same order as $q$, for $\wt{\Xi}$) we separate them as $\wt{\Gamma}=\wt{\Gamma}_0+\wt{\Gamma}_1$ and $\wt{\Xi}=\wt{\Xi}_0+\wt{\Xi}_1$.

The EM $U(1)$ gauge invariance of the bare Lagrangian requires
\begin{eqnarray}
&& q_\mu \: i\wt{\Gamma}^\mu(p) = iG_{bare}^{-1}(p-q/2) - iG_{bare}^{-1}(p+q/2),
\end{eqnarray}
\begin{eqnarray}
&& q_\mu \: i\wt{\Xi}^{\mu\nu}(p; q, q') = i\wt{\Gamma}^\nu(p-q/2; q') - i\wt{\Gamma}^\nu(p+q/2; q'), \nonumber \\[.2cm]
&& q'_\nu \: i\wt{\Xi}^{\mu\nu}(p; q, q') = i\wt{\Gamma}^\mu(p-q'/2; q) - i\wt{\Gamma}^\mu(p+q'/2; q).
\end{eqnarray}
These lead to
\begin{eqnarray}
i\wt{\Gamma}_0^\mu = -\partial_p^\mu iG_{bare}^{-1}, \ \ \ \ \ i\wt{\Gamma}_1^\mu = i\hat{\mu}^{\mu\nu} \ iq_\nu,
\label{wtGamma_gaugeinv}
\end{eqnarray}
\begin{eqnarray}
i\wt{\Xi}_0^{\mu\nu}(p) = -\partial_p^\nu i\wt{\Gamma}_0^\mu(p) = -\partial_p^\mu i\wt{\Gamma}_0^\nu(p), \ \ \ \ \ \ i\wt{\Xi}_1^{\mu\nu}(p; q, q') = -\partial_p^\nu i\wt{\Gamma}_1^\mu(p; q) - \partial_p^\mu i\wt{\Gamma}_1^\nu(p; q').
\label{wtXi_gaugeinv}
\end{eqnarray}
Here $(\hat{\mu}^\alpha_{\ \delta})^{\mu\nu}(p)$ is the bare EM dipole matrix (such as that in the Pauli term) that is antisymmetric in $\mu\nu$ and Hermitian in $^\alpha_{\ \delta}$.

\subsubsection{Full Electromagnetic Vertex}

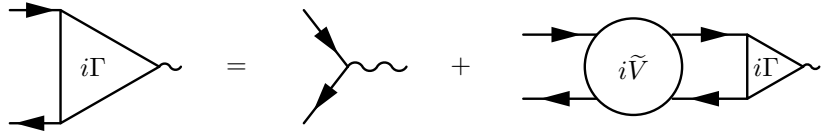
\begin{eqnarray*}
\parbox{30mm}{
\begin{fmffile}{zzz-Gamma-recursion}
\begin{fmfgraph*}(25, 15)
\fmfleftn{l}{3}\fmfrightn{r}{3}
\fmf{phantom}{r1,G1,l1}
\fmf{phantom}{l3,G2,r3}
\fmf{fermion}{G1,l1}
\fmf{fermion}{l3,G2}
\fmffreeze
\fmfpoly{empty,label=$i\Gamma$}{G2,G1,G3}
\fmf{photon,tension=4}{r2,G3}
\end{fmfgraph*}
\end{fmffile}
}
= \hspace{.5cm}
\parbox{20mm}{
\begin{fmffile}{zzz-Gamma-recursion-wtGamma}
\begin{fmfgraph*}(15, 15)
\fmfleftn{l}{3}\fmfrightn{r}{3}
\fmf{fermion}{v,l1}
\fmf{fermion}{l3,v}
\fmf{photon,tension=1.5}{r2,v}
\end{fmfgraph*}
\end{fmffile}
}
+\hspace{.5cm}
\parbox{45mm}{
\begin{fmffile}{zzz-Gamma-recursion-wtV-Gamma}
\begin{fmfgraph*}(40, 15)
\fmfleftn{l}{5}\fmfrightn{rr}{5}
\fmf{phantom}{l5,r5}\fmf{phantom,tension=3.5}{r5,rr5}
\fmf{phantom}{l4,r4}\fmf{phantom,tension=3.5}{r4,rr4}
\fmf{phantom}{l3,r3}\fmf{phantom,tension=3.5}{r3,rr3}
\fmf{phantom}{l2,r2}\fmf{phantom,tension=3.5}{r2,rr2}
\fmf{phantom}{l1,r1}\fmf{phantom,tension=3.5}{r1,rr1}
\fmffreeze
\fmf{phantom}{l4,lv4,rv4,r4}
\fmf{phantom}{r2,rv2,lv2,l2}
\fmf{fermion,tension=0}{l4,lv4}
\fmf{fermion,tension=0}{r2,rv2}
\fmf{fermion,tension=0}{rv4,r4}
\fmf{fermion,tension=0}{lv2,l2}
\fmf{phantom}{l3,o,r3}
\fmfv{d.sh=circle,d.f=empty,d.si=0.32w}{o}
\fmf{phantom,label.dist=0,label=$i\wt{V}$}{l3,r3}
\fmffreeze
\fmfpoly{empty,label=$i\Gamma$}{r4,r2,g}
\fmf{photon}{rr3,g}
\end{fmfgraph*}
\end{fmffile}
}
\end{eqnarray*}
Diagrammatically, one can see the full $A\psi^\dagger\psi$ EM vertex is given by the recursion relation
\begin{eqnarray}
i\Gamma^\mu = i\wt{\Gamma}^\mu + iV \ \left\{ iG \ i\wt{\Gamma}^\mu \ iG \right\} = i\wt{\Gamma}^\mu + i\wt{V} \ \left\{ iG \ i\Gamma^\mu \ iG \right\}
\end{eqnarray}
as drawn above. The recursion relation at zeroth order in $q$ is
\begin{eqnarray}
i\Gamma_0^\nu &=& i\wt{\Gamma}_0^\nu + i\wt{V}_0 \left(i\Delta_0 + i\Delta'_0 \right) i\Gamma_0^\nu \nonumber \\[.2cm]
&=& i\bar{\Gamma}_0^\nu + i\bar{V}_0 \ i\Delta'_0 \ i\Gamma_0^\nu,
\label{Gamma0_recursion}
\end{eqnarray}
where we defined
\begin{eqnarray}
i\bar{\Gamma}_0^\nu \equiv \left(\mathbf{1}+i\bar{V}_0 \ i\Delta_0\right) i\wt{\Gamma}_0^\nu.
\label{Gammabar_recursion}
\end{eqnarray}
The purpose of the second equality of \eqref{Gamma0_recursion} is that, now the effect of $\Delta'_0$, to be related to the deformation of the FS later, is singled out, and $i\bar{\Gamma}_0$ is independent of $q$.

The recursion relation at first order in $q$ is
\begin{eqnarray}
i\Gamma_1^\nu - i\bar{V}_0 \ i\Delta'_0 \ i\Gamma_1^\nu &=& \left(\mathbf{1}+i\bar{V}_0 \ i\Delta_0\right) i\wt{\Gamma}_1^\nu + i\bar{V}_0 \left(i\Delta'_1+i\Delta^r_1+i\Delta^s_1 - C_1\right) i\Gamma_0^\nu \phantom{,} \nonumber \\[.2cm]
&& + \left(\mathbf{1}+i\bar{V}_0 \ i\Delta_0\right) i\wt{V}_1 \left( i\Delta_0 + i\Delta'_0 \right) i\Gamma_0^\nu.
\label{Gamma1_recursion_1}
\end{eqnarray}
Of course the recursion \eqref{Gamma1_recursion_1} can be expressed in many equivalent ways; we have chosen to express it such that on the right-hand-side there is no $\Delta'_0$ (either explicit ones or those hidden in $\Gamma_0^\nu$) to the left of any quantity of order $q$. For the purpose of deriving the Boltzmann equation, we want to further rewrite \eqref{Gamma1_recursion_1} so that each $\Gamma_0^\nu$ has $\Delta'_0$ or $\Delta'_1$ or $C_1$ on its immediate left. We can achieve so by substituting \eqref{Gamma0_recursion} for those $\Gamma_0$'s in \eqref{Gamma1_recursion_1} whose immediate left are not yet $\Delta'_0$ or $\Delta'_1$ or $C_1$. The result is
\begin{eqnarray}
i\Gamma_1^\nu = i\bar{\Gamma}_1^\nu + i\bar{V}_1 \ i\Delta'_0 \ i\Gamma_0^\nu + i\bar{V}_0 \ i\Delta'_1 \ i\Gamma_0^\nu + i\bar{V}_0 \ (-C_1) \ i\Gamma_0^\nu + i\bar{V}_0 \ i\Delta'_0 \ i\Gamma_1^\nu,
\label{Gamma1_recursion}
\end{eqnarray}
where
\begin{eqnarray}
i\bar{V}_1 \equiv \left(\mathbf{1}+i\bar{V}_0 \ i\Delta_0\right) i\wt{V}_1 \left(\mathbf{1}+i\bar{V}_0 \ i\Delta_0\right)^T + i\bar{V}_0 \left(i\Delta^r_1+i\Delta^s_1\right) i\bar{V}_0,
\label{barV1}
\end{eqnarray}
\begin{eqnarray}
i\bar{\Gamma}_1^\nu &\equiv& \left(\mathbf{1}+i\bar{V}_0 \ i\Delta_0\right) i\wt{\Gamma}_1^\nu + i\bar{V}_1 \ i\Delta_0 \ i\wt{\Gamma}_0^\nu + i\bar{V}_0 \left(i\Delta^r_1+i\Delta^s_1\right) i\wt{\Gamma}_0^\nu \nonumber \\[.2cm]
&=& \left(\mathbf{1}+i\bar{V}_0 \ i\Delta_0\right) \left(i\wt{\Gamma}_1^\nu + i\wt{V}_1 \ i\Delta_0 \ i\bar{\Gamma}_0^\nu \right) + i\bar{V}_0 \left(i\Delta^r_1+i\Delta^s_1\right) i\bar{\Gamma}_0^\nu
\label{barGamma1}
\end{eqnarray}
are partial sums at first order in $q$ that involve no factor of $\Delta'$ or $C_1$. By construction, $i\bar{V}_1$ and $i\bar{\Gamma}_1$ are analytic in $q$ as $q\rightarrow 0$. Thus, in \eqref{Gamma1_recursion} we singled out the $\Delta'_0, \Delta'_1$ and $C_1$ -- the only factors that are non-analytic as $q\rightarrow 0$ -- in the recursion, which are to be related to the quasiparticle excitations.

We do not need to consider the ``full $AA\psi^\dagger \psi$ vertex''. In fact, the only place $\wt{\Xi}$ shows up in our proof is the expression of the current, in which we will immediately use \eqref{wtXi_gaugeinv} to eliminate $\wt{\Xi}$.

\subsubsection{Ward-Takahashi Identity}
\label{sssect_QFT_WTId}

Later in Section \ref{ssect_QFT_BE} we will show how the Boltzmann equation \eqref{BFL_Boltzmann_Eq_0}\eqref{BFL_Boltzmann_Eq_1} follow exactly from \eqref{Gamma0_recursion} and \eqref{Gamma1_recursion}. Before that, we need to answer a question: In QFT, it is the matrix $\Gamma^\nu$ governing the coupling to $A$, while in the kinetic formalism, it is the velocity $v^\nu$ (plus order $q$ couplings such as EM dipole). How to relate $\Gamma^\nu$ to $v^\nu$? The answer is the generalized Ward-Takahashi identity~\cite{Takahashi:1957xn}:
\begin{eqnarray}
\left\{iG(p+q/2) \ i\Gamma^\nu(p; q) \ iG(p-q/2) \right\} q_\nu = iG(p-q/2) - iG(p+q/2).
\label{WTId}
\end{eqnarray}
We want to extract its implications at leading and sub-leading orders in $q$ in the presence of FS.

At leading order in $q$, the Ward-Takahashi identity reads
\begin{eqnarray}
\left( i\Delta_0 + i\Delta'_0 \right) \Gamma_0^\nu \ q_\nu = -i\Delta_0 \ \partial_p^\nu G^{-1} \ q_\nu + i(Zuu^\dagger) \delta_{FS} \delta^\nu_i v^i q_\nu
\label{WTId_1_raw}
\end{eqnarray}
using \eqref{G_derivative}. This is equivalent to
\begin{eqnarray}
\Gamma_0^\nu \ q_\nu = -\partial_p^\nu G^{-1} \ q_\nu.
\label{WTId_1}
\end{eqnarray}
One can easily verify the equivalence by contracting $i\Delta_0+i\Delta'_0$ on the left of \eqref{WTId_1}, with the aid of \eqref{v_in_QFT}, to recover \eqref{WTId_1_raw}. We will see the result \eqref{WTId_1} is related to the gauge invariance of $\delta f_0$.

We can extract more detailed information from \eqref{WTId_1} -- we gain an identity similar to the original Ward identity~\cite{Ward:1950xp}, but in the presence of FS. For this purpose let us treat $|\q|/q^0$ as an independent small expansion parameter, and expand \eqref{WTId_1} to its zeroth and first order. This gives us two equations, about $\bar{\Gamma}^0_0$ and $\bar{\Gamma}^i_0$ respectively. Solving them with the help of \eqref{Gamma0_recursion} and the explicit expression for $\Delta'_0$, we find the Ward identity in the presence of FS:
\begin{eqnarray}
\bar{\Gamma}_0^\nu = -\partial_p^\nu G^{-1} + \bar{V}_0 \ (Zuu^\dagger) \delta_{FS} \delta^\nu_i v^i.
\label{Gamma0_WTId}
\end{eqnarray}
We can equivalently express \eqref{Gamma0_WTId} as
\begin{eqnarray}
i\Delta_0 \ i\bar{\Gamma}_0^\nu = -\partial_p^\nu iG + \left(\mathbf{1} + i\Delta_0 \ i\bar{V}_0 \right) (Zuu^\dagger) \delta_{FS} \delta^\nu_i v^i
\label{Gamma0_WTId_1}
\end{eqnarray}
using \eqref{G_derivative}. As we will see later, this result will help us relate the EM vertex in QFT to the velocity in the kinetic formalism.

At sub-leading order in $q$, the Ward-Takahashi identity reads
\begin{eqnarray}
\left(\Delta_0+\Delta'_0\right) \Gamma_1^\nu \ q_\nu + \left(\Delta^r_1+\Delta^s_1+\Delta'_1+iC_1\right) \Gamma_0^\nu \ q_\nu = 0.
\end{eqnarray}
The $C_1 \Gamma_0^\nu q_\nu$ term vanishes on its own, due to \eqref{WTId_1}, \eqref{v_in_QFT} and \eqref{Collision_WTId}. For the remaining terms, we can conclude
\begin{eqnarray}
\Delta_0 \ \Gamma_1^\nu \ q_\nu + \Delta^r_1 \ \Gamma_0^\nu \ q_\nu = 0 = \Delta'_0 \ \Gamma_1^\nu \ q_\nu + \left(\Delta^s_1+\Delta'_1\right) \Gamma_0^\nu \ q_\nu.
\label{WTId_2}
\end{eqnarray}
The two sides must vanish separately because the right-hand-side involves the singular factor $\delta(\xi(\p))$, while the left-hand-side does not. Later we will see \eqref{WTId_2} is related to the gauge invariance of $\delta f_1$.

\subsubsection{Luttinger's Theorem}
\label{sssect_LuttTh}

To demonstrate the usefulness of the formalism we have developed so far, we present a simple proof to the celebrated Luttinger's theorem \cite{Luttinger:1960zz} in Fermi liquid:
\begin{align}
\langle J^0 \rangle = \int_\p \theta(\epsilon_F-E(\p)).
\end{align}
This is a statement that is obvious for Fermi gas but highly non-trivial for interacting Fermi liquid. To show this, we take the chemical potential derivative of the charge density:
\begin{align}
\partial^F \langle i J^0 \rangle &= \int_p \partial^F\left( -\tr \{ \wt\Gamma_0^0(p) \, iG(p) \} \right) = -\partial^F \left((i\wt\Gamma_0^0)^T \, iG\right) = -(i\wt\Gamma_0^0)^T \partial^F iG.
\end{align}
Then we apply \eqref{G_FSderivative} and \eqref{Ginv_FSderivative}, and then \eqref{Gammabar_recursion}, \eqref{Gamma0_WTId} and \eqref{v_in_QFT}:
\begin{align}
\frac{\partial}{\partial\epsilon_F} \langle i J^0 \rangle &= (i\wt\Gamma_0^0)^T \left( \mathbf{1} + i\Delta_0 \ i\bar{V}_0 \right) (Zuu^\dagger) (1-\partial^F E) \delta_{FS} \nonumber \\[.2cm]
&= -(\partial_p^0 iG^{-1})^T (Zuu^\dagger) (1-\partial^F E) \delta_{FS} = i (v^0)^T \, (1-\partial^F E) \delta_{FS} = i \frac{\partial}{\partial\epsilon_F} \int_\p \theta(\epsilon_F-E(\p))
\end{align}
(note the facts that $(\bar{V}_0)^T = \bar{V}_0$ and $(\Delta_0)^T = \Delta_0$). Thus, we have proven the chemical potential derivative of the theorem, and hence the theorem itself assuming appropriate boundary conditions for the $\epsilon_F$ integral. Our proof is a simplified and more transparent variant of Landau's proof presented in \cite{abrikosov1975methods}; this proof is limited to Fermi liquids (which is all we consider in this paper) and is not as generalizable as Luttinger's proof \cite{Luttinger:1960zz, abrikosov1975methods}.

\subsection{Boltzmann Equation}
\label{ssect_QFT_BE}

Having extracted \eqref{Gamma0_WTId} from the Ward-Takahashi identity, we are ready to prove the Boltzmann equation \eqref{BFL_Boltzmann_Eq_0}\eqref{BFL_Boltzmann_Eq_1} from the recursion relations \eqref{Gamma0_recursion} and \eqref{Gamma1_recursion}. The distribution of excitations $\delta f$ will be defined in terms of QFT quantities, and as one should expect, our definition agrees with the Wigner function approach. Our derivation also provides the microscopic expressions for $\mathcal{V}^\nu$ and $\mu^{\mu\nu}$.

When an external EM field of small $q$ is present, the propagation of a quasiparticle is no longer translationally invariant -- the two-point propagator now depends on both $p$ and $q$. More precisely, 
\begin{eqnarray}
iG(p) \ \ \ \longrightarrow \ \ \ iG(p) + \left\{iG(p+q/2) \ i\Gamma^\nu(p; q) \ iG(p-q/2) \right\} A_\nu(q)
\label{prop_shift_by_A}
\end{eqnarray}
at linear response. We will focus on the shifted piece.

\subsubsection{Zeroth Order in $q$}

At zeroth order in $q$, using the identity \eqref{Gamma0_WTId}, we can express the recursion relation \eqref{Gamma0_recursion} as
\begin{eqnarray}
i\Gamma_0^\nu \ A_\nu = -i\partial_p^\nu G^{-1} \ A_\nu - i\bar{V}_0 (Zuu^\dagger) \ \delta W_0,
\label{Gamma0_in_W0}
\end{eqnarray}
where $\delta W_0$ is a quantity restricted on the FS:
\begin{eqnarray}
\delta W_0 \equiv \frac{(uu^\dagger)^T}{Z} \ \Delta'_0 \ \Gamma_0^\nu \ A_\nu - \delta_{FS} v^i A_i.
\label{W0_in_Gamma0}
\end{eqnarray}
We can see $\delta W_0$ is gauge invariant from \eqref{WTId_1} and \eqref{v_in_QFT}. Substituting \eqref{Gamma0_in_W0} into \eqref{W0_in_Gamma0}, we find the recursion relation for $\delta W_0$:
\begin{eqnarray}
\delta W_0 = \delta_{FS} \ \frac{v^i}{v^\mu q_\mu-i\epsilon\: \sgn(q^0)} \left(-iF_{i0} - q_i \ \mathcal{U} \ \delta W_0 \right).
\label{Boltzmann_Eq_0}
\end{eqnarray}
This proves the Boltzmann equation \eqref{BFL_Boltzmann_Eq_0} at zeroth order in $q$, if we make the identification
\begin{eqnarray}
2\pi \delta(p^0-(E(\p)-\epsilon_F)) \ \delta f (\p; q) \equiv \delta W(p; q)
\end{eqnarray}
to factor out the on-shell condition. Note that the computation above is time-ordered, therefore the $i\epsilon$ prescription depends on $\sgn(q^0)$; when computing the physical quasiparticle distribution in kinetic theory, retarded boundary condition should be used, which corresponds to removing the $\sgn(q^0)$ factor in the $i\epsilon$ prescription. This proof is a generalization to that in \cite{abrikosov1975methods}, with multi-component spinor / Bloch state $u^\alpha$ and the presence of external EM field, and without rotational symmetry.

The definition \eqref{W0_in_Gamma0} of $\delta W_0$ agrees with the quasiparticle Wigner function to first order in $A$ and zeroth order in $q$. The first term of \eqref{W0_in_Gamma0} corresponds to the singular part of \eqref{prop_shift_by_A} projected onto the $u$ band (at zeroth order in $q$), which we identify as the distribution of excited quasiparticles; the factor of $Z$ difference is the quasiparticle wave function renormalization. The second term of \eqref{W0_in_Gamma0} is due to the Peierl's substitution in the equilibrium part $\theta(\epsilon_F-E)$ of the Wigner function; Fourier transforming to the position space, it corresponds to the Wilson line in the Wigner function expanded at first order in $A$.

\subsubsection{First Order in $q$}

At first order in $q$, we assert we should define
\begin{eqnarray}
\delta W_1 \equiv \frac{(uu^\dagger)^T}{Z} \left(\Delta'_0 \ \Gamma_1^\nu A_\nu + \Delta'_1 \ \Gamma_0^\nu A_\nu + iC_1 \ \Gamma_0^\nu A_\nu\right).
\label{W1_in_Gamma}
\end{eqnarray}
Its gauge invariance follows from \eqref{WTId_2} and \eqref{Collision_WTId}. It also agrees with the order $q$ singular part of the Wigner function -- as can be seen from \eqref{prop_shift_by_A} -- projected onto the $u$ band. In particular, the projection onto the $u$ band should be done by the momentum space Wilson line
\begin{eqnarray}
\lim_{n\rightarrow \infty} \ \left(u^\alpha u_{\alpha_1 \phantom{\beta\!\!\!\!}}^\dagger\right)\left(p+q/2\right) \ \left(u^{\alpha_1} u_{\alpha_2 \phantom{\beta\!\!\!\!}}^\dagger\right)\left(p+q(n-1)/2n\right) \cdots \left(u^{\alpha_{2n}} u_\beta^\dagger\right)\left(p-q/2\right).
\end{eqnarray}
It equals $u^\alpha(p) u^\dagger_\beta(p) + \mathcal{O}(q^2)$, so we can just use $(uu^\dagger)^T(p)$ at first order in $q$.

Now we derive the kinetic recursion relation for $\delta W_1$. Substituting \eqref{Gamma0_recursion} and \eqref{Gamma1_recursion} into \eqref{W1_in_Gamma}, we have
\begin{eqnarray}
\delta W_1 &=& \frac{(uu^\dagger)^T}{Z} \left[-\Delta'_0 \ \bar{V}_1 \ \Delta'_0 \ \Gamma_0^\nu - \Delta'_0 \ \bar{V}_0 \ \Delta'_0 \ \Gamma_1^\nu + iC_1 \ \Gamma_0^\nu - \Delta'_0 \ \bar{V}_0 \ iC_1 \ \Gamma_0^\nu \right. \nonumber \\[.2cm]
&& \hspace{2cm} \left. - \ \left(\Delta'_0 \ \bar{V}_0 \ \Delta'_1 + \Delta'_1 \ \bar{V}_0 \ \Delta'_0\right) \Gamma_0^\nu + \left(\Delta'_0 \ \bar{\Gamma}_1^\nu + \Delta'_1 \ \bar{\Gamma}_0^\nu \right) \right] \ A_\nu.
\end{eqnarray}
We use the identity
\begin{eqnarray}
\Delta'_1 &=& (uu^\dagger) (uu^\dagger)^T \Delta'_1 + \Delta'_1 (uu^\dagger) (uu^\dagger)^T \nonumber \\[.2cm]
&=& (uu^\dagger) (uu^\dagger)^T \Delta'_1 + iq_\lambda \mathcal{A}^\lambda \ \Delta'_0 = \Delta'_0 \ iq_\lambda \mathcal{A}^\lambda + \Delta'_1 (uu^\dagger) (uu^\dagger)^T
\label{Delta_prime_1_Id}
\end{eqnarray}
and the fact $(uu^\dagger)^T \Delta'_1 (uu^\dagger) = 0$ to rewrite $\delta W_1$ as
\begin{eqnarray}
\delta W_1 &=& \frac{\Delta' (uu^\dagger)^T}{Z} \left[-\left( \bar{V}_1 + iq_\lambda \mathcal{A}^\lambda \ \bar{V}_0 + \bar{V}_0 \ iq_\lambda \mathcal{A}^\lambda \right) \Delta'_0 \ \Gamma_0^\nu + \left(\bar{\Gamma}_1^\nu + iq_\lambda \mathcal{A}^\lambda \ \bar{\Gamma}_0^\nu \right) \right. \nonumber \\[.2cm]
&& \hspace{4.3cm} \left. - \ \bar{V}_0 (uu^\dagger) (uu^\dagger)^T \left(\Delta'_0 \ \Gamma_1^\nu + \Delta'_1 \ \Gamma_0^\nu + iC_1 \ \Gamma_0^\nu\right) \right] A_\nu \nonumber \\[.2cm]
&& + \ \frac{(uu^\dagger)^T}{Z} iC_1 \ \Gamma_0^\nu A_\nu
\end{eqnarray}
The second line can be easily identified as $(1/Z^2) \Delta' \ \mathcal{U} \ \delta W_1$. In the first line, we substitute \eqref{W0_in_Gamma0} for $\Delta'_0 \Gamma_0^\nu A_\nu$. Then we define the gradient interaction potential via
\begin{eqnarray}
&& \hspace{-1cm} iq_\mu \: \mathcal{V}^\mu(\p, \k) \equiv \nonumber \\[.2cm]
&& \hspace{-.5cm} \left. (Zuu^\dagger)^T(p) \left[ \bar{V}_1(p, k; q) +iq_\mu \left( \mathcal{A}^\mu(p) \bar{V}_0(p, k) + \bar{V}_0(p, k) \mathcal{A}^\mu(k) \right) \right] (Zuu^\dagger)(k) \right|_{p, k \ on \ FS}
\label{grad_int_def}
\end{eqnarray}
(note that even if the microscopic interaction is contact interaction, in kinetic theory $\mathcal{V}^\mu$ is still non-zero) and define the EM dipole moment via
\begin{eqnarray}
iq_\mu \: \mu^{\mu\nu}(\p) \equiv \left. (Zuu^\dagger)^T \left( \bar{\Gamma}_1^\nu + iq_\mu \mathcal{A}^\mu \ \bar{\Gamma}_0^\nu \right) \right|_{p \ on \ FS} - iq_\mu \: \mathcal{V}^\mu \ \delta_{FS} \delta^\nu_i v^i.
\label{EM_dipole_def}
\end{eqnarray}
As we will show explicitly below, $\mu^{\mu\nu}$ is antisymmetric in $\mu\nu$. With these definitions, the recursion relation for $\delta W_1$ becomes
\begin{eqnarray}
\delta W_1 = \delta_{FS} \ \frac{v^i \: q_i}{v^\mu q_\mu - i\epsilon \: \sgn(q^0)} \left(\mu^{\nu\lambda} iq_\nu A_\lambda - \mathcal{U} \ \delta W_1 - \mathcal{V}^\nu \ iq_\nu \ \delta W_0 \right) \ + \ \frac{(uu^\dagger)^T}{Z} iC_1 \ \Gamma_0^\nu A_\nu.
\label{Boltzmann_Eq_1_raw}
\end{eqnarray}
(The gauge invariance of $\delta W_1$ also implicitly requires the antisymmetry of $\mu^{\mu\nu}$.)

The last step is to rewrite the $C_1$ term (valid for $d\geq 3$ only):
\begin{eqnarray}
\frac{(uu^\dagger)^T}{Z} iC_1 \ \Gamma_0^\nu A_\nu &=& \frac{i|q^0| q^0}{v^\mu q_\mu} \ \delta_{FS} \ \mathcal{C} \ \delta_{FS} \left(\frac{v^i q_i}{v^\mu q_\mu} -1\right) (Z uu^\dagger)^T \ \Gamma_0^\nu A_\nu \nonumber \\[.2cm]
&=& \frac{i|q^0| q^0}{v^\mu q_\mu} \ \delta_{FS} \ \mathcal{C} \left[(\delta W_0 + \delta_{FS} v^i A_i) \phantom{(Z uu^\dagger)^T} \right. \nonumber \\[.2cm] 
&& \hspace{2cm} \left. - \delta_{FS} (Z uu^\dagger)^T \left(-\partial_p^\nu G^{-1} A_\nu - \bar{V}_0 (Zuu^\dagger) \ \delta W_0\right) \right] \nonumber \\[.2cm]
&=& \frac{i|q^0| q^0}{v^\mu q_\mu} \ \delta_{FS} \ \mathcal{C} \left(\delta W_0 + \delta_{FS} \ \mathcal{U} \ \delta W_0 \right)
\end{eqnarray}
where in the second equality we used \eqref{W0_in_Gamma0} and \eqref{Gamma0_in_W0}, and in the third equality we used \eqref{v_in_QFT} and \eqref{Collision_WTId}.

Now we have
\begin{eqnarray}
\delta W_1 &=& \frac{\delta_{FS}}{v^\mu q_\mu - i\epsilon \: \sgn(q^0)} \left[ v^i q_i \left(\mu^{\nu\lambda} \frac{F_{\nu\lambda}}{2} - \mathcal{U} \ \delta W_1 - \mathcal{V}^\nu \ iq_\nu \ \delta W_0 \right) \right. \nonumber \\[.2cm]
&& \hspace{5cm} \left. + \ i|q^0| q^0 \ \mathcal{C} \left(\delta W_0 + \delta_{FS} \ \mathcal{U} \ \delta W_0 \right) \phantom{\frac{1}{1}} \right].
\label{Boltzmann_Eq_1}
\end{eqnarray}
The computation done here is time-ordered. When computing physical quasiparticle distribution, we should use retarded boundary condition, which corresponds to using the retarded versions of $\Delta'$ and $C$ -- that is, to remove the $\sgn(q^0)$ on the $i\epsilon$ prescription, and remove the absolute value on $|q^0|$ in the collision term. This proves \eqref{BFL_Boltzmann_Eq_1}.

\subsubsection{Electromagnetic Dipole Moment}
\label{sssect_QFT_EM_dipole}

The definition \eqref{EM_dipole_def} of $\mu^{\mu\nu}$ is unusual, and its antisymmetry in $\mu\nu$ is not manifest. Now we present it in a more familiar form that is explicitly antisymmetric. Using \eqref{Gamma0_WTId}, \eqref{Gamma0_WTId_1} and the explicit expressions of $\bar{\Gamma}_1$ and $\bar{V}_1$, we can express the EM dipole moment as
\begin{eqnarray}
\mu^{\mu\nu} = \mu_{bare}^{\mu\nu} + \mu_{band}^{\mu\nu} + \mu_{anom.}^{\mu\nu}
\end{eqnarray}
which we explain term by term below.

The bare EM dipole moment is due to the bare EM dipole matrix (e.g. the Pauli term):
\begin{eqnarray}
\mu_{bare}^{\mu\nu} \equiv \left. (Zuu^\dagger)^T \left(\mathbf{1} - \bar{V}_0 \Delta_0 \right) \hat{\mu}^{\mu\nu} \right|_{p \ on \ FS}
\end{eqnarray}
where $\hat{\mu}^{\mu\nu}$ has been introduced in \eqref{wtGamma_gaugeinv} and is antisymmetric in $\mu\nu$.

The band EM dipole moment, due to the $p$ dependence of $u$, is
\begin{eqnarray}
\mu_{band}^{\mu\nu} \equiv -\left. (Zuu^\dagger)^T \mathcal{A}^\mu \ \partial_p^\nu G^{-1} \right|_{p \ on \ FS} = \left. -i Z \ \left\{ \partial_p^{[\mu} u^\dagger \ G^{-1} \ \partial_p^{\nu]} u \right\} \right|_{p \ on \ FS}.
\end{eqnarray}
In the second equality we used the trick \eqref{IBPtrick}. It is explicitly antisymmetric in $\mu\nu$. In non-interacting theory, $u$ depends only on $\p$ but not $p^0$, so $\mu_{band}$ would be purely magnetic (e.g. the $g=2$ magnetic dipole of free Dirac fermion). In interacting theory, $u$ may depend on $p^0$, so $\mu_{band}$ may have electric dipole components \cite{shindou2008gradient}.

The anomalous EM dipole moment, due to interactions, is defined via
\begin{eqnarray}
iq_\mu \: \mu_{anom.}^{\mu\nu} &\equiv& -\left. (Zuu^\dagger)^T \bar{V}_0 \left( -\left(\Delta^r_1 + \Delta^s_1\right) \partial_k^\nu G^{-1} + iq_\mu \mathcal{A}^\mu \ (Zuu^\dagger) \delta_{FS} \delta^\nu_i v^i \right) \right|_{p \ on \ FS} \nonumber \\[.2cm]
&& -\left. (Zuu^\dagger)^T \left(\mathbf{1} - \bar{V}_0 \Delta_0 \right) \wt{V}_1 \ \partial_k^\nu iG \right|_{p \ on \ FS}.
\end{eqnarray}
To get a better understanding of $\mu_{anom.}^{\mu\nu} $, we do the following. For the term with $\Delta^r_1$, we use the explicit expression of $\Delta^r_1$. For the term with $\Delta^s_1$, we use the identity
\begin{eqnarray}
\Delta^s_1 \ \partial_k^\nu G^{-1} = \mathcal{A}^\nu \ (Zuu^\dagger) \delta_{FS} \delta^\mu_i v^i \ iq_\mu
\label{Deltas1_dGinv}
\end{eqnarray}
which again follows from the trick \eqref{IBPtrick}.
Now, the anomalous EM dipole moment reads
\begin{eqnarray}
\mu_{anom.}^{\mu\nu} &=& -\left. (Zuu^\dagger)^T \bar{V}_0 \left( \left\{G (\partial_p^{[\mu} G^{-1}) G (\partial_p^{\nu]} G^{-1}) G\right\} + 2\mathcal{A}^{[\mu} \ (Zuu^\dagger) \delta_{FS} \delta^{\nu]}_i v^i \right) \right|_{p \ on \ FS} \nonumber \\[.2cm]
&& + \left. (Zuu^\dagger)^T \left(\mathbf{1} - \bar{V}_0 \Delta_0 \right) \partial_q^\mu \left( i\wt{V}_1 \ \partial_k^\nu iG \right) \right|_{p \ on \ FS}.
\label{EM_dipole_anom}
\end{eqnarray}
The antisymmetry in $\mu\nu$ is manifest in the first line. Gauge invariance of \eqref{Boltzmann_Eq_1_raw} requires the second line above to be antisymmetric in $\mu\nu$ too; more explicitly we show this from diagrams in Appendix C.

In general, $\mu_{anom.}^{i0}\neq 0$, so even when there is no bare electric dipole matrix, the quasiparticle will still acquire an electric dipole moment due to interactions. This gives rise to the second term in \eqref{BFL_AHE_uniform} which is absent in usual Fermi gas.

\subsection{Current}
\label{ssect_QFT_current}

We now prove the expression of the current \eqref{BFL_current_0}\eqref{BFL_current_1}. Previously we have defined $\delta W$, $\mathcal{U}$, $\mathcal{V}^\nu$ and $\mu^{\mu\nu}$ from QFT, but we have not shown they are real in the position space. But these immediately follow once we have \eqref{BFL_current}, because in position space the quantum expectation of the current must be real for arbitrary $A$, $q$, interaction strength and initial / boundary conditions of $\delta W$.

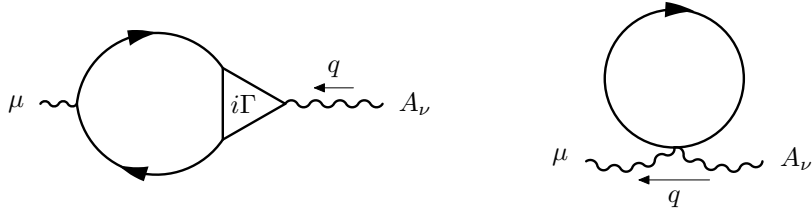
\begin{center}
\begin{fmffile}{zzz-current-1}
\begin{fmfgraph*}(45, 25)
\fmfleftn{l}{3}\fmfrightn{r}{3}
\fmf{photon,tension=8}{l2,g}
\fmf{fermion,left=0.7}{G1,g}
\fmf{fermion,left=0.7}{g,G2}
\fmfpoly{empty,label=$i\Gamma$,tension=1}{G2,G1,G3}
\fmf{photon,tension=3,label.side=right,label.dist=10,label=$q$}{r2,G3}
\fmflabel{$A_\nu$}{r2}
\fmflabel{$\mu$}{l2}
\momentumarrow{a}{up}{6}{r2,G3}
\end{fmfgraph*}
\end{fmffile}
\hspace{2cm}
\begin{fmffile}{zzz-current-2}
\begin{fmfgraph*}(25, 25)
\fmfleftn{l}{6}\fmfrightn{r}{6}
\fmf{phantom}{l6,m,r6}
\fmf{phantom_arrow,left=1.22}{l3,r3}
\fmffreeze
\fmf{photon,left=0.3}{o,l2}
\fmf{photon,left=0.3}{r2,o}
\fmf{plain,left,tension=0.1}{o,m}
\fmf{plain,left,tension=0.1}{m,o}
\fmf{phantom}{l2,lm,l3}
\fmf{phantom}{r2,rm,r3}
\fmflabel{$A_\nu$}{rm}
\fmflabel{$\mu$}{lm}
\fmf{phantom,label.side=left,label.dist=11,label=$q$}{r2,l2}
\momentumarrow{a}{down}{7}{r2,l2}
\end{fmfgraph*}
\end{fmffile}
\end{center}

As drawn above, the expectation of the current induced by $A$ at linear response is given by
\begin{eqnarray}
i\delta J^\mu(q) &=& - \int_p \tr \left\{ i\wt{\Gamma}^\mu(p; -q) \ iG(p+q/2) \ i\Gamma^\nu(p; q) \ iG(p-q/2) \right\} A_\nu(q). \nonumber \\[.2cm]
&& - \int_p \tr \left\{ {i\wt{\Xi}^{\mu\nu}}(p; -q, q) \ iG(p) \right\} A_\nu(q)
\label{current_QFT}
\end{eqnarray}
where the negative sign is due to the fermion loop. In the second line, we use \eqref{wtXi_gaugeinv} and integrate $p$ by parts to eliminate $\wt{\Xi}$. Below we work in double fermion notation, at zeroth and first order in $q$ separately.

We emphasize that here we are computing the time-ordered correlation of $\delta J$ and $A$, while in linear response we should compute the retarded correlation. This difference only shows up in the recursion relation that $\delta W$ satisfies, i.e. the Boltzmann equation, and in the above we have already handled this difference. The expression of $\delta J$ in terms of $\delta W$ is the same for time-ordered and retarded correlation.

\subsubsection{Zeroth Order in $q$}

At zeroth order in $q$, 
\begin{eqnarray}
i\delta J_0^\mu = -(i\wt{\Gamma}_0^\mu)^T \left(i\Delta_0 \ i\Gamma_0^\nu + i\Delta'_0 \ i\Gamma_0^\nu + \partial_p^\nu iG\right) A_\nu
\end{eqnarray}
where the integration over $p$ is understood. For the $i\Gamma_0^\nu$ in the first term, whose immediate left is not $\Delta'_0$, we apply the recursion relation \eqref{Gamma0_recursion}, and get
\begin{eqnarray}
i\delta J_0^\mu = -(i\wt{\Gamma}_0^\mu)^T \left(\mathbf{1} + i\Delta_0 \ i\bar{V}_0\right) i\Delta'_0 \ i\Gamma_0^\nu A_\nu - (i\wt{\Gamma}_0^\mu)^T \left( i\Delta_0 \ i\bar{\Gamma}_0^\nu + \partial_p^\nu iG\right) A_\nu
\end{eqnarray}
Due to \eqref{Gamma0_WTId_1} and the facts $(\bar{V}_0)^T = \bar{V}_0$, $(\Delta_0)^T = \Delta_0$, the above reduces to
\begin{eqnarray}
\delta J_0^\mu = (\bar{\Gamma}_0^\mu)^T \left( \Delta'_0 \ \Gamma_0^\nu \ A_\nu - (Zuu^\dagger) \ \delta_{FS} \delta^\nu_i v^i\right) A_\nu = (\bar{\Gamma}_0^\mu)^T (Zuu^\dagger) \ \delta W_0.
\end{eqnarray}
Finally, applying \eqref{Gamma0_WTId}, we obtain
\begin{eqnarray}
\delta J_0^\mu = (v^\mu)^T \ \delta W_0 + \left(\delta^\mu_i v^i \delta_{FS}\right)^T \mathcal{U} \ \delta W_0.
\end{eqnarray}
The transpose on the left implies integration over $p$. This is \eqref{BFL_current_0}.

\subsubsection{First Order in $q$}

At first order in $q$, 
\begin{eqnarray}
i\delta J_1^\mu &=& -(i\wt{\Gamma}_1^\mu(-q))^T \left(i\Delta_0 + i\Delta'_0\right) i\Gamma_0^\nu A_\nu - (i\wt{\Gamma}_0^\mu)^T \left(i\Delta_0 + i\Delta'_0\right) i\Gamma_1^\nu(q) A_\nu \nonumber \\[.2cm]
&& - \ (i\wt{\Gamma}_0^\mu)^T \left(i\Delta'_1 + i\Delta^r_1 + i\Delta^s_1 - C_1 \right)(q) \ i\Gamma_0^\nu A_\nu \nonumber \\[.2cm]
&& - \ (i\wt{\Gamma}_1^\mu(-q))^T \ \partial_p^\nu iG \ A_\nu - (i\wt{\Gamma}_1^\nu(q))^T \ \partial_p^\mu iG \ A_\nu.
\end{eqnarray}
We rewrite this according to the following: If the immediate left of an $i\Gamma_0^\nu$ is not $\Delta'_0$ or $\Delta'_1$ or $C_1$, we apply the recursion relations \eqref{Gamma0_recursion} to it; similarly, if the immediate left of an $i\Gamma_1^\nu$ is not $\Delta'_0$, we apply \eqref{Gamma1_recursion} to it. We find
\begin{eqnarray}
i\delta J_1^\mu &=& - (i\bar{\Gamma}_0^\mu)^T \left(i\Delta'_0 \ i\Gamma_1^\nu(q) + i\Delta'_1(q) \ i\Gamma_0^\nu - C_1(q) \ i\Gamma_0^\nu\right) A_\nu - (i\bar{\Gamma}_1^\mu(-q))^T \ i\Delta'_0 \ i\Gamma_0^\nu A_\nu \nonumber \\[.2cm]
&& + \ \mbox{(terms regular in $q$)}.
\label{deltaJ1_mid}
\end{eqnarray}
We will take care of the terms in the second line of \eqref{deltaJ1_mid} later. To terms in the first line, we apply the identity \eqref{Delta_prime_1_Id}, and get
\begin{eqnarray}
(i\bar{\Gamma}_0^\mu)^T (Zuu^\dagger) \ \delta W_1 + \left( (i\bar{\Gamma}_1^\mu(-q))^T + (i\bar{\Gamma}_0^\mu)^T iq_\lambda \mathcal{A}^\lambda \right) (Zuu^\dagger) \left(\delta W_0 + \delta_{FS} v^j A_j \right) 
\end{eqnarray}
Now use \eqref{Gamma0_WTId} in the first term, and \eqref{EM_dipole_def} and the facts $\mathcal{A}^\lambda=-(\mathcal{A}^\lambda)^T$, $\mathcal{V}^\nu(\p, \k)=-\mathcal{V}^\nu(\k, \p)$ in the second term, the first line of \eqref{deltaJ1_mid} becomes
\begin{eqnarray}
i(v^\mu)^T \delta W_1 + \left(\delta^\mu_i v^i \delta_{FS}\right)^T i\mathcal{U} \ \delta W_1+ \ \left((i\mu^{\mu\nu})^T+ (\delta_{FS}\delta^\mu_i v^i)^T i\mathcal{V}^\nu \right) iq_\nu \left(\delta W_0 + \delta_{FS} v^j A_j \right).
\label{deltaJ1_mid_1st}
\end{eqnarray}
Notice that the $\delta W$ dependence agrees with \eqref{BFL_current_1}.

The second line of \eqref{deltaJ1_mid} -- terms regular in $q$ -- can be read-off diagrammatically:
\begin{eqnarray}
&& - (i\wt{\Gamma}_1^\mu(-q))^T \left(i\Delta_0 \ i\bar{\Gamma}_0^\nu + i\partial_p^\nu G\right) A_\nu - \left((i\Delta_0 \ i\bar{\Gamma}_0^\mu)^T + (i\partial_p^\mu G)^T \right) i\wt{\Gamma}_1^\nu(q) A_\nu \nonumber\\[.2cm]
&& -(i\bar{\Gamma}_0^\mu)^T \left(i\Delta^r_1+i\Delta^s_1\right)(q) \ i\bar{\Gamma}_0^\nu A_\nu - \left( i\Delta_0 \ i\bar{\Gamma}_0^\mu \right)^T i\wt{V}_1(q) \left( i\Delta_0 \ i\bar{\Gamma}_0^\nu \right) A_\nu
\end{eqnarray}
where the $\partial_p G$ terms follow from the $\wt{\Xi}$ terms in \eqref{current_QFT}. There are many equivalent expressions; we have chosen to express it so that it appears ``symmetric'' to read from left to right and from right to left. Now, substitute \eqref{Gamma0_WTId} into the $(\Delta^r_1+\Delta^s_1)$ term, and substitute \eqref{Gamma0_WTId_1} into the rest; next, for each of the two terms in the second line above, we expand like $-a \, c \, b = -a \, c\, b_2 - a_2 \, c \, b + a_2 \, c\, b_2 - a_1 \, c\, b_1$ for $a=a_1+a_2$, $b=b_1+b_2$. The result is
\begin{eqnarray}
&& - (i\bar{\Gamma}_1^\mu(-q))^T \ (Zuu^\dagger) \delta_{FS} \delta^\nu_j v^j A_\nu \nonumber \\[.2cm]
&& - (\delta_{FS}\delta^\mu_i v^i)^T (Zuu^\dagger)^T \ i\bar{\Gamma}_1^\nu(q) \ A_\nu + (\delta_{FS}\delta^\mu_i v^i)^T (Zuu^\dagger)^T \ i\bar{V}_1(q) \ (Zuu^\dagger) \delta_{FS} \delta^\nu_j v^j A_\nu \nonumber \\[.2cm]
&& - (i\partial_p^\mu G^{-1})^T \left(i\Delta^r_1+i\Delta^s_1\right)(q) \ i\partial_p^\nu G^{-1} \ A_\nu - (i\partial_p^\mu G)^T \ i\wt{V}_1(q) \ i\partial_k^\nu G \ A_\nu.
\label{sigma_Y_ref}
\end{eqnarray}
The last term vanishes in a highly non-trivial manner, as we show diagrammatically and combinatorially in Appendix D. The remaining terms, inspecting the definitions of $\mu^{\mu\nu}$ and $\mathcal{V}^\nu$, can be expressed line by line as
\begin{eqnarray}
&& -\left((i\mu^{\mu\lambda})^T+ (\delta_{FS}\delta^\mu_i v^i)^T i\mathcal{V}^\lambda - (i\bar{\Gamma}_0^\mu)^T \mathcal{A}^\lambda \ (Zuu^\dagger) \right) iq_\lambda \ \delta_{FS} \delta^\nu_j v^j A_\nu \nonumber \\[.2cm]
&& - (\delta_{FS}\delta^\mu_i v^i)^T \left( i\mu^{\lambda\nu} - (Zuu^\dagger)^T \mathcal{A}^\lambda \ i\bar{\Gamma}_0^\nu + (Zuu^\dagger)^T \left( \mathcal{A}^\lambda i\bar{V_0} - i\bar{V_0} \mathcal{A}^\lambda \right) (Zuu^\dagger) \delta_{FS} \delta^\nu_j v^j \right) iq_\lambda A_\nu \nonumber \\[.2cm]
&& - (i\partial_p^\mu G^{-1})^T \left(i\Delta^r_1+i\Delta^s_1\right)(q) \ i\partial_p^\nu G^{-1} \ A_\nu.
\end{eqnarray}
Substituting \eqref{Gamma0_WTId} for $\bar{\Gamma}_0$, and using the definition of $\mu_{band}^{\mu\nu}$, we find
\begin{eqnarray}
&& -\left((i\mu^{\mu\lambda})^T+ (\delta_{FS}\delta^\mu_i v^i)^T i\mathcal{V}^\lambda\right) iq_\lambda \ \delta_{FS} \delta^\nu_j v^j A_\nu - (\delta_{FS} \delta^\mu_i v^i)^T i\mu^{\lambda\nu} iq_\lambda A_\nu \nonumber \\[.2cm]
&& + \: (i\mu_{band}^{\mu\lambda})^T\delta_{FS} \delta^\nu_j v^j \ iq_\lambda A_\nu + (\delta_{FS} \delta^\mu_i v^i)^T i\mu_{band}^{\lambda\nu} \ iq_\lambda A_\nu \nonumber \\[.2cm]
&& - \: (i\partial_p^\mu G^{-1})^T \left(i\Delta^r_1+i\Delta^s_1\right)(q) \ i\partial_p^\nu G^{-1} \ A_\nu.
\end{eqnarray}
Finally, for the $\Delta^r_1$ term, use its explicit expression, and for the $\Delta^s_1$ term, use \eqref{Deltas1_dGinv} and the definition of $\mu_{band}^{\mu\nu}$. We arrive at
\begin{eqnarray}
-\left((i\mu^{\mu\nu})^T+ (\delta_{FS} \delta^\mu_i v^i)^T i\mathcal{V}^\nu \right) iq_\nu \ \delta_{FS} v^j A_j - (\delta_{FS} \delta^\mu_i v^i)^T i\mu^{\nu\lambda} F_{\nu\lambda}/2 + i\sigma^{\mu\nu\lambda} F_{\nu\lambda}/2,
\label{deltaJ1_mid_2nd}
\end{eqnarray}
where the Hall conductivity tensor $\sigma^{\mu\nu\lambda}$, totally antisymmetric in $\mu\lambda\nu$, is defined by $\sigma^{\mu\nu\lambda} \equiv \sigma_r^{\mu\nu\lambda}+\sigma_s^{\mu\nu\lambda}$, with
\begin{eqnarray}
&& \sigma_r^{\mu\nu\lambda} \equiv \int_p \: \tr\left\{ (\partial_p^{[\mu} iG^{-1}) \: iG \: (\partial_p^\nu iG^{-1}) \: iG \: (\partial_p^{\lambda]} iG^{-1}) \: iG \right\}, \nonumber \\[.2cm]
&& \sigma_s^{\mu\nu\lambda} \equiv 3 \int_p \delta_{FS} \ v^i \delta^{[\mu}_i \mu_{band}^{\nu\lambda]}.
\label{sigma_def}
\end{eqnarray}
One may notice the similarity between the definitions of $\sigma^{\mu\nu\lambda}$ and $\mu_{anom.}^{\nu\lambda}$ (except in $\sigma^{\mu\nu\lambda}$, the $\wt{V}_1$ term vanishes due to the proof in Appendix D).

After combining \eqref{deltaJ1_mid_1st} and \eqref{deltaJ1_mid_2nd} into \eqref{deltaJ1_mid}, we arrive at
\begin{eqnarray}
\delta J_1^\mu &=& (v^\mu)^T \delta W_1 + (\mu^{\mu\nu})^T iq_\nu \delta W_0 + (\delta_{FS} \delta^\mu_i v^i)^T \left( - \mu^{\nu\lambda} F_{\nu\lambda}/2 + \mathcal{U} \ \delta W_1 + \mathcal{V}^\nu \ iq_\nu \delta W_0 \right) \nonumber \\[.2cm]
&& + \ \sigma^{\mu\nu\lambda} F_{\nu\lambda}/2.
\end{eqnarray}
This is \eqref{BFL_current_1}.

\subsubsection{Coleman-Hill's Theorem}

Interestingly, our derivation for $\delta J_1^\mu$ above, most crucially the cancellation in Appendix D, provides an alternative diagrammatic proof to Coleman-Hill's theorem \cite{Coleman:1985zi}, for QFTs restricted to our assumptions (which are less general than in the original proof). The theorem states that in a gapped fermionic system, the Hall conductivity is unaffected by the interactions. When the system is gapped, i.e. in the absence of FS, our result reduces to $\delta J_1^\mu = \sigma_r^{\mu\nu\lambda} F_{\nu\lambda}/2$, that is, the full Hall conductivity is equal to $\sigma_r^{\mu\nu\lambda}$. Let $g$ be some interaction strength, we have (denoting $\partial^g\equiv \partial/\partial g$)
\begin{eqnarray}
\partial^g \sigma_r^{\mu\nu\lambda} &=& - \int_p \: \partial^g \ \tr\left\{ (\partial_p^{[\mu} G^{-1}) \: G \: (\partial_p^\nu G^{-1}) \: G \: (\partial_p^{\lambda]} G^{-1}) \: G \right\} \nonumber \\[.2cm]
&=& - \int_p \: 4\partial^{[g} \ \tr\left\{ (\partial_p^\mu G^{-1}) \: G \: (\partial_p^\nu G^{-1}) \: G \: (\partial_p^{\lambda]} G^{-1}) \: G \right\}.
\end{eqnarray}
In the second equality we added some total derivative terms so to antisymmetrize the $\partial^g$ altogether with the three $\partial_p$'s. But because of the antisymmetrization, the integrand actually vanishes. This means the full Hall conductivity is independent of interaction strength. This proves Coleman-Hill's theorem \cite{Coleman:1985zi}, for QFTs restricted to our assumptions. In asserting ``the integrand vanishes'', we implicitly made use of the fact that $\partial iG = \{ iG \ \partial iG^{-1} \ iG \}$ in the absence of FS, the fact that $G_{bare}^{-1}$ by definition is independent of $g$, and the physical assumption that the dependence of the self-energy $\Sigma$ on $g$ is non-singular.

\subsubsection{Chemical Potential Dependence of the Hall Conductivity Tensor}
\label{sssect_QFT_dsigma_dEF}

Now we prove \eqref{BFL_dsigma_dEF_detail}, the important result relating the Hall conductivity to the Berry curvature on the FS.

We first consider the $\epsilon_F$ dependence of $\sigma_r^{\mu\nu\lambda}$:
\begin{eqnarray}
\partial^F \sigma_r^{\mu\nu\lambda} = - \int_p \: 4\partial^{[F} \ \tr\left\{ (\partial_p^\mu G^{-1}) \: G \: (\partial_p^\nu G^{-1}) \: G \: (\partial_p^{\lambda]} G^{-1}) \: G \right\}.
\end{eqnarray}
The integrand is non-vanishing because in the presence of FS, the derivative outside the trace acts on the $p^0$ pole structure of the $iG$'s, and the pole structure depends on $p_i$ and $\epsilon_F$. In fact, by similar reasoning that led to the FS term in \eqref{G_derivative_indices}, here we are led to
\begin{eqnarray}
\partial^F \sigma_r^{\mu\nu\lambda} &=& 12 \int_p \: i\pi \delta(p^0-\xi) \ \partial^{[F}\sgn\, \xi \: \times \nonumber \\[.2cm]
&& \hspace{2cm} \sum_{w} \sum_{w'} \frac{Z}{\chi_w \chi_{w'}} \left\{ u^\dagger \: (\partial_p^\mu G^{-1}) \: w w^\dagger \: (\partial_p^\nu G^{-1}) \: w' w'^\dagger \: (\partial_p^{\lambda]} G^{-1}) \: u \right\} \nonumber \\[.2cm]
&& + \ 12 \int_p \: i\pi \left(-\partial_{p^0} \delta(p^0-\xi) \right) \ \partial^{[F}\sgn\, \xi \: \times \nonumber \\[.2cm]
&& \hspace{3cm} \sum_w \frac{Z^2}{\chi_w} \left\{ w^\dagger \: (\partial_p^\mu G^{-1}) \: uu^\dagger \: (\partial_p^\nu G^{-1}) \: uu^\dagger \: (\partial_p^{\lambda]} G^{-1}) \: w \right\}.
\end{eqnarray}
The first term arises from the single pole (appearing as $i\pi \delta(p^0-\xi) \: \sgn\, \xi$ in principle function decomposition) when one of the three $G$'s is in the $u$ band; the second term arises from the double pole (appearing as $i\pi (-\partial_{p^0} \delta(p^0-\xi)) \: \sgn\, \xi$) when two of the $G$'s are in the $u$ band. The triple pole contribution when all three $G$'s are in the $u$ band vanishes under the total antisymmetrization. We can evaluate the above using the trick \eqref{IBPtrick}. We find
\begin{eqnarray}
\partial^F \sigma_r^{\mu\nu\lambda} &=& 12 \int_p \: i\pi \delta(p^0) \ \partial^{[F} \sgn\,\xi \: \left\{ \partial_p^\mu u^\dagger \: (\mathbf{1}- uu^\dagger) \: Z(\partial_p^\nu G^{-1}) \: (\mathbf{1}- uu^\dagger) \: \partial_p^{\lambda]} u \right\} \nonumber \\[.2cm]
&& - \ 12 \int_p i\pi \delta(p^0) \ \partial^{[F}\sgn\,\xi \: \partial_{p^0} \left\{ \partial_p^\mu u^\dagger \ Z^2\partial_p^\nu \chi_u \sum_w \left(\chi_u - \chi_w\right)^2 \frac{ww^\dagger}{\chi_w} \ \partial_p^{\lambda]} u \right\}.
\end{eqnarray}
In the first line, $\mathbf{1}- uu^\dagger$ can be further replaced by $\mathbf{1}$ thanks to \eqref{IBPtrick} and the antisymmetrization. In the second line, we need the following relevant terms, according to \eqref{chiu_near_FS_good}:
\begin{eqnarray}
\left. -Z^2\partial_p^\nu \chi_u \ (\chi_u-\chi_w)^2 \right|_{p \ on \ FS} = Zv^\nu \chi_w^2
\end{eqnarray}
\begin{eqnarray}
\left. \partial_{p^0} \left(-Z^2\partial_p^\nu \chi_u \ (\chi_u-\chi_w)^2\right) \right|_{p \ on \ FS} = Zv^\nu \partial_{p^0} \chi_w^2 + \left(\partial_p^\nu Z + \, 2iZ\gamma \: \partial_p^\nu |p^0| \right) \chi_w^2 - 2v^\nu\chi_w.
\end{eqnarray}
Similar results hold when $\partial_p^\nu \chi_u$ is replaced with $\partial^F \chi_u$; recall that $v^F\equiv \partial^F E$. The remaining problem is, how to understand $\delta(p^0) \partial_{p^0} |p^0|$? We should understand it as $0$, because in this paper, the generalized function $\delta(p^0)$ always arises as the approximation to a narrow rectangular function over an interval centered at $p^0=0$ (more precisely, the rectangular function is $\theta(p^0-q^0/2) - \theta(p^0+q^0/2)$, see Appendix A), and $\delta(p^0) \partial_{p^0} |p^0|$ corresponds to taking the difference of $|p^0|$ between the two sides of the interval, which is obviously $0$. (Note that in $d=2$, the parametrization \eqref{chiu_near_FS_good} does not apply; however, general analytic properties of $\delta\Sigma$ still require it to be odd in $p^0$~\cite{Luttinger:1961zz, abrikosov1975methods}, and therefore the corresponding contribution here must still vanish.) Thus, at the end, we have
\begin{eqnarray}
\partial^F \sigma_r^{\mu\nu\lambda} &=& 12 \int_p \: i 2\pi \delta(p^0) \ \partial^{[F}\theta(\epsilon_F-E) \: \times \nonumber \\[.2cm]
&& \hspace{1.5cm} \left( -2v^\mu \left\{ \partial_p^\nu u^\dagger \: \partial_p^{\lambda]} u \right\} + \left(\partial_p^\mu + v^\mu \partial_{p^0} \right) \left\{ \partial_p^\nu u^\dagger \: ZG^{-1} \: \partial_p^{\lambda]} u \right\} \right).
\label{dsigma_r}
\end{eqnarray}
Note that the derivatives of $-2\theta(\epsilon-E)$ are always the same as those of $\sgn\, \xi$; we choose to express as the former because it admits the intuition as the ``Fermi sea'', at least near the FS.

Next, from the expression of $\mu_{band}^{\nu\lambda}$, we observe $\sigma_s^{\mu\nu\lambda}$ can be expressed as
\begin{eqnarray}
\sigma_s^{\mu\nu\lambda} = 3 \int_p \: 2\pi \delta(p^0-(E-\epsilon_F)) \ \partial_p^{[\mu} \theta(\epsilon_F-E) \ i \left\{ \partial_p^\nu u^\dagger \: ZG^{-1} \: \partial_p^{\lambda]} u \right\}.
\label{sigma_s_alt}
\end{eqnarray}
Now we take $\partial^F$ and find
\begin{eqnarray}
\partial^F \sigma_s^{\mu\nu\lambda} &=& 12 \int_p \partial^{[F} \left( 2\pi \delta(p^0-(E-\epsilon_F)) \ \partial_p^\mu \theta(\epsilon_F-E) \ i \left\{ \partial_p^\nu u^\dagger \: ZG^{-1} \: \partial_p^{\lambda]} u \right\} \right) \nonumber \\[.2cm]
&=& -12 \int_p \: i 2\pi \ \partial^{[F} \theta(\epsilon_F-E) \left( \delta(p^0) \partial_p^\mu - v^\mu \partial_{p^0} \delta(p^0) \right) \left\{ \partial_p^\nu u^\dagger \: ZG^{-1} \: \partial_p^{\lambda]} u \right\} \nonumber \\[.2cm]
&=& -12 \int_p \: i 2\pi \delta(p^0) \ \partial^{[F} \theta(\epsilon_F-E) \left( \partial_p^\mu + v^\mu \partial_{p^0} \right) \left\{ \partial_p^\nu u^\dagger \: ZG^{-1} \: \partial_p^{\lambda]} u \right\}
\label{dsigma_s}
\end{eqnarray}
In the second equality we used
\begin{eqnarray}
\partial_p^\mu \delta(p^0-(E-\epsilon_F)) = - v^\mu \partial_{p^0} \delta(p^0-(E-\epsilon_F))
\end{eqnarray}
and likewise for $\partial^F$.

Finally we combine \eqref{dsigma_r} and \eqref{dsigma_s} and obtain
\begin{eqnarray}
\partial^F \sigma^{\mu\nu\lambda} = 12 \int_p \: 2\pi \delta(p^0) \ \partial^{[F} \theta(\epsilon_F-E) \: v^\mu \: (-2i) \left\{ \partial_p^\nu u^\dagger \: \partial_p^{\lambda]} u \right\}.
\end{eqnarray}
Due to \eqref{u_derivative_on_shell} (and the similar version for $\partial^F$) and the antisymmetrization, we can replace $u$ with $\mathfrak{u}$, and perform the $p^0$ integral to obtain
\begin{eqnarray}
\partial^F \sigma^{\mu\nu\lambda} = 12 \int_\p \partial^{[F} \theta(\epsilon_F-E) \: v^\mu \: b^{\nu\lambda]}.
\label{sigma_FSderivaitve}
\end{eqnarray}
Expanding the antisymmetrization explicitly, this is \eqref{BFL_dsigma_dEF_detail}.

If $d=2$, or if $d>2$ and the Berry curvature is an exact 2-form on the FS, we can continuously define $\mathfrak{u}(\p)$ over the FS, and derive \eqref{BFL_dsigma_dEF_1} via integration by parts:
\begin{eqnarray}
\partial^F \sigma^{\mu\nu\lambda} &=& 24 \int_\p \partial^{[F} \left(\partial_p^\mu \theta(\epsilon_F-E) \: v^\nu \: a^{\lambda]} \right) \nonumber \\[.2cm]
&=& \partial^F \ 6 \int_\p \delta(\epsilon_F-E) \: \delta^{[\mu}_0 v^\nu \: a^{\lambda]}
\end{eqnarray}
where in the first equality we used the fact that $v^\mu$ and $v^F$ are respectively $-\partial_p^\mu$ and $-\partial^F$ acted on $p^0-(E-\epsilon_F)$, and in the second equality, total $\p$ derivatives vanish due to the $\p$ integral, while total $p_0$ derivative vanishes trivially as the integrand has no $p^0$ dependence. 

If $d>2$ and the Berry curvature is not an exact 2-form on the FS, we need to integrate by parts in another way. We can rewrite \eqref{sigma_FSderivaitve} using the $P^\lambda_k$ introduced below \eqref{BFL_sigma_const_FS}:
\begin{eqnarray}
\partial^F \sigma^{ij\lambda} = 12 \int_\p \partial^{[F} \theta(\epsilon_F-E) \: b^{ij} \: \partial_p^{k]} P^\lambda_k
\end{eqnarray}
where the total antisymmetrization is in indices $[Fijk]$. Using the assumption that there is no band degeneracy near the FS and hence $\partial_p^{[k}b^{ij]}=0$ near the FS, we have
\begin{eqnarray}
\partial^F \sigma^{ij\lambda} = 12 \int_\p \partial^{[F} \left( -\partial_p^k \theta(\epsilon_F-E) \: b^{ij]} \: P^\lambda_k \right).
\end{eqnarray}
When we proceed further, note that if the fermions are in a lattice and if the FS intersects the boundary of our choice of first Brillouin zone, then for $\lambda=0$ we cannot drop the total $p$-derivative terms because $p_k$ is not continuous when we identify the opposite boundaries of the first Brillouin zone~\cite{Haldane:2004zz}. We find
\begin{eqnarray}
\partial^F \sigma^{ij\lambda} = \partial^F \left( 3 \int_\p \delta(\epsilon_F-E) b^{[ij} v^{k]} \: P^\lambda_k + 6\int_\p \partial_p^{[k} \left( \delta(\epsilon_F-E) \: v^i \: a^{j]} \: P^\lambda_k\right) \right)
\end{eqnarray}
(we used the Bianchi identity $\partial^F b^{\nu\lambda} = 2 \partial_p^{[\lambda} b^{\nu] F}$). This proves \eqref{BFL_dsigma_dEF_e} and \eqref{BFL_dsigma_dEF_b}, and hence \eqref{BFL_sigma_const_FS}.

We remind that the derivation above fails at discrete values of $\epsilon_F$ around which the FS develops new disconnected components (e.g. \eqref{FS_discont_change}). Across those values of $\epsilon_F$ it is unclear whether $\sigma^{\mu\nu\lambda}$ may have a jump, as commented in Section \ref{ssect_kinetic_sigma_EF}.

\section{Discussion}
\label{sect_discussion}

We have constructed an extension of Landau's Fermi liquid theory to
systems with Berry curvatures. One can see several directions to
extend our theory. First, we assumed the Fermi level crosses only one band; 
one can generalize this to multiple bands. 
Two scenarios are of particular physical interest: either that the multiple bands crossing the Fermi level are 
completely degenerate~\cite{shindou2008gradient}, or that these multiple bands 
have completely disjoint Fermi surfaces. The generalizations of our theory to both
 scenarios are straightforward. Second, our discussion is limited to linear response. It
would be interesting to extend the scope of the kinetic theory to
include also nonlinear response, so that important effects such as the $(3+1)d$ chiral anomaly can be captured. 
Third, we assumed that the quantum field
theory describing the fermions does not have couplings of the type
$A\phi \psi^\dagger \psi$. It would be interesting to see if the 
kinetic theory can be extended to include couplings of this type in
the QFT. Also, one may try to understand if long-ranged
interactions can be included, to the extent that these interactions do
not destroy the Fermi liquid ground state.

Some interesting questions are raised in the context of our Berry
Fermi liquid theory. We have found that, beside a
$\epsilon_F$-dependent piece, the Hall conductivity contains a constant
piece $\sigma_o^{\mu\nu\lambda}$ in Section \ref{sssect_kinetic_sigma_EF_no_anomaly}.
Is this contribution topological and not renormalized by interactions?
Does it receive jumps at discrete values of $\epsilon_F$ around which
the FS develops new disconnected components (e.g. in the case
\eqref{FS_discont_change})? In gapped system,
$\sigma_o^{\mu\nu\lambda}=\sigma_r^{\mu\nu\lambda}$ in $d=2$ is topological
\cite{Ishikawa:1986wx}.

More broadly, one may ask: Is it possible to
have a notion of topologically equivalent / distinct Fermi liquids?
For example, are the Fermi liquids in normal metal and in Weyl metal
topologically distinct under some notion? In this paper, we see both a puzzle 
and a hint regarding such notion. The puzzle is the possible jump of
$\sigma^{\mu\nu\lambda}$ mentioned previously. The hint is the manifestation 
of anomaly-related transport effects in the distinction between \eqref{BFL_dsigma_dEF_b} 
and \eqref{BFL_dsigma_dEF_2b}. It would be interesting to study if such 
problems can be covered under a coherent framework.

We note also that the matching between the microscopic theory and the 
Fermi liquid theory is done here at the level of dynamical
equations. If there is a way to do the matching at the level of
action and path integral measure, like that in Berry Fermi gas 
\cite{Chen:2014cla}, it would provide a much more transparent derivation 
of the Berry Fermi liquid theory. It may also help to extend the kinetic 
theory beyond linear response.

Finally, given the generality of the assumptions, the formalism should
have broad applications in physical systems. It would be interesting
if predictions of the Berry Fermi liquid theory can be directly
compared to experiments.

\section*{Acknowledgements}
We thank Michael Geracie and Misha Stephanov for discussions. We especially thank
Max Metlitski for bringing up the issue of quasiparticle collisions. This 
work is supported, in part, by the US DOE Grant
No.\ DE-FG02-13ER41958, MRSEC Grant No.\ DMR-1420709 by the NSF, ARO
MURI Grant No.\ 63834-PH-MUR, and by a Simons Investigator grant from
the Simons Foundation.

\addcontentsline{toc}{section}{Appendix A}
\section*{Appendix A}

In this appendix we present the following:
\begin{itemize}
\item
First we present the Cutkosky cutting rule for our fermionic system.
\item
Then we introduce how to count the power of $q$ in a cut diagram, and show $D_1$, $C^{ph}_1$ and $C^{pp}_1$ are the only cut sub-diagrams that contribute at order $q$. Then we restrict to $d\geq 3$ (the case of $d=2$ will be discussed in Appendix B) and justify their parametrization \eqref{Decay_parametrization}, \eqref{Cph_parametrization} and \eqref{Cpp_parametrization}, and show the relation \eqref{Collision_WTId}.
\item
Finally, we show (including $d=2$) that collisions have no contribution to the antisymmetric part $\Pi^{[\mu\nu]}$ of the current-current correlation. So collision is ``uninteresting'' to the main focus of this paper.
\end{itemize}
Since only the $u$ band is involved in the present discussion, for convenience we will consider a single band system in this appendix, i.e. $u=1$; in multi-band systems, the restoration of the $u$ eigenvector is obvious.

Consider a two point correlation $\Pi$ with momentum $q$ (for definiteness, we let $q$ run from the right to the left of the diagram). In this appendix, unless otherwise specified, $\Pi$ refers to the current-current correlation $\Pi^{\mu\nu}(q)$, and $\Re\Pi$ and $\Im\Pi$ really mean the Hermitian and anti-Hermitian parts of $\Pi^{\mu\nu}$. The Cutkosky cutting rule gives the difference between the retarded correlation and the advanced correlation:
\begin{eqnarray}
\Pi_{cut} &\equiv& -i(\Pi_R-\Pi_A) = \Im\Pi_R-\Im\Pi_A = 2\Im\Pi_R = -2\Im\Pi_A \nonumber\\[.2cm]
&=& \Pi_{cut-} - \Pi_{cut+}.
\label{def_Cut}
\end{eqnarray}
The three equalities in the first line follow from general analytic properties of two-point correlations~\cite{Luttinger:1961zz, abrikosov1975methods}. In the second line, $\Pi_{cut+}$ is defined as the following. Consider a certain Feynman diagram in $\Pi$, with a certain Cutkosky cut -- a cut through a number of internal fermion propagators such that the Feynman diagram is disconnected into two parts, with one current insertion (or other operators, depending on what $\Pi$ is) contained in each part. Clearly the total momentum running from right to left across the cut is $q$. For those fermion propagators that are being cut, we place the fermions on-shell, which, according to the Cutkosky cutting rule, means to replace each cut propagator by
\begin{eqnarray}
iG(p) \ \ \ \longrightarrow \ \ 2\pi Z(\p) \ \delta(p^0-\xi_\p) \ \sgn(p^0) \theta(\mp p^0),
\end{eqnarray}
where $\theta(\mp p^0)$ is taken when the fermion runs across the cut from right to left / from left to right (given we have chosen $q$ to run from right to left). Then we sum over all possible ways of cutting over all Feynman diagrams, and the result is defined as $\Pi_{cut-}$. And $\Pi_{cut+}$ is defined in a similar manner, but with $\theta(\pm p^0)$ taken when the fermion runs across the cut from right to left / from left to right.

An important consequence is, by energy conservation, all these cut propagators must have energies between $\pm|q^0|$. This is because, those step functions require the on-shell quasiparticles' energies to appear in energy conservation in the form ``the sum of positive energies minus the sum of negative energies is equal to $\mp q^0$'' respectively in $\Pi_{cut \mp}$ (so $\Pi_{cut\mp}$ is non-vanishing only for negative / positive $q^0$ respectively, hence our $\mp$ subscript). In retrospect, this justifies why we could restrict to the $u$ band in a multi-band system, and why we could ignore the possibility of cutting through an interaction mediator: Because by the assumptions about our QFT, neither the other bands of the fermion nor the short-ranged interaction mediator(s) have any low energy on-shell excitation.

Below we discuss how to count the power of $q$ in a cut sub-diagram. We have to note that the power counting introduced below has missing piece. In $d=2$ the missing piece is order $q \ln q$ (less suppressed than order $q$ when $q$ is small) and lead to complications discussed in Appendix B. In this appendix we work with $d\geq 3$, where this missing piece is neglected as they are beyond order $q$. After introducing the power counting, we will argue the only cut sub-diagram that contributes at zeroth order in $q$ is a cut through double propagator, leading to $\Delta'$ in \eqref{Delta_prime}, and the only ones that contribute at first order in $q$ are those three pairs of cut sub-diagrams for $D_1$, $C^{ph}_1$ and $C^{pp}_1$, leading to \eqref{Decay_parametrization}, \eqref{Cph_parametrization} and \eqref{Cpp_parametrization}.

First consider a cut through $n>2$ internal fermion propagators. Generically they all have different internal momenta. As discussed above, all their energies are restricted by $|q^0|$, hence the integration over the $n$ internal energies yields a suppression of order $(q^0)^n$; on the other hand, the argument of the delta function of energy conservation is of order $q^0$. Therefore, the contribution of the $n$ cut propagators is of order $(q^0)^{n-1}$.

The case of $n=2$ is special. For $n=2$, when one on-shell fermion is at low energy and hence near the FS, the other on-shell fermion, due to spatial momentum conservation and the smallness of $\q$, is automatically near the FS too, and hence at low energy too (because of on-shell). This means the smallness of their energies provides only one constraint, instead of two independent constraints. This lowers the power counting of $q^0$ by one. Thus, the cut through double propagator is not first order but zeroth order in $q$. More exactly, let us compute the cut through the double propagator of momenta $p\pm q/2$. According to the cutting rule, the cut sub-diagram is equal to
\begin{eqnarray}
2\Im\Delta'_R(q) &=& 2\pi Z(\p+\q/2) \delta(p^0+q^0/2-\xi(\p+\q/2)) \ 2\pi Z(\p-\q/2) \delta(p^0-q^0/2-\xi(\p-\q/2)) \nonumber\\[.2cm]
&& (-1) \left(\theta(-(p^0+q^0/2)) \theta(p^0-q^0/2) - \theta(p^0+q^0/2) \theta(-(p^0-q^0/2)) \right) \nonumber \\[.2cm]
&=& (2\pi Z(\p))^2 \ q^0 \ \delta(p^0) \ \delta(p^0-\xi(\p)) \ \delta(v^\mu(\p) q_\mu) \ + \ \mathcal{O}(q^2).
\label{Delta_from_Cutkosky}
\end{eqnarray}
To relate this to the time-ordered $\Delta'$, we use the Kramers-Kronig dispersion relation of a general two-point correlation~\cite{Luttinger:1961zz, abrikosov1975methods}:
\begin{eqnarray}
i\Pi(q) = \frac{i}{2\pi} \int d\omega \frac{\Pi_{cut}(\omega, \q)}{-q^0+\omega-i\epsilon \: \sgn\, q^0} \ + \ i\, (\mbox{real terms unrelated to Cutkosky cut}).
\label{dispersion_relation}
\end{eqnarray}
(The $\Pi$ here is time-ordered; if retarded or advanced, the $\sgn\, q^0$ should be replaced with $\pm 1$.) Performing the integration yields the time ordered $i\Delta'(q)$ in \eqref{Delta_prime}, as desired. (The integration generally involves $\Pi_{cut}$ at non-small values of $\omega$. But $\Im\Delta'$ in particular is non-vanishing only when $\omega$ equals the small value $v^i q_i$.)

For $n>2$, the $n$ cut propagators contribute order $(q^0)^{n-1}$, and for current-current correlation $n$ must be even (with $n/2$ cut propagators running across the cut from right to left, and the other $n/2$ from left to right). So it seems the corrections from Cutkosky cut beyond $\Delta'$ (beyond $n=2$) are at least of order $q^3$. (This justifies our analytic expansion of the $q$-2PI interaction vertex to zeroth and first order in $q$.) How can there be order $q$ sub-diagrams? Consider the following situation. Given that all cut propagators are on-shell, if there is a pair of propagators, one cut and one uncut, whose momenta are dictated by momentum conservation to differ by $q$, then that uncut propagator will be nearly-on-shell (due to the smallness of $q$), and contributes a factor of order $1/q$. There can be at most one such nearly-on-shell propagator on either the left or the right of the cut, so there can be at most two of them in total. Therefore, there exist cut sub-diagrams at order $q$: Such cut sub-diagrams have four cut propagators, and two nearly-on-shell propagators, one on each side of the cut. These propagators can be organized in six different ways, which are the three pairs of cut sub-diagrams for $D_1$, $C^{ph}_1$ and $C^{pp}_1$ respectively, presented in Section \ref{sssect_QFT_Collision}.

Now we evaluate the sub-diagrams for $D_1$, $C^{ph}_1$ and $C^{pp}_1$ according to the cutting rule. First,
\begin{eqnarray}
2(D_R)_1(p; q) &=& \int_{k, l} \left(-\frac{1}{2}\right) \left(\frac{iZ(\p-\q/2)}{-q^0-\xi(\p-\q/2) + \xi(\p+\q/2)}\right)^2 \nonumber \\[.2cm]
&& \hspace{1cm} i^2 \left|V(p-q/2, k+l \rightarrow k-q/2, p+l)\right|^2 \nonumber \\[.2cm]
&& \hspace{1cm} (2\pi)^4 Z(\p+\q/2) Z(\k-\q/2) Z(\p+\l) Z(\k+\l) \nonumber \\[.2cm]
&& \hspace{1cm} \delta(p^0+q^0/2-\xi(\p+\q/2)) \ \delta(k^0-q^0/2-\xi(\k-\q/2)) \nonumber \\[.2cm]
&& \hspace{1cm} \delta(p^0+l^0-\xi(\p+\l)) \ \delta(k^0+l^0-\xi(\k+\l)) \nonumber \\[.2cm]
&& \hspace{1cm}\left[ \: \theta(-(p^0+q^0/2)) \theta(k^0-q^0/2) \theta(-(k^0+l^0)) \theta(p^0+l^0) \right. \nonumber \\[.2cm]
&& \hspace{2cm} \left. - \theta(p^0+q^0/2) \theta(-(k^0-q^0/2)) \theta(k^0+l^0) \theta(-(p^0+l^0)) \: \right] \nonumber \\[.2cm]
&& \ - \ (\mbox{with } q\leftrightarrow -q).
\label{DR1}
\end{eqnarray}
The $-1/2$ is due to fermionic statistics. The products of step functions restrict the energies $p^0, k^0$ and $l^0$ to order $q^0$; for example, the first product of step functions restricts $q^0/2<k^0<-l^0<p^0<-q^0/2$. We already argued that the $q$ suppression in the cut sub-diagrams is dictated by the step functions, so in the $Z$'s, the $V$'s and the on-shell delta functions, we can neglect the $q$-dependences, as well as the $p^0, k^0, l^0$ dependences. We are then led to
\begin{eqnarray}
2(D_R)_1(p; q) &=& -\frac{1}{2}\int_{k, l} \frac{i^2 Z(\p)^3 Z(\k) Z(\p+\l) Z(\k+\l)}{(v^\mu(\p) q_\mu)^2} \ i^2 \left|V(p, k+l \rightarrow k, p+l)\right|^2 \nonumber \\[.2cm]
&& \hspace{.8cm} (2\pi)^4 \delta(\xi(\p)) \delta(\xi(\k)) \delta(\xi(\p+\l)) \delta(\xi(\k+\l)) \nonumber \\[.2cm]
&& \hspace{.8cm} \left[ \phantom{+} \theta(-(p^0+q^0/2)) \theta(k^0-q^0/2) \theta(-(k^0+l^0)) \theta(p^0+l^0) \right. \nonumber \\[.2cm]
&& \hspace{1cm} - \theta(p^0+q^0/2) \theta(-(k^0-q^0/2)) \theta(k^0+l^0) \theta(-(p^0+l^0)) \nonumber \\[.2cm]
&& \hspace{1cm} + \theta(-(k^0+q^0/2)) \theta(p^0-q^0/2) \theta(-(p^0+l^0)) \theta(k^0+l^0) \nonumber \\[.2cm]
&& \hspace{1cm} \left. - \theta(k^0+q^0/2) \theta(-(p^0-q^0/2)) \theta(p^0+l^0) \theta(-(k^0+l^0)) \: \right].
\end{eqnarray}
Inspecting the $p, q$ dependence, together with power counting, this justifies the parametrization \eqref{Decay_parametrization} with non-negative $\gamma$ (what is remained to be shown is that the $\gamma$ here is the same $\gamma$ that appears in $\Im\Sigma$). In particular, the sign of the retarded $(D_R)_1$ is given by the sign of $q^0$, so the time-ordered $D_1=(D_R)_1 \sgn\, q^0$ is non-negative.

Next,
\begin{eqnarray}
2(C^{ph}_R)_1(p, k; q) &=& \int_l \ (-1) \ \frac{iZ(\p-\q/2)}{-q^0-\xi(\p-\q/2) + \xi(\p+\q/2)} \ \frac{iZ(\k+\q/2)}{q^0-\xi(\k+\q/2) + \xi(\k-\q/2)} \nonumber \\[.2cm]
&& \hspace{.5cm} iV(p-q/2, k+l \rightarrow k-q/2, p+l) \ iV(k+q/2, p+l \rightarrow p+q/2, k+l) \nonumber \\[.2cm]
&& \hspace{.5cm} (2\pi)^4 Z(\p+\q/2) Z(\k-\q/2) Z(\p+\l) Z(\k+\l) \nonumber \\[.2cm]
&& \hspace{.5cm} \delta(p^0+q^0/2-\xi(\p+\q/2)) \ \delta(k^0-q^0/2-\xi(\k-\q/2)) \nonumber \\[.2cm]
&& \hspace{.5cm} \delta(p^0+l^0-\xi(\p+\l)) \ \delta(k^0+l^0-\xi(\k+\l)) \nonumber \\[.2cm]
&& \hspace{.5cm}\left[ \: \theta(-(p^0+q^0/2)) \theta(k^0-q^0/2) \theta(-(k^0+l^0)) \theta(p^0+l^0) \right. \nonumber \\[.2cm]
&& \hspace{1.5cm} \left. - \theta(p^0+q^0/2) \theta(-(k^0-q^0/2)) \theta(k^0+l^0) \theta(-(p^0+l^0)) \: \right] \nonumber \\[.2cm]
&& \ + \ (\mbox{with } p\leftrightarrow k, \mbox{ except the arguments of the $V$'s kept unchanged}).
\label{CphR1}
\end{eqnarray}
Making the small $q$ approximations we made for $(D_R)_1$, we are led to
\begin{eqnarray}
2(C^{ph}_R)_1(p, k; q) &=& - \int_{l} \frac{i^2 Z(\p)^2 Z(\k)^2 Z(\p+\l) Z(\k+\l)}{-(v^\mu(\p) q_\mu)(v^\mu(\k) q_\mu)} \ i^2 \left|V(p, k+l \rightarrow k, p+l) \right|^2 \nonumber \\[.2cm]
&& \hspace{.8cm} (2\pi)^4 \delta(\xi(\p)) \delta(\xi(\k)) \delta(\xi(\p+\l)) \delta(\xi(\k+\l)) \nonumber \\[.2cm]
&& \hspace{.8cm} \left[ \phantom{+} \theta(-(p^0+q^0/2)) \theta(k^0-q^0/2) \theta(-(k^0+l^0)) \theta(p^0+l^0) \right. \nonumber \\[.2cm]
&& \hspace{1cm} - \theta(p^0+q^0/2) \theta(-(k^0-q^0/2)) \theta(k^0+l^0) \theta(-(p^0+l^0)) \nonumber \\[.2cm]
&& \hspace{1cm} + \theta(-(k^0+q^0/2)) \theta(p^0-q^0/2) \theta(-(p^0+l^0)) \theta(k^0+l^0) \nonumber \\[.2cm]
&& \hspace{1cm} \left. - \theta(k^0+q^0/2) \theta(-(p^0-q^0/2)) \theta(p^0+l^0) \theta(-(k^0+l^0)) \: \right].
\end{eqnarray}
This justifies the parametrization \eqref{Cph_parametrization} with non-negative $\lambda^{ph}$.

Last,
\begin{eqnarray}
2(C^{pp}_R)_1(p, k; q) &=& \int_l \left(-\frac{1}{2}\right) \frac{iZ(\p-\q/2)}{-q^0-\xi(\p-\q/2) + \xi(\p+\q/2)} \ \frac{iZ(\k-\q/2)}{-q^0-\xi(\k-\q/2) + \xi(\k+\q/2)} \nonumber \\[.2cm]
&& \hspace{.5cm} iV(p-q/2, k+q/2 \rightarrow k-l, p+l) \ iV(k-l, p+l \rightarrow p+q/2, k-q/2) \nonumber \\[.2cm]
&& \hspace{.5cm} (2\pi)^4 Z(\p+\q/2) Z(\k+\q/2) Z(\p+\l) Z(\k-\l) \nonumber \\[.2cm]
&& \hspace{.5cm} \delta(p^0+q^0/2-\xi(\p+\q/2)) \ \delta(k^0+q^0/2-\xi(\k+\q/2)) \nonumber \\[.2cm]
&& \hspace{.5cm} \delta(p^0+l^0-\xi(\p+\l)) \ \delta(k^0-l^0-\xi(\k-\l)) \nonumber \\[.2cm]
&& \hspace{.5cm}\left[ \: \theta(-(p^0+q^0/2)) \theta(-(k^0+q^0/2)) \theta(k^0-l^0) \theta(p^0+l^0) \right. \nonumber \\[.2cm]
&& \hspace{1.5cm} \left. - \theta(p^0+q^0/2) \theta(k^0+q^0/2) \theta(-(k^0-l^0)) \theta(-(p^0+l^0)) \: \right] \nonumber \\[.2cm]
&& \ - \ (\mbox{with } q\leftrightarrow -q, \mbox{ except the arguments of the $V$'s kept unchanged}).
\label{CppR1}
\end{eqnarray}
Making the small $q$ approximations we made for $(D_R)_1$, we are led to
\begin{eqnarray}
2(C^{pp}_R)_1(p, k; q) &=& -\frac{1}{2} \int_{l} \frac{i^2 Z(\p)^2 Z(\k)^2 Z(\p+\l) Z(\k-\l)}{(v^\mu(\p) q_\mu)(v^\mu(\k) q_\mu)} \ i^2 |V(p, k \rightarrow k-l, p+l)|^2 \nonumber \\[.2cm]
&& \hspace{.8cm} (2\pi)^4 \delta(\xi(\p)) \delta(\xi(\k)) \delta(\xi(\p+\l)) \delta(\xi(\k-\l)) \nonumber \\[.2cm]
&& \hspace{.8cm} \left[ \phantom{+} \theta(-(p^0+q^0/2)) \theta(-(k^0+q^0/2)) \theta(k^0-l^0) \theta(p^0+l^0) \right. \nonumber \\[.2cm]
&& \hspace{1cm} - \theta(p^0+q^0/2) \theta(k^0+q^0/2) \theta(-(k^0-l^0)) \theta(-(p^0+l^0)) \nonumber \\[.2cm]
&& \hspace{1cm} + \theta(k^0-q^0/2)) \theta(p^0-q^0/2) \theta(-(p^0+l^0)) \theta(-(k^0-l^0)) \nonumber \\[.2cm]
&& \hspace{1cm} \left. - \theta(-(k^0-q^0/2)) \theta(-(p^0-q^0/2)) \theta(p^0+l^0) \theta(k^0-l^0) \: \right].
\end{eqnarray}
This justifies the parametrization \eqref{Cpp_parametrization} with non-negative $\lambda^{pp}$. 

Inspecting the exact expressions \eqref{DR1}, \eqref{CphR1} and \eqref{CppR1} (these expressions are before we apply power counting, and hence hold in $d=2$ as well), we can observe the important relation \eqref{C_D_relation}. In particular, to see the second equality in \eqref{C_D_relation}, we shift $k\rightarrow k+l \mp q/2$ in the two cut sub-diagrams contributing to $(C^{pp}_R)_1$. This leads to \eqref{Collision_WTId}, which is required by the Ward-Takahashi identity.

It remains to show the $\gamma$ in the parametrization of $D_1$ is the same $\gamma$ that appears in $\Im\Sigma$. Now we apply the Cutkosky cutting rule to $-\Sigma(p)$ with small $p^0$. The leading cut diagram is order $(p^0)^2$, involving three cut propagators~\cite{Luttinger:1961zz}. More explicitly, at this order we have
\begin{eqnarray}
2\Im(-\Sigma_R)(p) &=& \int_{k, l} \left(-\frac{1}{2} \right) i^2 |V(p, k+l \rightarrow k, p+l)|^2 \ (2\pi)^3 Z(\k) Z(\k+\l) Z(\p+\l) \nonumber \\[.2cm]
&& \hspace{.75cm} \delta(k^0-\xi(\k)) \delta(k^0+l^0-\xi(\k+\l)) \delta(p^0+l^0-\xi(\p+\l)) \nonumber \\[.2cm]
&& \hspace{.75cm} \left[ \: (-1)^2 \ \theta(-k^0) \theta(k^0+l^0) \theta(-(p^0+l^0)) \phantom{{X^T}^T} \right. \nonumber \\[.2cm]
&& \hspace{1.5cm} \left. \phantom{{X^T}^T} - (-1) \ \theta(k^0) \theta(-(k^0+l^0)) \theta(p^0+l^0) \: \right].
\end{eqnarray}
We can see $-\Im\Sigma_R$ is explicitly positive, as it should~\cite{Luttinger:1961zz, abrikosov1975methods}. Combined with power counting, we justify the parametrization $-\Im\Sigma=-\Im\Sigma_R \, \sgn\, p^0 = \gamma' p^0 |p^0|$ at small $p^0$ with positive $\gamma'$. To verify $\gamma'=(3/2)\gamma$, we compare the evaluations of $-\Im\Sigma_R$ and $(D_R)_1$ and restrict to order $q$, we find
\begin{eqnarray}
2(D_R)_1(p; q) &=& \frac{i^2 Z^2}{(v^\mu q_\mu)^2} \ 2\pi Z \delta(\xi) \left(\theta(-(p^0+q^0/2)) \theta(p^0-q^0/2) + \theta(p^0+q^0/2) \theta(-(p^0-q^0/2)) \right) \nonumber \\[.2cm]
&& \left[ \: \sgn(p^0-q^0/2) (-2\Im\Sigma_R(p+q/2)) - \sgn(p^0+q^0/2) (-2\Im\Sigma_R(p-q/2)) \: \right] \nonumber \\[.2cm]
&=& 2\frac{Z^3 \: 2\pi \delta(\xi)}{(v^\mu q_\mu)^2}\: \sgn(q^0)\theta(|q^0|/2-|p^0|) \left(-\Im\Sigma_R(p-q/2) - \Im\Sigma_R(p+q/2) \right).
\label{D_Sigma_relation}
\end{eqnarray}
The last factor is equal to $\gamma'(\p) \left(2(p^0)^2 + (q^0)^2/2\right)$ to leading order. Now, we average $p^0$ between $\pm|q^0|/2$ before replacing the step function with $\delta(p^0) q^0$; this is equivalent to making the approximation
\begin{eqnarray}
\sgn(q^0)\: \theta(|q^0|/2-|p^0|) = 2\frac{\delta(p^0)}{1!}\: \frac{q^0}{2} + 2 \frac{\partial_{p^0}^2 \delta(p^0)}{3!} \left(\frac{q^0}{2}\right)^3 + \cdots,
\end{eqnarray}
where the second term is needed to take into account the $(p^0)^2$ from $\Im\Sigma_R$. This leads to \eqref{Decay_parametrization}, and in particular, verifies $\gamma'=(3/2)\gamma$.

We have computed the order $q$ contributions to $\Im\Pi_R$. Is there an associated part in $\Re\Pi$ through the Kramers-Kronig dispersion relation \eqref{dispersion_relation}? For $\Im\Sigma_R$, and hence the $(D_1)_R$ contribution to $\Im\Pi_R$, it is known that the associated real part amounts to a correction to $Z(\p)$; but $Z(\p)$ itself appeared in our cutting rule to start with, so this just means we must use the self-consistent, i.e. physical, value of $Z(\p)$. As long as we have done so, there is no further contribution to $\Re\Pi$ from $(D_1)_R$. From Ward identity we know $(C^{ph})_R$ and $(C^{pp})_R$ must have no further contribution to $\Re\Pi$ either. This differs from the scenario in the zeroth order contribution, where $\Delta'$ in \eqref{Delta_prime} has both real and imaginary parts.

Finally we want to show collisions are ``uninteresting" towards our main focus of this paper -- the collision term $C_1$ has no contribution to the antisymmetric part $\Pi^{[\mu\nu]}$ of the current-current correlation, and therefore has no contribution to the anomalous Hall effect or the chiral magnetic effect. First, we note that $C_1$ (despite the $1$ subscript, in $d=2$ it also involves terms of order $q\ln q$) has a special property
\begin{eqnarray}
C_1(p, k; q)=C_1(k, p; q).
\label{C1_pk_symm}
\end{eqnarray}
In particular, the $D_1$ term in $C_1$ has this property simply because it is proportional to $\delta^{d+1}(p-k)$. On the other hand, inspecting the exact expressions \eqref{CphR1} and \eqref{CppR1}, we see $C^{ph}_1$ and $C^{pp}_1$ are symmetric under $p\leftrightarrow k$ up to the $q$-dependences in the $V$'s. But the $q$-dependences in the $V$'s can be neglected, for their effects would be further suppressed by order $q$, i.e. contribute to order $q^2$ (and order $q^2 \ln q$ in $d=2$) which are beyond order $q$. Therefore \eqref{C1_pk_symm} holds to order $q$ -- including in $d=2$, since the argument here did not rely on power counting. The collision contribution to $i(i\Im\Pi^{\mu\nu})$ is given by
\begin{eqnarray}
&& -\int_{p, k} i\Re\Gamma_0^{\mu}(p; -q) \ (-C_1(p, k; q)) \ i\Re\Gamma_0^{\nu}(k; q) \nonumber \\[.2cm]
&& \hspace{2cm} -\int_{p, k} i \left(i\Im\Gamma_0^{\mu} \right)(p; -q) \ (-C_1(p, k; q)) \ i\left(i\Im\Gamma_0^{\nu}\right)(k; q).
\end{eqnarray}
The $q$-dependence in $\Gamma_0^\mu$ comes from $\Delta'$ which is even in $q$, so $\Gamma_0^{\mu}$ is also even in $q$. Thus, due to \eqref{C1_pk_symm}, the expression above is symmetric in $\mu\nu$. Now that $C_1$ does not contribute to $\Im\Pi^{[\mu\nu]}$, by the Kramers-Kronig dispersion relation it has no associated contribution to $\Re\Pi^{[\mu\nu]}$ either, and thus our claim is proven.

\

\addcontentsline{toc}{section}{Appendix B}
\section*{Appendix B}

As we mentioned in Appendix A, the power counting of $q$ presented there has missing piece. Here it is: The power counting relied on the assumption that the internal momenta carried by the cut propagators are generically not close to each other. However, in the integration over the internal momenta, there must be some region where this assumption does not hold -- the internal momenta, restricted near the FS, can appear collinear with one another. To estimate the scale of the the contribution from the collinear regime, one views this regime as a quasi-one-dimensional system~\cite{baym1978physics}. Take the self-energy $\Sigma$ for example. In $d=1$ dimension, the self-energy $\Sigma\sim p^0 \ln p^0$ -- this leads to the well-known non-Fermi liquid behavior. In higher dimensions, quasi-one-dimensional power counting estimates the contribution from the collinear regime to be $\sim (p^0)^d \ln p^0$. For $d=2$ this is less suppressed than the usual $(p^0)^2$.

Denote the collinear regime contribution to $\Sigma$ as $\delta\Sigma$; the leading contribution has three intermediate collinear on-shell fermions. Based on the quasi-one-dimensional power counting, along with the general analyticity requirements that $-\Sigma_R(\omega, \p)$ is analytic in the $\omega$ upper-half-plane and $-\Im\Sigma_R>0$~\cite{Luttinger:1961zz, abrikosov1975methods}, it is tempting to parametrize the collinear regime contribution as
\begin{eqnarray}
-\delta\Sigma_R(\omega, \p) = ia(\p) \, \omega^2 \ln(-i\omega/p_F) + (\mbox{higher order contributions})
\end{eqnarray}
for small $\omega$ in the upper-half-plane. Here $p_F$ is the size scale of the FS, and $a$ is positive; the branch cut of $\ln$ is placed along the negative real axis. Taking $\omega=p^0+i\epsilon$ gives $-\delta\Sigma_R(p)$. Unfortunately, when $\xi(\p)$ is of the same order as $p^0$, such parametrization is wrong. It is known~\cite{chubukov2003nonanalytic,chubukov2005singular} that $-\delta\Sigma_R$ has complicated dependence on $p^0$ and $p^0-\xi(\p)$, but still scales as $(p^0)^2 \ln p^0$ when $\xi(\p)$ and $p^0$ are of the same order. 

Now that there is no simple way to parametrize $-\delta\Im\Sigma$, there is no simple way to parametrize the decay factor $-D_1$ (despite the $1$ subscript, here it also involves terms of order $q \ln q$), because the latter can be expressed in terms of the former, with $p^0$ being order $q^0$ and $p^0-\xi$ being order $v^\mu q_\mu$. Similarly, the collinear regime contributions to $-C^{ph}_1$ and $-C^{pp}_1$ are terms of order $q \ln q$ proportional to $\delta^{d+1}(p\mp k)$ (respectively corresponding to forward and back scattering); these terms have no simple parametrization either.

Although the collision term in $d=2$ is not parametrized, we know \eqref{C_D_relation} must still hold, as it is required by the Ward-Takahashi identity. Also, we argued that \eqref{C1_pk_symm}, which makes collisions ``uninteresting'', still holds in $d=2$.

\

\addcontentsline{toc}{section}{Appendix C}
\section*{Appendix C}

In this appendix we explicitly show from Feynman diagrams the second line of \eqref{EM_dipole_anom} is antisymmetric in $\mu\nu$. Consider diagrams in the $q$-2PI sum $i\wt{V}^{\alpha \ \gamma}_{\ \delta, \ \beta}(p, k; q)$. We can separate these diagrams into two types:
%(a subtlety about whether a diagram in $i\wt{V}$ always belongs to one of these two types is addressed at the end of this Appendix):
\begin{itemize}
\item
Type I: The $q$-2PI diagram has a fermion line at the top, running in with momentum and index $(p-q/2, \delta)$ and running out with $(k-q/2, \gamma)$, and a fermion line at the bottom, running in with momentum and index $(k+q/2, \beta)$ and running out with $(p+q/2, \alpha)$. Type I diagrams are summed in $\wt{V}$ with plus sign. Some examples are shown below.

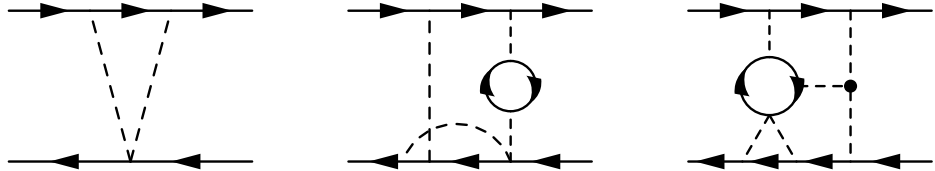
\begin{center}
\begin{fmffile}{zzz-A-TypeI-1}
\begin{fmfgraph*}(40, 20)
\fmfleftn{l}{2}\fmfrightn{r}{2}
\fmf{fermion}{r1,v1,l1}
\fmf{fermion}{l2,v2,v3,r2}
\fmf{dashes, tension=0}{v1,v2}
\fmf{dashes, tension=0}{v1,v3}
\end{fmfgraph*}
\end{fmffile}
\begin{fmffile}{zzz-A-TypeI-2}
\begin{fmfgraph*}(40, 20)
\fmfleftn{l}{2}\fmfrightn{r}{2}
\fmf{fermion}{l2,v3,v4,r2}
\fmf{fermion}{r1,v2,v1,l1}
\fmf{dashes,tension=0}{v1,v3}
\fmffreeze
\fmf{phantom}{v2,v7,v8,v4}
\fmf{dashes,tension=0}{v2,v7}
\fmf{dashes,tension=0}{v4,v8}
\fmf{fermion,left,tension=0}{v7,v8}
\fmf{fermion,left,tension=0}{v8,v7}
\fmffreeze
\fmf{phantom}{r1,v2,v1,v5,v6,l1}
\fmf{dashes, right=0.7, tension=0}{v2,v5}
\end{fmfgraph*}
\end{fmffile}
\begin{fmffile}{zzz-A-TypeI-3}
\begin{fmfgraph*}(40, 20)
\fmfleftn{l}{2}\fmfrightn{r}{2}
\fmf{fermion}{l2,v1,v2,r2}
\fmf{fermion}{r1,v3}
\fmf{fermion,tension=1.5}{v3,v4,v5,l1}
\fmffreeze
\fmf{phantom}{l1,v6,v3}
\fmffreeze
\fmf{phantom}{v1,v7,v9,v8,v6}
\fmf{fermion,left,tension=0.15,tag=1}{v7,v8}
\fmf{fermion,left,tension=0.15}{v8,v7}
\fmffreeze
\fmf{dashes,tension=1.5}{v2,v10,v3}
\fmf{dashes}{v1,v7}
\fmf{dashes}{v8,v4}
\fmf{dashes}{v8,v5}
\fmffreeze
\fmfposition
\fmfipath{p[]}
\fmfiset{p1}{vpath1(__v7,__v8)}
\fmfi{dashes}{point length(p1)/2 of p1 -- vloc(__v10)}
\fmfdot{v10}
\end{fmfgraph*}
\end{fmffile}
\end{center}

For Type I diagrams, we can assign momenta on the internal propagators so that $-q/2$ runs through the top fermion line, $+q/2$ runs through the bottom fermion line, and all other internal propagators are independent of $q$.

\item
Type II: The $q$-2PI diagram has a fermion line on the left, running in with momentum and index $(p-q/2, \delta)$ and running out with $(p+q/2, \alpha)$, and a fermion line on the right, running in with momentum and index $(k+q/2, \beta)$ and running out with $(k-q/2, \gamma)$. Type II diagrams are summed in $\wt{V}$ with minus sign, because of fermionic statistics. Some examples are shown below.

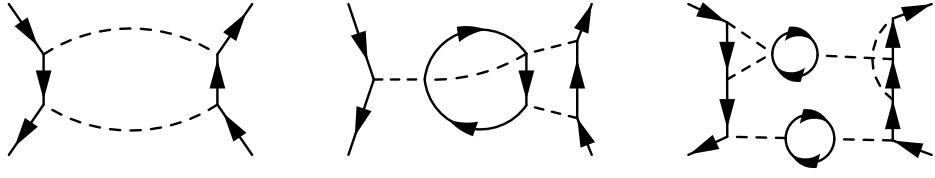
\begin{center}
\begin{fmffile}{zzz-A-TypeII-1}
\begin{fmfgraph*}(40, 20)
\fmfleftn{l}{2}\fmfrightn{r}{2}
\fmf{fermion,tension=2}{r1,v1,v2,r2}
\fmf{fermion,tension=2}{l2,v4,v3,l1}
\fmf{dashes,left=0.3,tension=0.4}{v1,v3}
\fmf{dashes,left=0.3,tension=0.4}{v4,v2}
\end{fmfgraph*}
\end{fmffile}
\begin{fmffile}{zzz-A-TypeII-2}
\begin{fmfgraph*}(40, 20)
\fmfleftn{l}{2}\fmfrightn{r}{2}
\fmf{fermion,tension=2}{l2,v1,l1}
\fmf{fermion,tension=1.5}{r1,v2,v3,r2}
\fmf{phantom,tension=2}{r1,v2}
\fmf{phantom,tension=2}{v3,r2}
\fmf{dashes,tension=2}{v1,v4}
\fmf{phantom}{v2,v5,v6,v4}
\fmf{phantom}{v3,v7,v8,v4}
\fmf{dashes,tension=0}{v2,v5}
\fmf{dashes,tension=0}{v3,v7}
\fmffreeze
\fmf{fermion,left=0.7}{v4,v7}
\fmf{fermion}{v7,v5}
\fmf{fermion,left=0.7}{v5,v4}
\fmf{dashes,left=0.15}{v7,v4}
\end{fmfgraph*}
\end{fmffile}
\begin{fmffile}{zzz-A-TypeII-3}
\begin{fmfgraph*}(40, 20)
\fmfleftn{l}{2}\fmfrightn{r}{2}
\fmf{fermion,tension=1}{r1,v1,v12,v8,v2,r2}
\fmf{fermion,tension=1}{l2,v4,v9,v3,l1}
\fmf{dashes,tension=2}{v1,v6}
\fmf{dashes,tension=2}{v7,v3}
\fmf{fermion,left,tension=1.2}{v6,v7}
\fmf{fermion,left,tension=1.2}{v7,v6}
\fmf{phantom,tension=0.7}{v2,v4}
\fmf{phantom,tension=2}{r1,v1}
\fmf{phantom,tension=2}{r2,v2}
\fmf{phantom,tension=2}{l1,v3}
\fmf{phantom,tension=2}{l2,v4}
\fmffreeze
\fmf{dashes,tension=1.2}{v8,v11}
\fmf{dashes,tension=1}{v9,v10}
\fmf{dashes,tension=1}{v4,v10}
\fmf{fermion,left,tension=1}{v11,v10}
\fmf{fermion,left,tension=1}{v10,v11}
\fmf{dashes,left=0.5,tension=0.4}{v12,v2}
\end{fmfgraph*}
\end{fmffile}
\end{center}

For Type II diagrams, however we assign the internal momenta, $q$ will in general appear in some propagator(s) on both the left and right fermion lines, as well as on some internal propagators in the middle.

\end{itemize}
We want to show $\int_k \left(i\wt{V}_1\right)^{\alpha \ \gamma}_{\ \delta, \ \beta}(p, k; q) \ \partial_k^\nu iG^\beta_{\ \gamma}(k)$ is equal to $q_\mu$ times a quantity antisymmetric in $\mu\nu$. We consider the Type I and Type II contributions separately.

For Type I diagrams in $i\wt{V}(p, k; q)$, when expanded to linear order in $q$, we pick one propagator $iG(l\pm q/2)$ on the bottom (top) fermion line and replace it with $(\pm q_\mu/2) \partial_l^\mu iG(l)$, and in all other propagators set $q$ to zero; we sum up all possible ways of such expansion, and sum up all possible Type I diagrams. (One may wonder why the interaction vertices are independent of $q$. We explain this at the end of this Appendix.) As a result, Type I contribution to $\int_k \left(i\wt{V}_1\right)^{\alpha \ \gamma}_{\ \delta, \ \beta}(p, k; q) \ \partial_k^\nu iG^\beta_{\ \gamma}(k)$ can be summarized as
\begin{eqnarray}
\frac{q_\mu}{2} \int_k \int_l \left(i\wt{Y}\right)^{\alpha \: \ \xi \: \ \gamma}_{\ \delta, \ \zeta, \ \beta}(p, l, k) \ \partial_l^\mu iG^\zeta_{\ \xi}(l) \ \partial_k^\nu iG^\beta_{\ \gamma}(k),
\label{tildeY}
\end{eqnarray}
where $i\wt{Y}$ is a sum of connected diagrams:
\begin{itemize}
\item
Diagrams satisfying the following conditions are summed in $i\wt{Y}$ with plus sign:

There is a fermion line running in with momentum and index $(p, \delta)$ and running out with $(k, \gamma)$, a fermion line running in with momentum and index $(k, \beta)$ and running out with $(l, \xi)$, and a fermion line running in with momentum and index $(l, \zeta)$ and running out with $(p, \alpha)$.

Moreover, among the internal fermion propagators on these three fermion lines, none of them is dictated by momentum conservation to have momentum $p, k$ or $l$, and no pair of them is dictated by momentum conservation to have same momenta. Equivalently, among those fermion propagators, one cannot cut any one or two of them to disconnect the diagram.
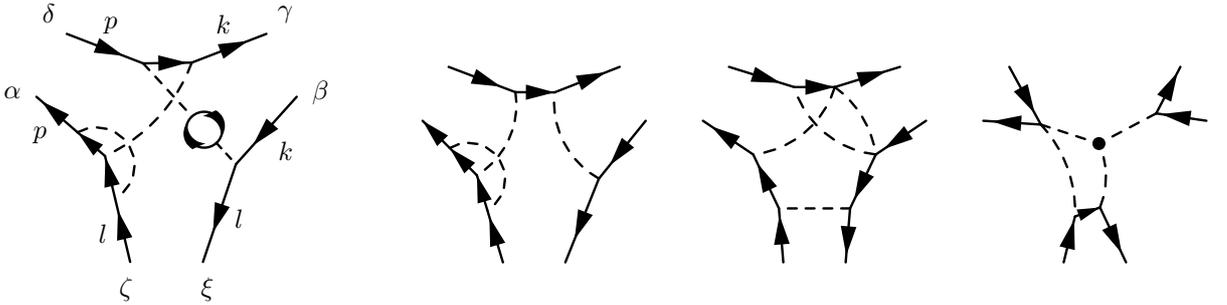
\vspace{.2cm}
\begin{center}
\begin{fmffile}{zzz-wtY}
\begin{fmfgraph*}(35, 35)
\fmfsurroundn{v}{24}
\fmflabel{$\alpha$}{v12}
\fmflabel{$\delta$}{v10}
\fmflabel{$\xi$}{v20}
\fmflabel{$\zeta$}{v18}
\fmflabel{$\gamma$}{v4}
\fmflabel{$\beta$}{v2}
\fmf{fermion,label.side=left,label=$k$}{pk2,v4}
\fmf{fermion,label.side=left,label=$p$}{v10,pk1}
\fmf{fermion}{pk1,pk2}
\fmf{fermion,label.side=left,label=$p$}{lp3,v12}
\fmf{fermion,label.side=left,label=$l$}{v18,lp1}
\fmf{fermion}{lp1,lp2}
\fmf{fermion}{lp2,lp3}
\fmf{fermion,label.side=left,label=$l$}{kl1,v20}
\fmf{fermion,label.side=left,label=$k$}{v2,kl1}
\fmf{dashes,left=0.2,tension=0.3}{pk2,lp2}
\fmf{phantom,left=0.3,tension=0.3}{kl1,pk1}
\fmf{dashes,left,tension=0}{lp3,lp1}
\fmf{phantom,left,tension=0.5}{lp1,lp2,lp3}
\fmffreeze
\fmf{dashes,tension=1}{pk1,o1}
\fmf{dashes,tension=2.5}{o2,kl1}
\fmf{fermion,left}{o1,o2}
\fmf{fermion,left}{o2,o1}
\end{fmfgraph*}
\end{fmffile}
\hspace{1cm}
\begin{fmffile}{zzz-not-wtY-1}
\begin{fmfgraph*}(30, 30)
\fmfsurroundn{v}{24}
\fmf{fermion}{pk2,v4}
\fmf{fermion}{v10,pk1}
\fmf{fermion,tension=2}{pk1,pk2}
\fmf{fermion}{lp3,v12}
\fmf{fermion}{v18,lp1}
\fmf{fermion}{lp1,lp2}
\fmf{fermion}{lp2,lp3}
\fmf{fermion}{kl1,v20}
\fmf{fermion}{v2,kl1}
\fmf{dashes,left=0.3,tension=0.3}{pk1,lp2}
\fmf{dashes,left=0.3,tension=0.3}{kl1,pk2}
\fmf{dashes,left,tension=0}{lp3,lp1}
\fmf{phantom,left,tension=0.5}{lp1,lp2}
\end{fmfgraph*}
\end{fmffile}
\hspace{0.1cm}
\begin{fmffile}{zzz-not-wtY-2}
\begin{fmfgraph*}(30, 30)
\fmfsurroundn{v}{24}
\fmf{fermion}{pk2,v4}
\fmf{fermion}{v10,pk1}
\fmf{fermion}{pk1,pk2}
\fmf{fermion}{lp2,v12}
\fmf{fermion}{v18,lp1}
\fmf{fermion}{lp1,lp2}
\fmf{fermion}{kl2,v20}
\fmf{fermion}{kl1,kl2}
\fmf{fermion}{v2,kl1}
\fmf{dashes,left=0.3,tension=0.3}{pk2,lp2}
\fmf{dashes,left=0.3,tension=0.3}{kl1,pk1}
\fmf{dashes,left=0.2,tension=0}{pk2,kl1}
\fmf{dashes,tension=0.3}{kl2,lp1}
\end{fmfgraph*}
\end{fmffile}
\hspace{0.1cm}
\begin{fmffile}{zzz-not-wtY-3}
\begin{fmfgraph*}(30, 30)
\fmfsurroundn{v}{24}
\fmf{fermion}{v2,kk,v4}
\fmf{fermion}{v10,pp,v12}
\fmf{fermion}{v18,ll1,ll2,v20}
\fmf{dashes,tension=1.3}{kk,o}
\fmf{dashes,tension=1.3}{pp,o}
\fmf{dashes,right=0.2}{ll2,o}
\fmfdot{o}
\fmf{dashes,left=0.2,tension=0.4}{pp,ll1}
\end{fmfgraph*}
\end{fmffile}
\end{center}
\vspace{.5cm}
For example, the diagram on the left contributes to $i\wt{Y}$, while the three on the right do not.

\item
Diagrams satisfying the conditions above, but with $(k, {}^\gamma_{\ \beta})$ and $(l, {}^\xi_{\ \zeta})$ switched, are summed in $i\wt{Y}$ with minus sign.
\end{itemize}
Notice that $i\wt{Y}$ is a totally antisymmetric 3-tensor in the double fermion linear space, i.e. it is antisymmetric under the exchange of any two of $(p, {}^\alpha_{\ \delta})$, $(k, {}^\gamma_{\ \beta})$ and $(l, {}^\xi_{\ \zeta})$. Thus, Type I contribution is antisymmetric in $\mu\nu$.

For Type II diagram contribution, we use a diagrammatic technique developed by Ward~\cite{Ward:1950xp}. Consider, for instance, the diagram below.
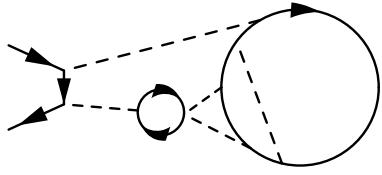
\begin{center}
\begin{fmffile}{zzz-A-prototype-no-symm}
\begin{fmfgraph*}(50, 20)
\fmfleftn{l}{5}\fmfrightn{r}{5}
\fmf{fermion,tension=5}{l4,v4,v2,l2}
\fmf{phantom}{v2,r1}
\fmf{phantom}{v4,r5}
\fmf{phantom}{v2,v3,v4}
\fmffreeze
\fmf{phantom}{v3,v6}
\fmf{phantom,tension=1}{r3,v6}
\fmffreeze
\fmf{fermion,left,tension=0,tag=1}{v6,r3}
\fmf{plain,left,tension=0,tag=2}{r3,v6}
\fmf{dashes,tension=2.5}{v7,v2}
\fmf{fermion,left,tension=2}{v7,v8}
\fmf{fermion,left,tension=2}{v8,v7}
\fmf{phantom,tension=1}{v8,r2}
\fmffreeze
\fmfipath{p[]}
\fmfiset{p1}{vpath1(__v6,__r3)}
\fmfiset{p2}{vpath2(__r3,__v6)}
\fmfi{dashes}{point length(p1)/3 of p1 .. vloc(__v4)}
\fmfi{dashes}{point 8length(p2)/8 of p2 .. vloc(__v8)}
\fmfi{dashes}{point 5length(p2)/7 of p2 .. vloc(__v8)}
\fmfi{dashes}{point 4length(p1)/7 of p2 .. point length(p1)/5 of p1}
\end{fmfgraph*}
\end{fmffile}
\end{center}
Let us call this a ``prototype diagram''. Let us call the loop on the right the $k$-loop, whose loop momentum is assigned $k$; it consists of five fermion propagators. A Type II diagram contribution to $\int_k i\wt{V}_1(p, k; q) \ \partial_k^\nu iG(k)$ is obtained by picking one propagator $iG$ on the $k$-loop -- in the example above there are five ways to do so -- and replacing it with $\partial_k^\nu iG$, and then letting the external momentum $q$ flow-in through it. The sum of all these five resulting diagrams forms a ``prototype class'' associated with the above prototype diagram. For each prototype class, we can fix one interaction propagator (it maybe an auxiliary propagator), through which $q$ in the $k$-loop flows out towards the left; for instance, in the example above, we may assign internal momenta so that $q$ always flows out the $k$-loop along the dashed line on the top. Now, for each Type II diagram in this prototype class, we expand $q$ at first order (since we are looking at $\wt{V}_1$). This corresponds to picking one internal propagator (fermion or interaction) that has $q$ in its argument, replacing it with $q_\mu$ times its momentum derivative, and then in all other internal propagators set $q$ to zero. Now:
\begin{itemize}
\item
If our picked $q$-dependent propagator is on the $k$-loop, then we have a $\partial_k^\nu iG(k)$ and a $\partial_k^\mu iG(k)$ on the $k$-loop, and summing up all such possibilities in the prototype class yields a quantity antisymmetric in $\mu\nu$, in a manner similar to the Type I diagram contribution.
\item
If our picked $q$-dependent propagator is not on the $k$-loop, then there is only one $\partial_k^\nu iG(k)$ on the $k$-loop, and summing up all such possibilities is equivalent to taking a total $k$-derivative on the $k$-loop (and $k$ is integrated over later). So the sum of such possibilities vanishes.
\end{itemize}
Thus, Type II diagram contribution to $\int_k (\partial_q^\mu i\wt{V}_1(p, k; q)) \ \partial_k^\nu iG(k)$ is also antisymmetric in $\mu\nu$. (There is a small caveat in the use of prototype diagrams. For example the prototype below
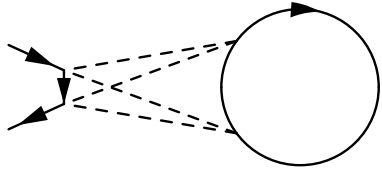
\begin{center}
\begin{fmffile}{zzz-A-prototype-symm}
\begin{fmfgraph*}(50, 20)
\fmfleftn{l}{5}\fmfrightn{r}{5}
\fmf{fermion,tension=5}{l4,v4,v2,l2}
\fmf{phantom}{v2,r1}
\fmf{phantom}{v4,r5}
\fmf{phantom}{v2,v3,v4}
\fmffreeze
\fmf{phantom}{v3,v6}
\fmf{phantom,tension=1}{r3,v6}
\fmffreeze
\fmf{fermion,left,tension=0,tag=1}{v6,r3}
\fmf{plain,left,tension=0,tag=2}{r3,v6}
\fmffreeze
\fmfposition
\fmfipath{p[]}
\fmfiset{p1}{vpath1(__v6,__r3)}
\fmfiset{p2}{vpath2(__r3,__v6)}
\fmfi{dashes}{point length(p1)/5 of p1 -- vloc(__v4)}
\fmfi{dashes}{point length(p1)/5 of p1 -- vloc(__v2)}
\fmfi{dashes}{point 4length(p1)/5 of p2 -- vloc(__v2)}
\fmfi{dashes}{point 4length(p1)/5 of p2 -- vloc(__v4)}
\end{fmfgraph*}
\end{fmffile}
\end{center}
has a symmetry of exchanging the two propagators on the $k$-loop. So when relating it to Type II contribution by replacing one $iG$ on the $k$-loop with $\partial_k^\nu iG$, we need an extra factor of $1/2$. Clearly this does not affect the final antisymmetry in $\mu\nu$.)

A left-over subtlety has to be addressed: When we were expanding the momentum running along a fermion line, we did not have contribution from the interaction vertices on the fermion line. Why is that? Recall our assumption (for simplicity, not for principle) about the QFT that any bare interaction vertex has no coupling to $A$, i.e. there is no e.g. $A\phi\psi^\dagger \psi$ bare vertex or $A\psi^\dagger \psi \psi^\dagger \psi$ bare vertex. By EM $U(1)$ gauge invariance, this also means the bare interaction vertices cannot depend on the momentum running along the charged fermion line. So a bare interaction vertex at most depends on the momentum running along the neutral interaction lines (which maybe auxiliary), for example in $(\psi^\dagger \psi) \: \partial_x^2 \phi$ or in $(\psi^\dagger \psi)^2\partial_x^2(\psi^\dagger \psi)$.

\

\addcontentsline{toc}{section}{Appendix D}
\section*{Appendix D}

In this appendix we show
\begin{eqnarray}
\sigma_Y^{\mu\nu\lambda} \equiv (i\partial_p^\mu G)^T \ \partial_q^\nu i\wt{V}_1(q) \ i\partial_k^\lambda G,
\end{eqnarray}
which appears in \eqref{sigma_Y_ref}, vanishes. According to Appendix C, we separate the contributions of Type I diagrams and Type II diagrams in $i\wt{V}_1$. In $\sigma_Y$, since $\partial iG$ is contracted on both sides, it is easy to see Type II contribution vanishes using Ward's method presented in Appendix C -- at least one of the $k$-loop and the $p$-loop involves a total derivative. We are left with Type I contribution to $\sigma_Y$. By \eqref{tildeY} we can express it as
\begin{eqnarray}
\sigma_Y^{\mu\nu\lambda} = \frac{1}{2} \int_k \int_l \int_p \left(i\wt{Y}\right)^{\alpha \: \ \xi \: \ \gamma}_{\ \delta, \ \zeta, \ \beta}(p, l, k) \ \partial_p^\mu iG^\delta_\alpha(p) \ \partial_l^\nu iG^\zeta_{\ \xi}(l) \ \partial_k^\lambda iG^\beta_{\ \gamma}(k).
\end{eqnarray}
We can describe diagrams in $\sigma_Y^{\mu\nu\lambda}$ as the following:
\begin{itemize}
\item
The diagram has a fermion loop, which we call the outer loop (formed by connecting the three fermion lines in $i\wt{Y}$ with the three differentiated propagators). We redefine $p$ so that it is now the loop momentum running around the outer loop. Interaction vertices separate the outer loop into $n\geq 3$ segments. Three of the segments are differentiated propagators $\partial_p^\mu iG$, $\partial_p^\nu iG$ and $\partial_p^\lambda iG$, the remaining $n-3$ segments are propagators $iG$.

Moreover, the interaction lines inside make the outer loop 2PI; that is, among all segments on the outer loop, no pair of them are dictated by momentum conservation to have the same momentum.

If $\partial_p^\mu$, $\partial_p^\nu$, $\partial_p^\lambda$ appear on the outer loop in the cyclic order against the fermion arrow, then the diagram is summed in $\sigma_Y^{\mu\nu\lambda}$ with coefficient $-1/2$ (the minus sign is because now we have an extra fermion loop -- the outer loop -- compared to $\wt{Y}$). If they appear on the outer loop in the cyclic order along the fermion arrow, then the diagram is summed with coefficient $+1/2$.
\end{itemize}
To show $\sigma_Y^{\mu\nu\lambda}=0$, below we introduce four notions.

First, let us be blind between $iG$ and $\partial_p iG$ on the outer loop. Then we are led to consider prototype diagrams like this one
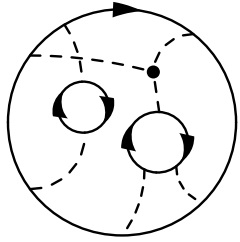
\begin{center}
\begin{fmffile}{zzz-B-prototype}
\begin{fmfgraph*}(30, 30)
\fmfleftn{l}{1}\fmfrightn{r}{1}
\fmf{fermion,left,tag=1}{l1,r1}
\fmf{plain,left,tag=2}{r1,l1}
\fmffreeze
\fmfsurroundn{v}{24}
\fmf{phantom}{v12,ml1,mr1,v3}
\fmfdot{mr1}
\fmf{phantom}{v14,ml2,mr2,v2}
\fmffreeze
\fmf{phantom}{mr2,mm,v20}
\fmf{fermion,left,tension=0}{ml2,ml1}
\fmf{fermion,left,tension=0}{ml1,ml2}
\fmf{fermion,left,tension=0,tag=3}{mr2,mm}
\fmf{fermion,left,tension=0,tag=4}{mm,mr2}
\fmffreeze
\fmfposition
\fmfipath{p[]}
\fmfiset{p1}{vpath1(__l1,__r1)}
\fmfiset{p2}{vpath2(__r1,__l1)}
\fmfiset{p3}{vpath3(__mr2,__mm)}
\fmfiset{p4}{vpath4(__mm,__mr2)}
\fmfi{dashes}{point 1length(p1)/5 of p1{right} .. vloc(__mr1)}
\fmfi{dashes}{point 5length(p1)/7 of p1{left} .. vloc(__mr1)}
\fmfi{dashes}{point 0length(p3)/7 of p3 -- vloc(__mr1)}
\fmfi{dashes}{point 1length(p1)/3 of p1.. {down}vloc(__ml1)}
\fmfi{dashes}{point 4length(p2)/5 of p2{right} .. vloc(__ml2)}
\fmfi{dashes}{point 1length(p4)/8 of p4{down} .. point 1length(p2)/2 of p2}
\fmfi{dashes}{point 7length(p3)/9 of p3{down} .. point 1length(p2)/4 of p2}
\end{fmfgraph*}
\end{fmffile}
.
\end{center}
A prototype diagram defines a prototype class: Diagrams contributing to $\sigma_Y$ are in the same prototype class if, after ignoring the distinction between $\partial_p iG$ and $iG$ on the outer loop, they reduce to the same prototype diagram. In fact, the sum $S^{\mu\nu\lambda}$ (we drop the $\mu\nu\lambda$ indices from here on) of diagrams (with coefficients $\pm 1/2$ assigned as before) within a prototype class vanishes, as we will show later. This leads to $\sigma_Y=0$, because clearly diagrams in $\sigma_Y$ are partitioned into prototype classes.

Second, let us fix a prototype class, and consider the placement of the three $\partial_p iG$'s on the outer loop. For simplicity, in the below we will restrict to prototype classes with no symmetry factor (there will be a symmetry factor of $1/n_s$ if the prototype diagram has a $\mathbb{Z}_{n_s}$ cyclic symmetry with respect to the outer loop, where $n_s$ divides $n$); we will return to the case with symmetry factor later. Now consider for example the diagram
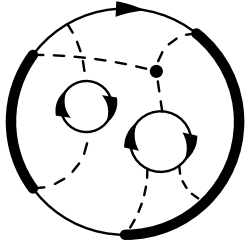
\begin{center}
\begin{fmffile}{zzz-B-prototype-antisymmetrization}
\begin{fmfgraph*}(30, 30)
\fmfleftn{l}{1}\fmfrightn{r}{1}
\fmf{fermion,left,tag=1}{l1,r1}
\fmf{plain,left,tag=2}{r1,l1}
\fmffreeze
\fmfsurroundn{v}{24}
\fmf{phantom}{v12,ml1,mr1,v3}
\fmfdot{mr1}
\fmf{phantom}{v14,ml2,mr2,v2}
\fmffreeze
\fmf{phantom}{mr2,mm,v20}
\fmf{fermion,left,tension=0}{ml2,ml1}
\fmf{fermion,left,tension=0}{ml1,ml2}
\fmf{fermion,left,tension=0,tag=3}{mr2,mm}
\fmf{fermion,left,tension=0,tag=4}{mm,mr2}
\fmffreeze
\fmfposition
\fmfipath{p[]}
\fmfiset{p1}{vpath1(__l1,__r1)}
\fmfiset{p2}{vpath2(__r1,__l1)}
\fmfiset{p3}{vpath3(__mr2,__mm)}
\fmfiset{p4}{vpath4(__mm,__mr2)}
\fmfi{dashes}{point 1length(p1)/5 of p1{right} .. vloc(__mr1)}
\fmfi{dashes}{point 5length(p1)/7 of p1{left} .. vloc(__mr1)}
\fmfi{dashes}{point 0length(p3)/7 of p3 -- vloc(__mr1)}
\fmfi{dashes}{point 1length(p1)/3 of p1.. {down}vloc(__ml1)}
\fmfi{dashes}{point 4length(p2)/5 of p2{right} .. vloc(__ml2)}
\fmfi{dashes}{point 1length(p4)/8 of p4{down} .. point 1length(p2)/2 of p2}
\fmfi{dashes}{point 7length(p3)/9 of p3{down} .. point 1length(p2)/4 of p2}
\fmffreeze
\fmfi{plain,width=4}{subpath (5length(p1)/7,length(p1)) of p1}
\fmfi{plain,width=4}{subpath (0length(p2)/4,1length(p2)/4) of p2}
\fmfi{plain,width=4}{subpath (1length(p2)/4,1length(p2)/2) of p2}
\fmfi{plain,width=4}{subpath (4length(p2)/5,5length(p2)/5) of p2}
\fmfi{plain,width=4}{subpath (0length(p1)/5,1length(p1)/5) of p1}
\end{fmfgraph*}
\end{fmffile}
.
\end{center}
This diagram represents the sum (with coefficients $\pm 1/2$ assigned as before) of all diagrams in the given prototype class such that the three $\partial_p iG$'s appear on the three thickened segments. This sum is manifestly antisymmetric in $\mu\nu\lambda$. Let us call the set of diagrams contributing to such sum an ``antisymmetrization class''. Obviously a prototype class can be partitioned into antisymmetrization classes. The purpose of introducing this notion is only for introducing the next notion.

The third notion to introduce is a partitioning finer than a prototype class (still, we restrict to those without symmetry factor) but coarser than an antisymmetrization class. In an antisymmetrization class, the three $\partial_p iG$'s are separated by a number of $iG$'s, for example, in the previous antisymmetrization diagram, the three $\partial_p iG$'s are separated by $0, 1$ and $2$ $iG$'s. But there are other antisymmetrization diagrams whose three $\partial_p iG$'s are also separated by $0, 1$ and $2$ $iG$'s. Let us introduce the notation $(012)$, which represents the sum of them:
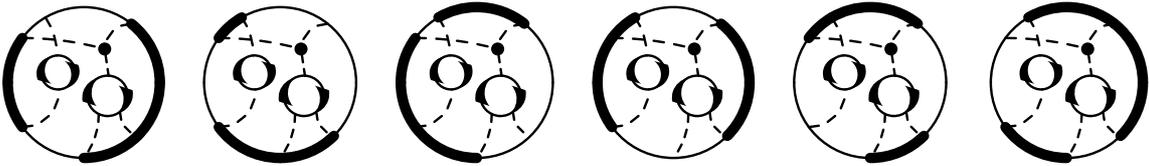
\begin{center}
\begin{fmffile}{zzz-B-prototype-separation-1}
\begin{fmfgraph*}(20, 20)
\fmfleftn{l}{1}\fmfrightn{r}{1}
\fmf{plain,left,tag=1}{l1,r1}
\fmf{plain,left,tag=2}{r1,l1}
\fmffreeze
\fmfsurroundn{v}{24}
\fmf{phantom}{v12,ml1,mr1,v3}
\fmfdot{mr1}
\fmf{phantom}{v14,ml2,mr2,v2}
\fmffreeze
\fmf{phantom}{mr2,mm,v20}
\fmf{fermion,left,tension=0}{ml2,ml1}
\fmf{fermion,left,tension=0}{ml1,ml2}
\fmf{fermion,left,tension=0,tag=3}{mr2,mm}
\fmf{fermion,left,tension=0,tag=4}{mm,mr2}
\fmffreeze
\fmfposition
\fmfipath{p[]}
\fmfiset{p1}{vpath1(__l1,__r1)}
\fmfiset{p2}{vpath2(__r1,__l1)}
\fmfiset{p3}{vpath3(__mr2,__mm)}
\fmfiset{p4}{vpath4(__mm,__mr2)}
\fmfi{dashes}{point 1length(p1)/5 of p1{right} .. vloc(__mr1)}
\fmfi{dashes}{point 5length(p1)/7 of p1{left} .. vloc(__mr1)}
\fmfi{dashes}{point 0length(p1)/7 of p3 -- vloc(__mr1)}
\fmfi{dashes}{point 1length(p1)/3 of p1.. {down}vloc(__ml1)}
\fmfi{dashes}{point 4length(p1)/5 of p2{right} .. vloc(__ml2)}
\fmfi{dashes}{point 1length(p1)/8 of p4{down} .. point 1length(p1)/2 of p2}
\fmfi{dashes}{point 7length(p1)/9 of p3{down} .. point 1length(p1)/4 of p2}
\fmffreeze
\fmfi{plain,width=4}{subpath (5length(p1)/7,length(p1)) of p1}\fmfi{plain,width=4}{subpath (0length(p2)/4,1length(p2)/4) of p2}
\fmfi{plain,width=4}{subpath (1length(p2)/4,1length(p2)/2) of p2}
%\fmfi{plain,width=4}{subpath (1length(p2)/2,4length(p2)/5) of p2}
\fmfi{plain,width=4}{subpath (4length(p2)/5,5length(p2)/5) of p2}\fmfi{plain,width=4}{subpath (0length(p1)/5,1length(p1)/5) of p1}
%\fmfi{plain,width=4}{subpath (1length(p1)/5,1length(p1)/3) of p1}
%\fmfi{plain,width=4}{subpath (1length(p1)/3,5length(p1)/7) of p1}
\end{fmfgraph*}
\end{fmffile}
\hspace{0cm}
\begin{fmffile}{zzz-B-prototype-separation-2}
\begin{fmfgraph*}(20, 20)
\fmfleftn{l}{1}\fmfrightn{r}{1}
\fmf{plain,left,tag=1}{l1,r1}
\fmf{plain,left,tag=2}{r1,l1}
\fmffreeze
\fmfsurroundn{v}{24}
\fmf{phantom}{v12,ml1,mr1,v3}
\fmfdot{mr1}
\fmf{phantom}{v14,ml2,mr2,v2}
\fmffreeze
\fmf{phantom}{mr2,mm,v20}
\fmf{fermion,left,tension=0}{ml2,ml1}
\fmf{fermion,left,tension=0}{ml1,ml2}
\fmf{fermion,left,tension=0,tag=3}{mr2,mm}
\fmf{fermion,left,tension=0,tag=4}{mm,mr2}
\fmffreeze
\fmfposition
\fmfipath{p[]}
\fmfiset{p1}{vpath1(__l1,__r1)}
\fmfiset{p2}{vpath2(__r1,__l1)}
\fmfiset{p3}{vpath3(__mr2,__mm)}
\fmfiset{p4}{vpath4(__mm,__mr2)}
\fmfi{dashes}{point 1length(p1)/5 of p1{right} .. vloc(__mr1)}
\fmfi{dashes}{point 5length(p1)/7 of p1{left} .. vloc(__mr1)}
\fmfi{dashes}{point 0length(p1)/7 of p3 -- vloc(__mr1)}
\fmfi{dashes}{point 1length(p1)/3 of p1.. {down}vloc(__ml1)}
\fmfi{dashes}{point 4length(p1)/5 of p2{right} .. vloc(__ml2)}
\fmfi{dashes}{point 1length(p1)/8 of p4{down} .. point 1length(p1)/2 of p2}
\fmfi{dashes}{point 7length(p1)/9 of p3{down} .. point 1length(p1)/4 of p2}
\fmffreeze
%\fmfi{plain,width=4}{subpath (5length(p1)/7,length(p1)) of p1}\fmfi{plain,width=4}{subpath (0length(p2)/4,1length(p2)/4) of p2}
\fmfi{plain,width=4}{subpath (1length(p2)/4,1length(p2)/2) of p2}
\fmfi{plain,width=4}{subpath (1length(p2)/2,4length(p2)/5) of p2}
%\fmfi{plain,width=4}{subpath (4length(p2)/5,5length(p2)/5) of p2}\fmfi{plain,width=4}{subpath (0length(p1)/5,1length(p1)/5) of p1}
\fmfi{plain,width=4}{subpath (1length(p1)/5,1length(p1)/3) of p1}
%\fmfi{plain,width=4}{subpath (1length(p1)/3,5length(p1)/7) of p1}
\end{fmfgraph*}
\end{fmffile}
\hspace{0cm}
\begin{fmffile}{zzz-B-prototype-separation-3}
\begin{fmfgraph*}(20, 20)
\fmfleftn{l}{1}\fmfrightn{r}{1}
\fmf{plain,left,tag=1}{l1,r1}
\fmf{plain,left,tag=2}{r1,l1}
\fmffreeze
\fmfsurroundn{v}{24}
\fmf{phantom}{v12,ml1,mr1,v3}
\fmfdot{mr1}
\fmf{phantom}{v14,ml2,mr2,v2}
\fmffreeze
\fmf{phantom}{mr2,mm,v20}
\fmf{fermion,left,tension=0}{ml2,ml1}
\fmf{fermion,left,tension=0}{ml1,ml2}
\fmf{fermion,left,tension=0,tag=3}{mr2,mm}
\fmf{fermion,left,tension=0,tag=4}{mm,mr2}
\fmffreeze
\fmfposition
\fmfipath{p[]}
\fmfiset{p1}{vpath1(__l1,__r1)}
\fmfiset{p2}{vpath2(__r1,__l1)}
\fmfiset{p3}{vpath3(__mr2,__mm)}
\fmfiset{p4}{vpath4(__mm,__mr2)}
\fmfi{dashes}{point 1length(p1)/5 of p1{right} .. vloc(__mr1)}
\fmfi{dashes}{point 5length(p1)/7 of p1{left} .. vloc(__mr1)}
\fmfi{dashes}{point 0length(p1)/7 of p3 -- vloc(__mr1)}
\fmfi{dashes}{point 1length(p1)/3 of p1.. {down}vloc(__ml1)}
\fmfi{dashes}{point 4length(p1)/5 of p2{right} .. vloc(__ml2)}
\fmfi{dashes}{point 1length(p1)/8 of p4{down} .. point 1length(p1)/2 of p2}
\fmfi{dashes}{point 7length(p1)/9 of p3{down} .. point 1length(p1)/4 of p2}
\fmffreeze
%\fmfi{plain,width=4}{subpath (5length(p1)/7,length(p1)) of p1}\fmfi{plain,width=4}{subpath (0length(p2)/4,1length(p2)/4) of p2}
%\fmfi{plain,width=4}{subpath (1length(p2)/4,1length(p2)/2) of p2}
\fmfi{plain,width=4}{subpath (1length(p2)/2,4length(p2)/5) of p2}
\fmfi{plain,width=4}{subpath (4length(p2)/5,5length(p2)/5) of p2}\fmfi{plain,width=4}{subpath (0length(p1)/5,1length(p1)/5) of p1}
%\fmfi{plain,width=4}{subpath (1length(p1)/5,1length(p1)/3) of p1}
\fmfi{plain,width=4}{subpath (1length(p1)/3,5length(p1)/7) of p1}
\end{fmfgraph*}
\end{fmffile}
\hspace{0cm}
\begin{fmffile}{zzz-B-prototype-separation-4}
\begin{fmfgraph*}(20, 20)
\fmfleftn{l}{1}\fmfrightn{r}{1}
\fmf{plain,left,tag=1}{l1,r1}
\fmf{plain,left,tag=2}{r1,l1}
\fmffreeze
\fmfsurroundn{v}{24}
\fmf{phantom}{v12,ml1,mr1,v3}
\fmfdot{mr1}
\fmf{phantom}{v14,ml2,mr2,v2}
\fmffreeze
\fmf{phantom}{mr2,mm,v20}
\fmf{fermion,left,tension=0}{ml2,ml1}
\fmf{fermion,left,tension=0}{ml1,ml2}
\fmf{fermion,left,tension=0,tag=3}{mr2,mm}
\fmf{fermion,left,tension=0,tag=4}{mm,mr2}
\fmffreeze
\fmfposition
\fmfipath{p[]}
\fmfiset{p1}{vpath1(__l1,__r1)}
\fmfiset{p2}{vpath2(__r1,__l1)}
\fmfiset{p3}{vpath3(__mr2,__mm)}
\fmfiset{p4}{vpath4(__mm,__mr2)}
\fmfi{dashes}{point 1length(p1)/5 of p1{right} .. vloc(__mr1)}
\fmfi{dashes}{point 5length(p1)/7 of p1{left} .. vloc(__mr1)}
\fmfi{dashes}{point 0length(p1)/7 of p3 -- vloc(__mr1)}
\fmfi{dashes}{point 1length(p1)/3 of p1.. {down}vloc(__ml1)}
\fmfi{dashes}{point 4length(p1)/5 of p2{right} .. vloc(__ml2)}
\fmfi{dashes}{point 1length(p1)/8 of p4{down} .. point 1length(p1)/2 of p2}
\fmfi{dashes}{point 7length(p1)/9 of p3{down} .. point 1length(p1)/4 of p2}
\fmffreeze
\fmfi{plain,width=4}{subpath (5length(p1)/7,length(p1)) of p1}\fmfi{plain,width=4}{subpath (0length(p2)/4,1length(p2)/4) of p2}
%\fmfi{plain,width=4}{subpath (1length(p2)/4,1length(p2)/2) of p2}
%\fmfi{plain,width=4}{subpath (1length(p2)/2,4length(p2)/5) of p2}
\fmfi{plain,width=4}{subpath (4length(p2)/5,5length(p2)/5) of p2}\fmfi{plain,width=4}{subpath (0length(p1)/5,1length(p1)/5) of p1}
\fmfi{plain,width=4}{subpath (1length(p1)/5,1length(p1)/3) of p1}
%\fmfi{plain,width=4}{subpath (1length(p1)/3,5length(p1)/7) of p1}
\end{fmfgraph*}
\end{fmffile}
\hspace{0cm}
\begin{fmffile}{zzz-B-prototype-separation-5}
\begin{fmfgraph*}(20, 20)
\fmfleftn{l}{1}\fmfrightn{r}{1}
\fmf{plain,left,tag=1}{l1,r1}
\fmf{plain,left,tag=2}{r1,l1}
\fmffreeze
\fmfsurroundn{v}{24}
\fmf{phantom}{v12,ml1,mr1,v3}
\fmfdot{mr1}
\fmf{phantom}{v14,ml2,mr2,v2}
\fmffreeze
\fmf{phantom}{mr2,mm,v20}
\fmf{fermion,left,tension=0}{ml2,ml1}
\fmf{fermion,left,tension=0}{ml1,ml2}
\fmf{fermion,left,tension=0,tag=3}{mr2,mm}
\fmf{fermion,left,tension=0,tag=4}{mm,mr2}
\fmffreeze
\fmfposition
\fmfipath{p[]}
\fmfiset{p1}{vpath1(__l1,__r1)}
\fmfiset{p2}{vpath2(__r1,__l1)}
\fmfiset{p3}{vpath3(__mr2,__mm)}
\fmfiset{p4}{vpath4(__mm,__mr2)}
\fmfi{dashes}{point 1length(p1)/5 of p1{right} .. vloc(__mr1)}
\fmfi{dashes}{point 5length(p1)/7 of p1{left} .. vloc(__mr1)}
\fmfi{dashes}{point 0length(p1)/7 of p3 -- vloc(__mr1)}
\fmfi{dashes}{point 1length(p1)/3 of p1.. {down}vloc(__ml1)}
\fmfi{dashes}{point 4length(p1)/5 of p2{right} .. vloc(__ml2)}
\fmfi{dashes}{point 1length(p1)/8 of p4{down} .. point 1length(p1)/2 of p2}
\fmfi{dashes}{point 7length(p1)/9 of p3{down} .. point 1length(p1)/4 of p2}
\fmffreeze
%\fmfi{plain,width=4}{subpath (5length(p1)/7,length(p1)) of p1}\fmfi{plain,width=4}{subpath (0length(p2)/4,1length(p2)/4) of p2}
\fmfi{plain,width=4}{subpath (1length(p2)/4,1length(p2)/2) of p2}
%\fmfi{plain,width=4}{subpath (1length(p2)/2,4length(p2)/5) of p2}
%\fmfi{plain,width=4}{subpath (4length(p2)/5,5length(p2)/5) of p2}\fmfi{plain,width=4}{subpath (0length(p1)/5,1length(p1)/5) of p1}
\fmfi{plain,width=4}{subpath (1length(p1)/5,1length(p1)/3) of p1}
\fmfi{plain,width=4}{subpath (1length(p1)/3,5length(p1)/7) of p1}
\end{fmfgraph*}
\end{fmffile}
\hspace{0cm}
\begin{fmffile}{zzz-B-prototype-separation-6}
\begin{fmfgraph*}(20, 20)
\fmfleftn{l}{1}\fmfrightn{r}{1}
\fmf{plain,left,tag=1}{l1,r1}
\fmf{plain,left,tag=2}{r1,l1}
\fmffreeze
\fmfsurroundn{v}{24}
\fmf{phantom}{v12,ml1,mr1,v3}
\fmfdot{mr1}
\fmf{phantom}{v14,ml2,mr2,v2}
\fmffreeze
\fmf{phantom}{mr2,mm,v20}
\fmf{fermion,left,tension=0}{ml2,ml1}
\fmf{fermion,left,tension=0}{ml1,ml2}
\fmf{fermion,left,tension=0,tag=3}{mr2,mm}
\fmf{fermion,left,tension=0,tag=4}{mm,mr2}
\fmffreeze
\fmfposition
\fmfipath{p[]}
\fmfiset{p1}{vpath1(__l1,__r1)}
\fmfiset{p2}{vpath2(__r1,__l1)}
\fmfiset{p3}{vpath3(__mr2,__mm)}
\fmfiset{p4}{vpath4(__mm,__mr2)}
\fmfi{dashes}{point 1length(p1)/5 of p1{right} .. vloc(__mr1)}
\fmfi{dashes}{point 5length(p1)/7 of p1{left} .. vloc(__mr1)}
\fmfi{dashes}{point 0length(p1)/7 of p3 -- vloc(__mr1)}
\fmfi{dashes}{point 1length(p1)/3 of p1.. {down}vloc(__ml1)}
\fmfi{dashes}{point 4length(p1)/5 of p2{right} .. vloc(__ml2)}
\fmfi{dashes}{point 1length(p1)/8 of p4{down} .. point 1length(p1)/2 of p2}
\fmfi{dashes}{point 7length(p1)/9 of p3{down} .. point 1length(p1)/4 of p2}
\fmffreeze
\fmfi{plain,width=4}{subpath (5length(p1)/7,length(p1)) of p1}\fmfi{plain,width=4}{subpath (0length(p2)/4,1length(p2)/4) of p2}
%\fmfi{plain,width=4}{subpath (1length(p2)/4,1length(p2)/2) of p2}
\fmfi{plain,width=4}{subpath (1length(p2)/2,4length(p2)/5) of p2}
%\fmfi{plain,width=4}{subpath (4length(p2)/5,5length(p2)/5) of p2}\fmfi{plain,width=4}{subpath (0length(p1)/5,1length(p1)/5) of p1}
%\fmfi{plain,width=4}{subpath (1length(p1)/5,1length(p1)/3) of p1}
\fmfi{plain,width=4}{subpath (1length(p1)/3,5length(p1)/7) of p1}
\end{fmfgraph*}
\end{fmffile}
.
\end{center}
In general, we call the set of diagrams contributing to $(abc)$ a ``cyclic class''. The name is because $(abc)$ is by definition the same object as $(bca)$ and $(cab)$. Clearly, a prototype class is partitioned into cyclic classes, and a cyclic class is partitioned into antisymmetrization classes. It is easy to see the sum $S$ of diagrams in the prototype class can be expressed as
\begin{eqnarray}
S=\sum'_{a+b+c=n-3} n \cdot s_{abc} \cdot (abc), \ \ \ \ \ \ \ \ s_{abc}=\left\{ \begin{array}{cl} 1/3 \ \ & \mbox{ if } a=b=c \\ 1 \ \ & \mbox{ otherwise} \end{array} \right.
\label{B_S_and_abc}
\end{eqnarray}
where $a,b,c$ are non-negative integers, and the prime on the sum means we only count $(abc)$, $(bca)$ and $(cab)$ once because they are the same object. We will see soon that by introducing the notion of cyclic class, we boil the Feynman diagram cancellation problem to a combinatorial problem.

In general $(abc)$, the sum of diagrams in a cyclic class, is not equal to zero. We need to introduce the fourth notion that bridges between prototype class and cyclic class; but unlike the three notions above, this fourth notion is not a partitioning. Given the prototype class, consider the following diagrams with two $\partial_p iG$'s: along the fermion arrow on the outer loop, we have $\partial_p^\mu iG$, then $m$ $iG$'s, then $\partial_p^\nu iG$, and then the remaining $(n-2-m)$ $iG$'s. We assume $m\leq n-2-m$. Sum up all such diagrams. But before we integrate over the outer loop momentum $p$, we take a total $\partial_p^\lambda$ derivative. Then we totally antisymmetrize between $\mu\nu\lambda$ and multiply by $3$. We denote the result by $\langle m \rangle$. By construction, $\langle m \rangle=0$ due to the total derivative. But on the other hand, it is easy to see $\langle m \rangle$ is a sum of cyclic classes:
\begin{eqnarray}
&& \hspace{4cm} \langle m \rangle = \sum'_{a+b+c=n-3} s^m_{abc} \cdot (abc), \label{B_m_and_abc} \\[.2cm]
&& s^m_{abc}=\left\{ \begin{array}{ll} 1 \ \ & \mbox{ if only one of $a, b, c$ is $m$, and the other two do not sum up to $m-1$} \\ 1+1=2 \ \ & \mbox{ if two of $a, b, c$ are $m$, the other is not $m$} \\ 1 \ \ & \mbox{ if $a=b=c=m$} \\ -1 \ \ & \mbox{ if two of $a, b, c$ sum up to $m-1$, and the other is not $m$} \\ 1-1=0 & \mbox{ if one of $a, b, c$ is $m$, and the other two sum up to $m-1$} \\ 0 \ \ & \mbox{ otherwise}. \end{array} \right. \nonumber 
\end{eqnarray}
We used the fact that the vertices on the outer loop are independent of $p$, whose reason is explained at the end of Appendix C. Diagrams contributing to a fixed $\langle m \rangle$ do not form an equivalence class, because clearly a given $(abc)$ can appear in several different $\langle m \rangle$'s.

Now that every $\langle m \rangle$ is equal to $0$, it would be desirable to express $S$ as a linear combination of $\langle m \rangle$'s. Indeed, now we shall show that
\begin{eqnarray}
S = \sum_{m=0}^{m\leq (n-2)/2} \left(n-2-2m\right) \cdot \langle m \rangle = 0.
\label{B_S_and_m}
\end{eqnarray}
This is a simple combinatorial problem that can be shown by matching the coefficient of each $(abc)$ on both sides, using equations \eqref{B_S_and_abc} and \eqref{B_m_and_abc}. Due to the cyclic property of $(abc)$, we can assume $a\leq b$, $a\leq c$. Then we discuss over 8 possibilities:
\begin{enumerate}

\item $a<b<c<(n-2)/2$: On the right-hand-side, $(abc)$ appears in three different $\langle m \rangle$'s:
\begin{itemize}
\item $m=a$: The coefficient of $(abc)$ is $(n-2-2a)\cdot 1$.
\item $m=b$: The coefficient of $(abc)$ is $(n-2-2b)\cdot 1$.
\item $m=c$: The coefficient of $(abc)$ is $(n-2-2c)\cdot 1$.
\end{itemize}
The sum of these coefficients is $n$, matches with the coefficient on the left-hand-side.

The case $a<c<b<(n-2)/2$ works in the same manner.

\item $a<b<c=(n-2)/2$: On the right-hand-side, $(abc)$ appears in three different $\langle m \rangle$'s:
\begin{itemize}
\item $m=a$: The coefficient of $(abc)$ is $(n-2-2a)\cdot 1$.
\item $m=b$: The coefficient of $(abc)$ is $(n-2-2b)\cdot 1$.
\item $m=c$: The coefficient of $(abc)$ is $0$.
\end{itemize}
The sum of these coefficients is $n$, matches with the coefficient on the left-hand-side.

The case $a<c<b=(n-2)/2$ works in the same manner.

\item $a<b<(n-2)/2<c$: On the right-hand-side, $(abc)$ appears in three different $\langle m \rangle$'s:
\begin{itemize}
\item $m=a$: The coefficient of $(abc)$ is $(n-2-2a)\cdot 1$.
\item $m=b$: The coefficient of $(abc)$ is $(n-2-2b)\cdot 1$.
\item $m-1=a+b$: The coefficient of $(abc)$ is $(n-2-2(a+b+1))\cdot(-1)$.
\end{itemize}
The sum of these coefficients is $n$, matches with the coefficient on the left-hand-side.

The case $a<c<(n-2)/2<b$ works in the same manner.

\item $a=b<c<(n-2)/2$: On the right-hand-side, $(abc)$ appears in two different $\langle m \rangle$'s:
\begin{itemize}
\item $m=a=b$: The coefficient of $(abc)$ is $(n-2-a-b)\cdot 2$.
\item $m=c$: The coefficient of $(abc)$ is $(n-2-2c)\cdot 1$.
\end{itemize}
The sum of these coefficients is $n$, matches with the coefficient on the left-hand-side.

The case $a=c<b<(n-2)/2$ works in the same manner.

\item $a=b<c=(n-2)/2$: On the right-hand-side, $(abc)$ appears in two different $\langle m \rangle$'s:
\begin{itemize}
\item $m=a=b$: The coefficient of $(abc)$ is $(n-2-a-b)\cdot 2$.
\item $m=c$: The coefficient of $(abc)$ is $0$.
\end{itemize}
The sum of these coefficients is $n$, matches with the coefficient on the left-hand-side.

The case $a=c<b=(n-2)/2$ works in the same manner.

\item $a=b<(n-2)/2<c$: On the right-hand-side, $(abc)$ appears in two different $\langle m \rangle$'s:
\begin{itemize}
\item $m=a=b$: The coefficient of $(abc)$ is $(n-2-a-b)\cdot 2$.
\item $m-1=a+b$: The coefficient of $(abc)$ is $(n-2-2(a+b+1))\cdot(-1)$.
\end{itemize}
The sum of these coefficients is $n$, matches with the coefficient on the left-hand-side.

The case $a=c<(n-2)/2<b$ works in the same manner.

\item $a<b=c<(n-2)/2$: On the right-hand-side, $(abc)$ appears in two different $\langle m \rangle$'s:
\begin{itemize}
\item $m=a$: The coefficient of $(abc)$ is $(n-2-2a)\cdot 1$.
\item $m=b=c$: The coefficient of $(abc)$ is $(n-2-b-c)\cdot 1$.
\end{itemize}
The sum of these coefficients is $n$, matches with the coefficient on the left-hand-side.

\item $a=b=c$: On the right-hand-side, $(abc)$ appears only when $m=a=b=c$, with coefficient $(n-2-2(n-3)/3)\cdot 1 = n/3$. This matches with the coefficient on the left-hand-side.

\end{enumerate}
This completes our proof, for prototype classes that have no symmetry factor.

For a prototype class whose prototype diagram has a $\mathbb{Z}_{n_s}$ symmetry with respect to the outer loop, we can pick one segment on the outer loop to be ``the special segment''; in the sum of diagrams, this leads to over-counting by a factor of $n_s$. But now that there is no $\mathbb{Z}_{n_s}$ symmetry any more, we can show the sum is zero as before. The factor of $n_s$ has no effect on the zero.

\section*{References}
\bibliographystyle{elsarticle-num}
\bibliography{BerryFL_QFT}

\end{document}